\documentclass[prc,twocolumn,showpreprintnumbers,floatfix,nofootinbib]{revtex4}
\usepackage{amsmath,amssymb,graphicx,bm}
\usepackage{bbm}
\usepackage{hyperref}
\usepackage{color,ulem}
\usepackage{slashed}
\allowdisplaybreaks[4] 
\newcommand{\vect}[1]{\bm{\mathrm{#1}}}
\newcommand{\rv}{\vect{r}}
\newcommand{\Rv}{\vect{R}}
\newcommand{\kv}{\vect{k}}
\newcommand{\eff}{\text{eff}}
\newcommand{\SNM}{\text{SNM}}
\newcommand{\high}{\text{high}}
\newcommand{\sym}{\text{sym}}
\newcommand{\sat}{\text{sat}}
\definecolor{BLUE}{rgb}{0,0,1}
\definecolor{RED}{rgb}{1,0,0}
\definecolor{VIOLET}{rgb}{1,0,1}

\newcommand{\ncdot}{\!\cdot\!}

\renewcommand{\emph}[1]{\textit{#1}}
\newcommand{\includedmass}{m}
\newcommand{\mnucleon}{m_N}
\newcommand{\lap}{\Delta}

\begin{document}

\title{New Skyrme parametrizations to describe finite nuclei and neutron star matter with realistic effective masses} 

\author{Mingya Duan}
\affiliation{Universit\'e Paris-Saclay, CNRS/IN2P3, IJCLab, 91405 Orsay, France}
\author{Michael Urban} 
\email{michael.urban@ijclab.in2p3.fr}
\affiliation{Universit\'e Paris-Saclay, CNRS/IN2P3, IJCLab, 91405 Orsay, France}

\begin{abstract}
The phenomenological Skyrme energy density functional theory is one of the most popular theories for dealing with finite nuclei and infinite nuclear matter, including neutron star matter. However, the density dependence of the effective masses of common Skyrme parametrizations is completely different from the one found in microscopic calculations. This can have drastic consequences. For instance, in a recent study we reported that in many Skyrme functionals, the neutron Fermi velocity exceeds the speed of light at densities that exist in neutron-star cores. To solve this problem, we try to construct new Skyrme parametrizations by including constraints from microscopic calculations of the effective mass in addition to binding energies and charge radii of finite nuclei and different microscopic equations of state of pure neutron matter. We give the parameters of the new Skyrme forces and show that our new effective interactions can successfully describe properties of finite nuclei and nuclear matter (including pure neutron matter, symmetric nuclear matter, and neutron star matter).
\end{abstract}

\maketitle

\section{Introduction}\label{sec:introduction}
Neutron stars are well-known as compact astrophysical objects, the second most compact objects after black holes. They are believed to be the remnants of supernova explosions at the end stage of the massive star ($8-30 M_{\odot}$, $M_{\odot}$ is the solar mass) evolution \citep{Cerda-DuranElias-Rosa2018}. They are indispensable objects of study in nuclear physics and astrophysics. They provide scientists with a natural laboratory for the study of microphysics and astrophysics due to their extreme properties, such as high density, low temperature (stable neutron stars), strong magnetic field, and so on. Besides, their extreme physical conditions cannot be produced using ground-based facilities. Many astrophysical phenomena, such as gamma-ray bursts (GRBs), gravitational waves (GWs), fast radio bursts (FRBs), and so on, are related to neutron stars \citep{Kumar2015,Abbott2017,Zhang2020}. Neutron stars may also be cosmic ray sources \citep{Shklovsky1967}. Therefore, the study of neutron star properties is particularly important for multi-messenger astronomy.

The interior matter of a proto-neutron star is dense and hot. The neutron star will rapidly cool down after its birth. The cooling is dominantly produced by neutrino emission from the neutron star interior during about $10^5$ years after its birth \citep{Yakovlev2004}, in which the neutrino emission is controlled by the state of neutron star matter. The equation of state (EoS) of neutron star matter describes how the pressure $P$ of matter depends on its density $\rho$ and temperature $T$. It is also a crucial physical input for determining neutron star structure. But it is not constrained well so far. In the theoretical description, the selected nucleon-nucleon interaction determines the EoS of a neutron star.

Neutron star matter is composed mainly but not exclusively of neutrons. Except in the crust, it is assumed to be homogeneous, and since its density and composition vary only on length scales that are very large compared to nuclear scales, it can be treated as infinite asymmetric nuclear matter.

There are many models for addressing nuclear matter systems, including non-relativistic and relativistic methods. These methods can be categorized as microscopic methods and phenomenological models. Especially, the phenomenological Skyrme energy density functional theory was in the early stage introduced to study nuclear structure \citep{Vautherin1972,Beiner1975,Kohler1976}, and has later been extended to study infinite nuclear matter systems such as neutron stars \citep{Chabanat1997}. Skyrme interactions are easy to use because they are zero-range interactions. The first Skyrme interaction widely used to describe neutron stars is SLy4, constructed by Chabanat et al. \cite{Chabanat1998}. In 2010, Goriely et al. also developed Skyrme interactions BSk19, BSk20, and BSk21 \citep{Goriely2010} for astrophysical applications, followed by a series of BSk interactions up to BSk32 \citep{Goriely2013,Goriely2016}. A comparison of 240 different Skyrme parametrizations with nuclear-matter constraints can be found in Ref.~\cite{Dutra2012}.

However, many common Skyrme forces have effective masses that are at variance with those of microscopic Brueckner-Hartree-Fock (BHF) \cite{Baldo2014} and relativistic Dirac-BHF \cite{vanDalen2005,Wang2023} calculations. The concept of nucleon effective mass in medium was first proposed by Brueckner in 1955 \citep{Brueckner1955}. Nucleon effective masses are crucial in describing finite nuclei and compact astrophysical objects. There are many different definitions of effective masses \cite{Jaminon1989}. The relevant effective mass in the present context is the Landau effective mass (not to be confused with, e.g., the Dirac effective mass of relativistic models), which is defined by \cite{vanDalen2005,Baldo2014,Jaminon1989}
\begin{equation}
  m^* = \hbar^2k_F \Big(\frac{d\epsilon_k}{dk}\Big|_{k_F}\Big)^{-1}\,,
\end{equation}
where $\epsilon_k$ is the energy of a quasiparticle with momentum $k$, and $k_F$ is the Fermi momentum.

\begin{figure}
\begin{center}
\includegraphics[scale=0.54]{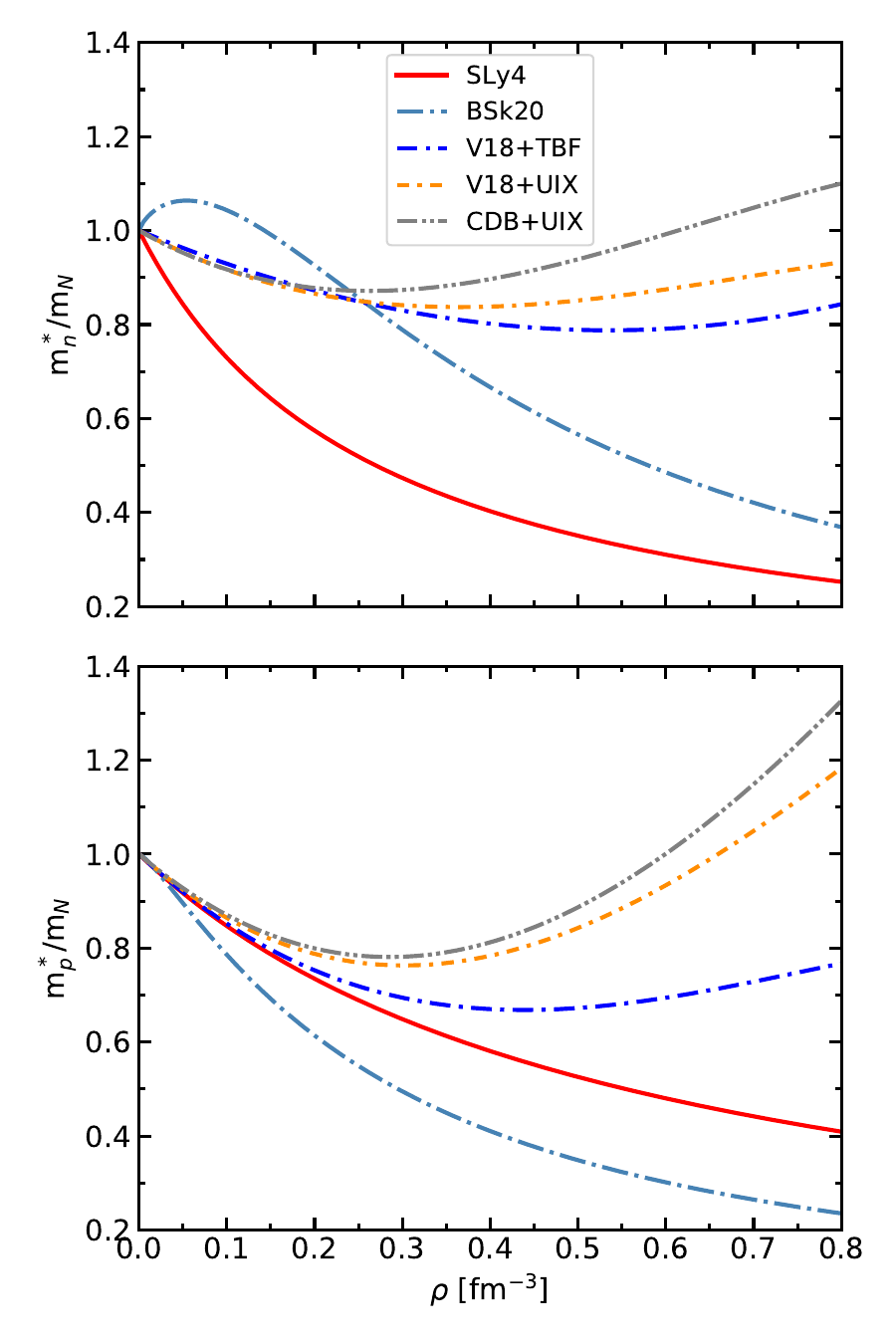} 
\caption{Nucleon effective masses computed using Skyrme interaction SLy4, the modified Skyrme interaction BSk20, and microscopic results V18+TBF, V18+UIX, and CDB+UIX of Ref. \cite{Baldo2014}, as functions of density in asymmetric nuclear matter with proton fraction $Y_p=0.1$. Upper panel: neutron effective mass; lower panel: proton effective mass.}
\label{fig:neutron-proton-effective-masses}
\end{center}
\end{figure}

In Fig.~\ref{fig:neutron-proton-effective-masses} we show different results for the neutron and proton effective masses (divided by the nucleon mass in vacuum, $m_N$) as functions of density in asymmetric nuclear matter with proton fraction $Y_p=0.1$. SLy4 and BSk20 are examples for traditional and modified Skyrme interactions, respectively. The other three curves (V18+TBF, V18+UIX, and CDB+UIX) come from microscopic BHF calculations \citep{Baldo2014} using different nucleon-nucleon interactions as input.
The SLy4 interaction predicts that the effective masses decrease with increasing density and become very small at large density. The neutron effective mass obtained using the BSk20 interaction initially increases but then it also strongly decreases at large densities. But both the neutron and proton effective masses computed using the microscopic method decrease first, then they increase with increasing density.

The effective mass is important because it determines the density of states at the Fermi surface and the Fermi velocity, $v_F = \hbar^{-1} (d\epsilon_k/dk)_{k_F} = \hbar k_F/m^*$. Our previous study \citep{Duan2023} reported that the SLy and BSk Skyrme interactions present the unphysical feature that the Fermi velocity of neutrons $v_{F,n}$ exceeds the speed of light $c$ already at relatively low densities, although they were widely used in study of astrophysical objects. The problem is directly related to the drastic decrease of the effective mass with increasing density in these functionals, while the microscopic effective masses stay large enough to give a nucleon Fermi velocity lower than the speed of light at all densities. The fact that the nucleon Fermi velocity exceeds the speed of light can of course only happen because the Skyrme theory is a non-relativistic theory. But we stress that not only in relativistic \cite{vanDalen2005}, but also in non-relativistic \cite{Baldo2014} BHF calculations, the effective masses are in fact large enough to ensure $v_{F,n}<c$ at all relevant densities. 

We hoped that the KIDS functionals considered in \cite{Hutauruk2022} would be an alternative option to solve the problem after checking the nucleon Fermi velocity. The effective masses in these models are larger than in the SLy4 or BSk functionals, but they do not follow the behavior of the BHF ones either.

Therefore, we suggest in this work two extended Skyrme parametrizations whose effective masses are fitted to the BHF results, similarly to  the Barcelona-Catania-Paris-Madrid (BCPM) functional \cite{Baldo2017}. Moreover, our new functionals are required to describe not only finite nuclei but also neutron star matter, and we therefore fit also the EoS of pure neutron matter to microscopic theories \cite{Akmal1998,Liu2022,Palaniappan2023}.

In this work, we first describe and motivate the selected form of the extended Skyrme functional in Sec. \ref{sec:Skyrme effective interactions}.  The fitting strategy and the parameters of the new Skyrme functionals follow in Sec. \ref{sec:new-Skyrme-forces}. Then we show the properties of finite nuclei, pure neutron matter (PNM), symmetric nuclear matter (SNM), and neutron star matter described by the new Skyrme forces in Sec. \ref{sec:results-all}. Finally, we summarize and conclude in Sec. \ref{sec:conclusion}.

\section{Skyrme force and energy functional} 
\label{sec:Skyrme effective interactions}
Skyrme interactions started with the original work of Skyrme \citep{Skyrme1956,Skyrme1958}. Since then, many Skyrme-like interactions were constructed to describe finite nuclei and infinite nuclear matter systems (such as neutron stars). Over the time, different forms of the Skyrme interaction have been given. The main difference of those forms is the number of the parameters of the corresponding terms \citep{Vautherin1972,Beiner1975,Kohler1976,Giannoni1980,Farine1997,Chabanat1997,Cochet2004,Chamel2009,Gil2019}. While the word Skyrme force is still in use, it has become more common to talk about Skyrme energy-density functionals \cite{Bender2003}. Since Skyrme forces contain at most second derivatives, i.e., second order in momenta, the contribution of higher powers in momenta, as well as many other effects (three-body force, short-range correlations, etc.) have to be simulated by making the coupling constants density dependent.

Standard and extended Skyrme interactions have often been written as density dependent two-body interactions. Then, the interaction used in this work would take the following form
\begin{equation}\label{eq:skyrme interaction-BSk}
\begin{aligned}
V(\rv)  = &  (t_0+y_0 P_{\sigma}) \delta (\rv) \\
&  + \frac{1}{2}(t_1+y_1 P_{\sigma})[\kv^{\prime\,2} \delta (\rv) + \delta (\rv) \kv^2] \\
&  +(t_2+y_2 P_{\sigma})\kv^{\prime} \cdot \delta (\rv) \kv\\
&  + \frac{1}{6} \sum_{i=1}^{3} (t_{3i}+y_{3i} P_{\sigma})[ \rho(\Rv)]^{\alpha_i} \delta (\rv) \\
&  + \frac{1}{2}(t_4+y_4 P_{\sigma})[\kv^{\prime\,2} [\rho(\Rv)]^{\beta} \delta (\rv) + \delta (\rv) [\rho(\Rv)]^{\beta} \kv^2] \\
&  +(t_5+y_5 P_{\sigma})\kv^{\prime} \cdot [\rho(\Rv)]^{\gamma} \delta (\rv) \kv\\
&  + iW_{0} \vect{\sigma} \cdot [\kv^{\prime} \times \delta(\rv) \kv],
\end{aligned}
\end{equation}
where $\kv=(\nabla_i-\nabla_j)/2i$ and $\kv^{\prime}=-(\nabla^{\prime}_{i} - \nabla^{\prime}_{j})/2i$, $\rv=\rv^{}_1-\rv^{}_2$ is the relative coordinate, $\Rv=\frac{1}{2} (\rv^{}_1 + \rv^{}_2)$, and $P_{\sigma}=(1+\vect{\sigma}_1 \cdot \vect{\sigma}_2)/2$ is the spin-exchange operator. The $t_4$ and $t_5$ terms, that add a density dependence to the standard $t_1$ and $t_2$ terms, were first introduced in the forces of the BSk family \citep{Chamel2009} to avoid the high-density ferromagnetic instability of neutron stars, and were later on also adopted by other groups, e.g., \cite{Zhang2016}. Furthermore, as in the KIDS interactions introduced in Ref. \cite{Gil2019}, the exponent $\alpha$ in the density-dependent contact term is not fixed as in the standard Skyrme interactions. There are different $\alpha$ values for the different $t_{3i}$ terms. We take $\alpha_{i} = \tfrac{i}{3}$ as in \cite{Gil2019}. We can see that Eq.~(\ref{eq:skyrme interaction-BSk}) employs not only the additional $t_{3i}$, $t_4$, and $t_5$ terms but also the combinations $y_i=t_ix_i$ \cite{Gil2019}, which are particularly convenient in cases such as BSk19-21 which set $t_2 = 0$ but $t_2x_2\neq 0$.

Within the above interaction, in the absence of currents and spin polarization, the energy density functional can be expressed as
\begin{equation}\label{eq:energy-density-functional-all}
\mathcal{H} = 
  \mathcal{K} 
  + \mathcal{H}_0
  + \mathcal{H}_3
  + \mathcal{H}_{\eff} 
  + \mathcal{H}_{\text{fin}} 
  + \mathcal{H}_{\text{SO}} 
  + \mathcal{H}_{\text{sg}}
  + \mathcal{H}_{\text{Coul}},
\end{equation}
where $\mathcal{K} = \frac{\hbar^2}{2\mnucleon} \tau$ is the kinetic-energy term, $\mathcal{H}_0$ is the zero-range term, $\mathcal{H}_3$ is the density-dependent term, $\mathcal{H}_{\eff}$ is the effective-mass term, $\mathcal{H}_{\text{fin}}$ is the finite-range term, $\mathcal{H}_{\text{SO}}$ is the spin-orbit term, $\mathcal{H}_{\text{sg}}$ is the term due to the tensor coupling with spin and gradient, and $\mathcal{H}_{\text{Coul}}$ is the energy density due to the Coulomb interaction:
\begin{equation}\label{eq:energy-density-functional-H03}
  \mathcal{H}_0 + \mathcal{H}_3 = C^\rho_0 \rho^2 + C^\rho_1(\rho_n-\rho_p)^2,
\end{equation}
\begin{equation}\label{eq:energy-density-functional-Heff}
 \mathcal{H}_{\eff} = C^\tau_0 \rho\tau 
   + C^\tau_1(\rho_n-\rho_p)(\tau_n-\tau_p),
\end{equation}
\begin{align}\label{eq:energy-density-functional-Hfin}
\mathcal{H}_{\text{fin}} =\;& C^{\nabla\rho}_0(\nabla\rho)^2 + C^{\nabla\rho}_1 [\nabla(\rho_n-\rho_p)]^2\nonumber \\
  &+ C^{\lap\rho}_1(\rho_n-\rho_p)\lap(\rho_n-\rho_p),
\end{align}
\begin{equation}\label{eq:energy-density-functional-HSO}
\mathcal{H}_{\text{SO}} = \frac{1}{4}W_0\, [3\,\vect{J}\ncdot\nabla\rho + (\vect{J}_n-\vect{J}_p)\ncdot\nabla(\rho_n-\rho_p)],
\end{equation}
\begin{equation}\label{eq:energy-density-functional-Hsg}
 \mathcal{H}_{\text{sg}} = -C^{sT}_0 \mathbb{J}^2-C^{sT}_1 (\mathbb{J}_n-\mathbb{J}_p)^2,
\end{equation}
\begin{align}\label{eq:energy-density-functional-Hcoul}
\mathcal{H}_{\text{Coul}} & =  \mathcal{H}_{\text{Coul}}^{\text{dir}}+ \mathcal{H}_{\text{Coul}}^{\text{exch}} \notag \\
&= \frac{1}{2} e^2 \rho_{p}(r) \int\! \frac{\rho_p(r^{\prime})d^3 r^{\prime}}{\vert \rv-\rv^{\prime} \vert} 
 - \frac{3}{4} \Big(\frac{3}{\pi} \Big)^{\!1/3} e^2 [\rho_p(r)]^{4/3} .
\end{align}
Here we used the common notations \cite{Bender2003} for the kinetic energy densities $\tau_q$ ($q=n$ for neutrons, and $q=p$ for protons), total number density $\rho=\rho_p+\rho_n$, kinetic-energy density $\tau=\tau_p+\tau_n$, spin-current tensor $\mathbb{J}=\mathbb{J}_p+\mathbb{J}_n$, and vector spin current density \cite{Lesinski2007} $\vect{J} = \vect{J}_p+\vect{J}_n$. The Coulomb exchange contribution $\mathcal{H}_{\text{Coul}}^{\text{exch}}$ is treated in Slater approximation.

In terms of the parameters of the force, the density dependent coefficients that appear in Eqs.~(\ref{eq:energy-density-functional-H03})-(\ref{eq:energy-density-functional-Hsg}) are given by
\begin{subequations}
\label{eq:Ccoef}
\begin{align}
\label{eq:Crho0}
C_0^{\rho}=& \frac{3}{8} t_0 +\sum_{i=1}^{3} \frac{1}{16} t_{3i} \rho^{\alpha_i},\\
C_1^{\rho}=& -\frac{1}{8} (t_0+2y_0) - \sum_{i=1}^3 \frac{1}{48} (t_{3i} +2y_{3i}) \rho^{\alpha_i}, \label{eq:Crho1}\\
C_0^{\tau}=& \frac{1}{16}[3t_1 + (5t_2+4y_2) + 3t_4 \rho^{\beta} + (5t_5 +4y_5) \rho^{\gamma}],\label{eq:Ctau0}\\
C_1^{\tau}=& \frac{1}{16} [-(t_1 +2y_1) +(t_2 +2y_2)\notag \\
&  - (t_4 +2y_4) \rho^{\beta} +(t_5 +2y_5) \rho^{\gamma}],\label{eq:Ctau1}\\
C^{\nabla\rho}_0 =\;&\frac{1}{64}[9t_1-(5t_2+4y_2) \nonumber \\
       &+(9+6\beta) t_4\rho^\beta-(5t_5+4y_5)\rho^\gamma],\label{eq:Cgrad0}\\
C^{\nabla\rho}_1 =\;&-\frac{1}{64}[3(t_1+2y_1)+(t_2+2y_2) \nonumber\\
       &+(t_4+2y_4)\rho^\beta + (t_5+2y_5)\rho^\gamma],\label{eq:Cgrad1}\\
C^{\lap\rho}_1 =\;& \frac{1}{32}(t_4+2y_4)\rho^\beta \label{eq:CLaplace1}\\
C_0^{sT}=& \frac{\eta_J}{16} [ -(t_1-2y_1) + (t_2 +2y_2) \label{eq:CsT0}\notag \\
& - (t_4-2y_4) \rho^{\beta} + (t_5 +2y_5) \rho^{\gamma}],\\
C_1^{sT}=& \frac{\eta_J}{16}(-t_1 + t_2 - t_4 \rho^{\beta} + t_5 \rho^{\gamma}).\label{eq:CsT1}
\end{align}
\end{subequations}

However, the two-body interaction in Eq. (\ref{eq:skyrme interaction-BSk}) should not be taken too seriously because it leads to a couple of terms which are afterwards not taken into account in the energy functional. For instance, Eq. (\ref{eq:skyrme interaction-BSk}) would imply that one should take $\eta_J=1$ in Eqs. (\ref{eq:CsT0}) and (\ref{eq:CsT1}). But in practice, it is quite common \cite{Bender2003} to neglect the $\vect{s}\cdot\vect{T}$ (in the notation of Ref. \cite{Bender2003}) and $\mathbb{J}^2$ terms in the functional, i.e., among others, the contributions of the $t_1$, $t_2$, $t_4$ and $t_5$ terms to the spin-orbit field. This choice, corresponding to $\eta_J=0$, was made, e.g., in SLy4 and in the BSk parametrizations from BSk19 onwards, and we will make it here, too.

Let us briefly motivate why we use this form with so many parameters. There are actually two different types of parameters. On the one hand, there are the exponents $\alpha_i$, $\beta$, and $\gamma$, which will be fixed from the beginning, as described in the next paragraph, and on the other hand, there are the $t_i$, $y_i$, and $W_0$, which are the 17 freely adjustable parameters of our functional.

First, notice that without the $t_4$ and $t_5$ terms, the effective masses would be strictly decreasing functions of density. If we want to reproduce the behavior of the effective masses as found in \cite{Baldo2014}, namely a decrease at low density followed by an increase at high density, additional density dependence in $\mathcal{H}_{\eff}$ is needed, which can be provided by the $t_4$ and $t_5$ terms. Next, notice that in most recent Skyrme parametrizations such as SLy or BSk, the effective mass term $\mathcal{H}_{\eff}$ is the one that determines the stiffness of the EoS at high density, because it grows like $\rho^{8/3}$ (or $\rho^{8/3+\beta}$ or $\rho^{8/3+\gamma}$ in the presence of $t_4$ and $t_5$), while the density dependent term $\mathcal{H}_{3}$ is only proportional to $\rho^{2+\alpha}$, and $\alpha < \tfrac{2}{3}$ in these parametrizations. So, if one requires that the EoS be stiff enough to support a $2M_\odot$ neutron star, $\mathcal{H}_{\eff}$ must be large enough, which in turn implies that the effective mass becomes too small at high density (resulting in a Fermi velocity that exceeds the speed of light, as mentioned in the Introduction). To overcome this problem, one would need $\alpha \geq \max(\tfrac{2}{3},\tfrac{2}{3}+\beta,\tfrac{2}{3}+\gamma)$. However, this leads to problems around saturation density, and this is the reason why in the KIDS functional the three $t_{3i}$ terms with three different exponents $\alpha_i = \tfrac{i}{3}$ were introduced \cite{Gil2019}, providing enough flexibility to adjust the EoS in the whole range of densities. (A different choice for the exponents $\alpha_i$ was recently proposed in Ref. \cite{Wang2024}.) As we do not wish to increase the number of parameters further, we choose $\beta = \gamma = \tfrac{1}{3}$ so that the $t_{33}$ with $\alpha_{33} = 1$ in $\mathcal{H}_3$ is sufficient to compensate the contribution of the $t_4$ and $t_5$ terms in $\mathcal{H}_{\eff}$ on the EoS. This choice of $\alpha_i$, $\beta$, and $\gamma$ corresponds to an expansion in powers of $k_F$ and therefore these exponents cannot be considered as free parameters.

Since the nucleon effective masses are independent of $t_{3i}$ terms, the expressions of the nucleon effective masses are given by Eq. (A10) of Ref. \cite{Chamel2009} after making the replacement $t_i x_i \rightarrow y_i$. Similarly, the nuclear mean field is given by Eq. (A11) of Ref. \cite{Chamel2009} if one replaces in addition the $t_3$ term there with the sum of three $t_{3i}$ terms.
Finally, the spin-orbit field is different from Eq. (A12) of Ref. \cite{Chamel2009} because, as mentioned before, we omit the contribution of the $t_1$, $t_2$, $t_4$ and $t_5$ terms ($\eta_J=0$). For completeness, the equations are given in Appendix \ref{app:hartreefock}.

\section{Fit of new Skyrme forces}
\label{sec:new-Skyrme-forces}
In this section, we describe how we determine the parameters of our new functionals.
To obtain realistic nucleon effective masses at high densities, we will determine some combinations of parameters by fitting the results from microscopic theory \citep{Baldo2014}. The other parameters are determined using microscopic calculations of the equation of state of uniform matter and properties of doubly magic nuclei.
\subsection{Determination of parameters related to infinite matter properties}\label{subsec:determination of parameters}
The effective mass $m^*_q$ [more precisely, $\hbar^2/(2m^*_q)$] of nucleon $q = n,p$ can be read off from the energy functional Eq.~\eqref{eq:energy-density-functional-all} by collecting the terms multiplying $\tau_q$, i.e.,
\begin{equation}\label{eq:m_star}
 \frac{\hbar^2}{2m^*_q} = \frac{\hbar^2}{2m_N} + C^\tau_0\rho+C^\tau_1 (2\rho_q-\rho)\,.
\end{equation}
The explicit general formula is given in Appendix \ref{app:hartreefock}. From Eq.~\eqref{eq:m_star}, one can get the isoscalar effective mass $m_s^*$ as the nucleon effective mass in symmetric nuclear matter (where the last term vanishes) and the isovector effective mass $m^*_v$ as the proton effective mass in pure neutron matter (or vice versa).
Notice that the three functions $1/m^*_n$, $1/m^*_p$, $1/m^*_s$ are related to each other by
$1/m^*_n+1/m^*_p = 2/m^*_s$, i.e., the inverse proton and neutron effective masses in asymmetric matter split symmetrically around the inverse effective mass in symmetric matter at the same density as noticed in \cite{Chamel2009}. But this relation is not fulfilled for the microscopic BHF effective masses. Therefore, we cannot expect to reproduce perfectly the BHF results even with our exended Skyrme parametrization.

After setting $\beta=\gamma=1/3$, the nucleon effective masses in asymmetric nuclear matter can be written as
\begin{equation}\label{eq:nucleon effective mass in ANM}
\frac{\hbar^2}{2m^*_q} =  \frac{\hbar^2}{2m_N} + (A_1+ B_1 \rho^{1/3})\rho\\
  + (A_2+B_2\rho^{1/3})\Big(\frac{\rho}{2}-\rho_q\Big),
\end{equation}
with
\begin{subequations}
\label{eq:A1-B2}
\begin{align}
A_1=&\frac{3}{16}t_1  +\frac{1}{4} \left (\frac{5}{4} t_2 +y_2 \right ),\\
B_1=&\frac{3}{16} t_4 + \frac{1}{4} \left (\frac{5}{4} t_5 +y_5 \right ),\\
A_2=& \frac{1}{8} [ (t_1+2y_1) -(t_2+2y_2)],\\
B_2=& \frac{1}{8} [(t_4+2y_4) - (t_5+2y_5)].
\end{align}
\end{subequations}
Then the values of $A_1$, $B_1$, $A_2$, and $B_2$ can be determined by fitting the neutron and proton effective masses for proton fractions $Y_p=0$, 0.1, 0.2, 0.3, 0.4, 0.5 of Ref. \cite{Baldo2014}. In the present work, we choose the results computed using V18+TBF. The resulting values of $A_1$, $B_1$, $A_2$ and $B_2$ are given in Table~\ref{table:A1-B1-A2-B2}, they are the same for both parametrizations shown in that table.

\begin{table}
\centering
\caption{Values of $A_1$, $B_1$, $A_2$, $B_2$, $C_1$, $C_2$, $C_3$, and $C_4$.} 
\label{table:A1-B1-A2-B2}
\begin{ruledtabular}
\begin{tabular}{lcc}
 & Sky3 & Sky4  \\
 \hline
$A_1$ (MeV fm$^5$)  & $   64.422$ & $   64.422$ \\ 
$B_1$ (MeV fm$^6$)  & $  -61.071$ & $  -61.071$ \\ 
$A_2$ (MeV fm$^5$)  & $   61.461$ & $   61.461$ \\ 
$B_2$ (MeV fm$^6$)  & $  -62.705$ & $  -62.705$ \\ 
$C_1$ (MeV fm$^3$)  & $ -563.808$ & $ -557.198$ \\ 
$C_2$ (MeV fm$^4$)  & $ 2089.140$ & $ 1760.903$ \\ 
$C_3$ (MeV fm$^5$)  & $-3970.980$ & $-2585.312$ \\ 
$C_4$ (MeV fm$^6$)  & $ 3000.670$ & $ 1597.951$ 
\end{tabular}
\end{ruledtabular}
\end{table}

Simultaneously, since in uniform matter $\tau_q=\frac{3}{5} (3\pi^2)^{2/3} \rho_q^{5/3}$, the energy per nucleon for pure neutron matter (neutron matter equation of state) can be written as (we specialize immediately to the case $\alpha_1=\beta=\gamma=\tfrac{1}{3}$, $\alpha_2=\tfrac{2}{3}$, and $\alpha_3=1$)
\begin{align}\label{eq:energy-per-nucleon-PNM}
\frac{E}{A}(\text{PNM})= & \frac{3\hbar^2}{10\mnucleon}(3\pi^2)^{2/3} \rho^{2/3} +C_1 \rho \notag \\
&  + C_2 \rho^{4/3} + C_3 \rho^{5/3} + C_4 \rho^2,
\end{align}
with 
\begin{subequations}
\label{eq:C1-C2-C3-C4s-PNM}
\begin{align}
 C_1= &\frac{1}{4} (t_0-y_0),\\
 C_2=& \frac{1}{24} (t_{31}-y_{31}),\\
 C_3=&\frac{3}{40} (3 \pi^2)^{2/3} (t_1 - y_1 + 3t_2 + 3y_2) + \frac{1}{24} (t_{32} - y_{32}),\\
 C_4=& \frac{1}{24} (t_{33} - y_{33}) + \frac{3}{40} (3 \pi^2)^{2/3} (t_4 - y_4 + 3 t_5+3 y_5).
\end{align}
\end{subequations}
Then the values of $C_1$, $C_2$, $C_3$, and $C_4$ can be determined by fitting the selected neutron matter equation of state. In this work, we choose the EoS computed using V18+TBF in \cite{Liu2022} (we will denote it by LZLWBS) and the EoS labeled ``A18+$\delta v$+UIX$^{*}$" in Ref.~\cite{Akmal1998} (we call it APR EoS here) to get two sets of parameters for the two new forces that we denote Sky3 and Sky4, respectively. The APR EoS is softer than the former at high densities. It is worth noting that data points of the LZLWBS EoS of Ref.~\cite{Liu2022} are only available at densities 0.06-0.90 fm$^{-3}$. To constrain the low-density EoS, we add some points provided by Ref. \cite{Palaniappan2023} at very low densities (0.01-0.03 fm$^{-3}$) to the LZLWBS EoS of \cite{Liu2022}. Notice that in both cases, Eq. (\ref{eq:energy-per-nucleon-PNM}) reproduces very precisely the respective EoS (see Fig. \ref{fig:EoS-PNM-SNM}), determining unambiguously the values of the $C_i$. The values of $C_1$, $C_2$, $C_3$, and $C_4$ can also be found in Table \ref{table:A1-B1-A2-B2}.

Concerning the EoS of symmetric matter, we do not constrain the saturation properties (saturation density $\rho_0$, energy per particle $E_0$, incompressibility $K_0$) from microscopic calculations because the microscopic predictions are usually quite bad, while these parameters have to be fine-tuned to reproduce the nuclear binding energies and radii. Therefore, these quantities will be precisely determined from the fit of finite nuclei described below. However, this is not true for the EoS at very high density. To fix the high-density EoS of symmetric matter, we impose that our EoS reproduces the microscopic calculation at the highest available density $\rho_{\high}$, i.e., $(\tfrac{E}{A})_{\SNM}(\rho_{\high}) = E_{\high}$. In the case of the LZLWBS EoS \cite{Liu2022}, we have $\rho_{\high} = 1$ fm$^{-3}$ and $E_{\high} = 291$ MeV, while in the case of APR EoS \cite{Akmal1998}, $\rho_{\high} = 0.96$ fm$^{-3}$ and $E_{\high} = 204.02$ MeV.

Altogether, after fixing $A_1$, $B_1$, $A_2$, $B_2$, $C_1\dots C_4$, and $E_{\high}$ to the  microscopic results, we are finally left with $N_p = 8$ free parameters to be determined by the fit of finite nuclei properties, less than in a standard Skyrme force (not counting $\alpha$, $\beta$ and $\gamma$ as free parameters).

\subsection{Determination of parameters related to properties of finite nuclei}
\label{sec:fit-finite-nuclei}
The new Skyrme forces should be suitable for computing finite nuclei. So it is necessary to fit the remaining free parameters to finite nuclei.

A major problem of this fit is the strong correlations among the different parameters. As we have already seen in the previous subsection, parameters always appear in certain combinations. In fact, it turns out to be helpful to pass from the set of 17 parameters $\{t_0,\allowbreak y_0,\dots, t_5,\allowbreak y_5,\allowbreak W_0\}$ to an equivalent one that consists of the quantities $\{A_1,\allowbreak B_1,\allowbreak A_2,\allowbreak B_2,\allowbreak C_1, \dots, C_4,\allowbreak E_{\high},\allowbreak \rho_0,\allowbreak E_0,\allowbreak K_0,\allowbreak C^{\nabla\rho}_{0\,\eff},\allowbreak C^{\nabla\rho}_{1\,\eff},\allowbreak W_0,\allowbreak C^{\nabla\rho}_{01},\allowbreak C^{\nabla\rho}_{11}\}$. Here, $C^{\nabla\rho}_{0\,\eff}$, $C^{\nabla\rho}_{1\,\eff}$, $C^{\nabla\rho}_{01}$, and $C^{\nabla\rho}_{11}$ are linear combinations of the parameters $t_1$, $y_1$, $t_2$, $y_2$, $t_4$, $y_4$, $t_5$, and $y_5$ that are defined such as to minimize their correlations with each other and with $A_1$, $B_1$, $A_2$, and $B_2$. Namely, $C^{\nabla\rho}_{0\,\eff}$, $C^{\nabla\rho}_{1\,\eff}$ are given by Eqs.~\eqref{eq:Cgrad0} and \eqref{eq:Cgrad1}, plus the contribution of $C^{\Delta\rho}_{11}$ after integration by parts [see Eqs.~\eqref{eq:effective gradient C}], taken at an empirical effective density at the surface of the nucleus where the gradient terms are most relevant, while $C^{\nabla\rho}_{01}$ and $C^{\nabla\rho}_{11}$ are the coefficients of the $\rho^{1/3}$ factor after setting $\beta=\gamma=\tfrac{1}{3}$ in Eqs.~\eqref{eq:Cgrad0} and \eqref{eq:Cgrad1}. More details are given in Appendix \ref{app:details-of-fits}.

For the fit of the unknown parameters $\rho_0$, $E_0$, \dots, $C^{\nabla\rho}_{11}$, we include experimental binding energies $E_b$ and charge radii $R_{\text{ch}}$ (if available) of the doubly closed shell nuclei listed in Table \ref{table:binding-energies-and-charge-radii}. Fitting the whole mass table, as done for the interactions of the BSk family, is a very difficult task and goes beyond the scope of this work.

The nuclear binding energies and charge radii in Hartree-Fock approximation are computed with the old but very fast Fortran code SKHAFO of \cite{Reinhard1991}, which we have updated to include the new terms (see Appendix \ref{app:hartreefock}) and also slightly modernized by replacing the common blocks with modules. This allows us to define the $\chi^2$ function
\begin{align}\label{eq:chi2}
\chi^2 &=  \chi_{E_b}^2 + \chi_{R_{\text{ch}}}^2 \notag \\
& = \sum_{i=1}^{N_B} \left[ \frac{E_b(i) - E_{b}^{\text{exp}}(i)}{\Delta E} \right]^2 + \sum_{i=1}^{N_{R}} \left[\frac{R_{\text{ch}} (i) - R_{\text{ch}}^{\text{exp}} (i)}{\Delta R} \right]^2,
\end{align}
where $N_B$ and $N_R$ are the numbers of experimental data points, $\Delta E$ and $\Delta R$ are the adopted errors. 
\begin{table}
\centering
\caption{Experimental binding energies and charge radii for doubly closed shell nuclei used in the fit.}
\label{table:binding-energies-and-charge-radii}
\begin{ruledtabular}
\begin{tabular}{lcc}
& $E_b$ (MeV) \footnote[1]{Taken from \cite{IAEA} unless stated otherwise.}& $ R_{\text{ch}}$ (fm) \footnotemark[1]\\
\hline
${}^{16}$O   & $-127.6193$     & $2.6991(52)$\\ 
${}^{40}$Ca  & $-342.0522$     & $3.4776(19)$\\ 
${}^{48}$Ca  & $-416.0012$     & $3.4771(20)$\\ 
${}^{56}$Ni  & $-483.9957(4)$  & $3.7226(3)$ \footnote[2]{Taken from Ref. \cite{Sommer2022}} \\ 
${}^{78}$Ni  & $-642.56(39)$   &           \\ 
${}^{100}$Sn & $-825.16(24)$   &           \\ 
${}^{132}$Sn & $-1102.8432(20)$& $4.7093(76)$\\ 
${}^{208}$Pb & $-1636.4302(12)$& $5.5012(13)$\\ 
\end{tabular}
\end{ruledtabular}
\end{table}
Following \cite{Chabanat1997}, we choose $\Delta E = 2$ MeV and $\Delta R = 0.02$ fm. We use the routine MIGRAD of Minuit
to minimize $\chi^2$ and to study the uncertainties of the parameters as well as their correlations. It turns out that the data do not allow us to determine all parameters. More constraints are needed.

\subsection{Constraints from finite-size instabilities}
\label{subsec:constraints from finite-size instabilities}

When $C^{\nabla\rho}_{0\,\eff}$ and $C^{\nabla\rho}_{1\,\eff}$ as defined in Appendix \ref{app:details-of-fits} are kept constant, $\chi^2$ depends only very weakly on the parameters $C^{\nabla\rho}_{01}$ and $C^{\nabla\rho}_{11}$ describing the density dependence of the gradient terms. Nevertheless, $\chi^2$ slightly decreases when $C^{\nabla\rho}_{01}$ and $C^{\nabla\rho}_{11}$ get more and more negative, until finally a spurious finite-size instability sets in, making the Hartree-Fock description of nuclei impossible \cite{Hellemans2013}.

This kind of instability can also be detected by computing the response functions of nuclear matter \cite{Hellemans2013}. In the random phase approximation (RPA), one can find the onset of the instability from the condition \citep{Pastore2014}
\begin{equation}\label{eq:onset-instability}
\frac{1}{\Pi^{(S,M,I)}_{\text{RPA}}(q,\omega=0)}=0
\end{equation}
($S$ is the spin, $M$ is the projection of spin along the direction of the momentum transfer $\vect{q}$, $I$ is the isospin, $\omega$ is the energy transfer) to detect poles in the response function $\Pi_{\text{RPA}}$, which we compute as explained in Ref.~\cite{Duan2023}.

Preliminary tests showed that $C^{\nabla \rho}_{01}$ and $C^{\nabla \rho}_{11}$ affect the appearance of the finite-size isoscalar and isovector instabilities, respectively. To describe finite nuclei, there should not be any instability (except the spinodal one) at densities $\rho\lesssim 0.2$ fm$^{-3}$ in symmetric nuclear matter. Furthermore, we want to use our interactions to describe neutrino scattering in supernova and (proto\mbox{-)}neutron star matter. Hence, also at higher densities, there should be no instability in the relevant range of momentum transfer $q$. Assuming a maximum temperature of $\sim 50$ MeV, neutrino energies will rarely exceed $\sim 250$ MeV, corresponding to neutrino momenta of $\sim 1.25$ fm$^{-1}$ and maximum momentum transfer of $\sim2.5$ fm$^{-1}$. Therefore, we fix $C^{\nabla \rho}_{01}$ and $C^{\nabla \rho}_{11}$ to the smallest values that satisfy these requirements, reducing the number of free parameters to $N_p=6$.
\subsection{Summary of the fitting procedure}
Let us now summarize the fitting procedure. From microscopic BHF calculations of nuclear matter, we get the nine parameters $A_1$, $B_1$, $A_2$, $B_2$, $C_1\dots C_4$, and $E_{\high}$. Then we fix $C^{\nabla \rho}_{01}$ and $C^{\nabla \rho}_{11}$ and fit the remaining six parameters $\rho_0$, $E_0$, $K_0$, $C^{\nabla\rho}_{0\,\eff}$, $C^{\nabla\rho}_{1\,\eff}$, and $W_0$ by minimizing $\chi^2$ defined in Eq.~\eqref{eq:chi2}. Finally, the resulting parametrization is checked for instabilities and the fit is repeated for different values of $C^{\nabla \rho}_{01}$ and $C^{\nabla \rho}_{11}$ until there are no spurious instabilities any more for $\rho\lesssim 2\rho_0$ and $q\lesssim 2.5$ fm$^{-1}$.

We present the parameters of Sky3 and Sky4 in Table \ref{table:parameters of the new skyrme force}. According to the employed microscopic EoS (LZLWBS for Sky3, APR for Sky4), Sky3 has a stiffer EoS than Sky4. In addition, we fix $C^{\nabla \rho}_{01} = -200$ MeV fm$^6$ and $C^{\nabla \rho}_{11}=0$ MeV fm$^6$ for Sky3, $C^{\nabla \rho}_{01} = -300$ MeV fm$^6$ and $C^{\nabla \rho}_{11}=100$ MeV fm$^6$ for Sky4, respectively, to eliminate the spurious finite-size instabilities. Both parametrizations fit the finite nuclei equally well: $\chi_{E_b}^2 \simeq 1.32$, $\chi_{R_{\text{ch}}}^2 \simeq 2.76$, and $\chi^2/\text{dof} \simeq 0.51$ for Sky3; $\chi_{E_b}^2 \simeq 2.32$, $\chi_{R_{\text{ch}}}^2 \simeq 2.75$, and $\chi^2/\text{dof} \simeq 0.63$ for Sky4 (dof = degrees of freedom = $N_B+N_{R}-N_p = 8+6-6=8$).

The mentioned values of $\chi^2$ depend of course on our choice for $\Delta E$ and $\Delta R$ mentioned below Eq.~\eqref{eq:chi2}. While this choice is irrelevant for the final parameter values, it is important for the estimation of the parameter uncertainties (see Appendix \ref{app:details-of-fits}). The fact that our $\chi^2/\text{dof}$  is of the order of $1$ indicates that the choices are reasonable.

\begin{table}
\centering
\caption{Parameters of the new Skyrme forces.} 
\label{table:parameters of the new skyrme force}
\begin{ruledtabular}
\begin{tabular}{lcc}

& Sky3& Sky4\\
\hline
$t_0$ (MeV fm$^3$)    & $-1800.864$  & $-1943.326$ \\ 
$t_1$ (MeV fm$^5$)    & $870.243$    & $1106.541$  \\
$t_2$ (MeV fm$^5$)    & $-36.893$    & $-1650.985$ \\
$t_{31}$ (MeV fm$^4$) & $13643.719$  & $15851.013$ \\
$t_{32}$ (MeV fm$^5$) & $-7467.193$  & $-7123.773$ \\
$t_{33}$ (MeV fm$^6$) & $7889.505$   & $5107.425$  \\
$t_4$ (MeV fm$^6$)    & $-984.081$   & $-1441.224$ \\
$t_5$ (MeV fm$^6$)    & $491.159$    & $3081.635$  \\
$y_0$ (MeV fm$^3$)    & $454.367$    & $285.466$   \\
$y_1$ (MeV fm$^5$)    & $-556.601$   & $358.595$   \\
$y_2$ (MeV fm$^5$)    & $-348.877$   & $1491.514$  \\
$y_{31}$ (MeV fm$^4$) & $-36495.639$ & $-26410.588$\\
$y_{32}$ (MeV fm$^5$) & $92479.691$  & $59567.033$ \\
$y_{33}$ (MeV fm$^6$) & $-68222.326$ & $-37339.367$\\
$y_4$ (MeV fm$^6$)    & $366.631$    & $-1004.797$ \\
$y_5$ (MeV fm$^6$)    & $-120.170$   & $-3015.408$ \\
$W_0$ (MeV fm$^{5}$)  & $111.486$    & $108.219$   \\
$\alpha_1$            & $1/3$        & $1/3$       \\
$\alpha_2$            & $2/3$        & $2/3$       \\
$\alpha_3$            & $1$          & $1$         \\
$\beta$               & $1/3$        & $1/3$       \\
$\gamma$              & $1/3$        & $1/3$       \\
$\eta_{J}$            & $0$          & $0$         \\
\end{tabular}
\end{ruledtabular}
\end{table}

\section{Results}
\label{sec:results-all}
In this section, we present the properties of finite nuclei and infinite nuclear matter (including pure neutron matter, symmetric nuclear matter, and neutron star matter) computed with the two new interactions.

\subsection{Description of finite nuclei}
\label{sec:description of finite nuclei}
\begin{figure}
\includegraphics[scale=0.44]{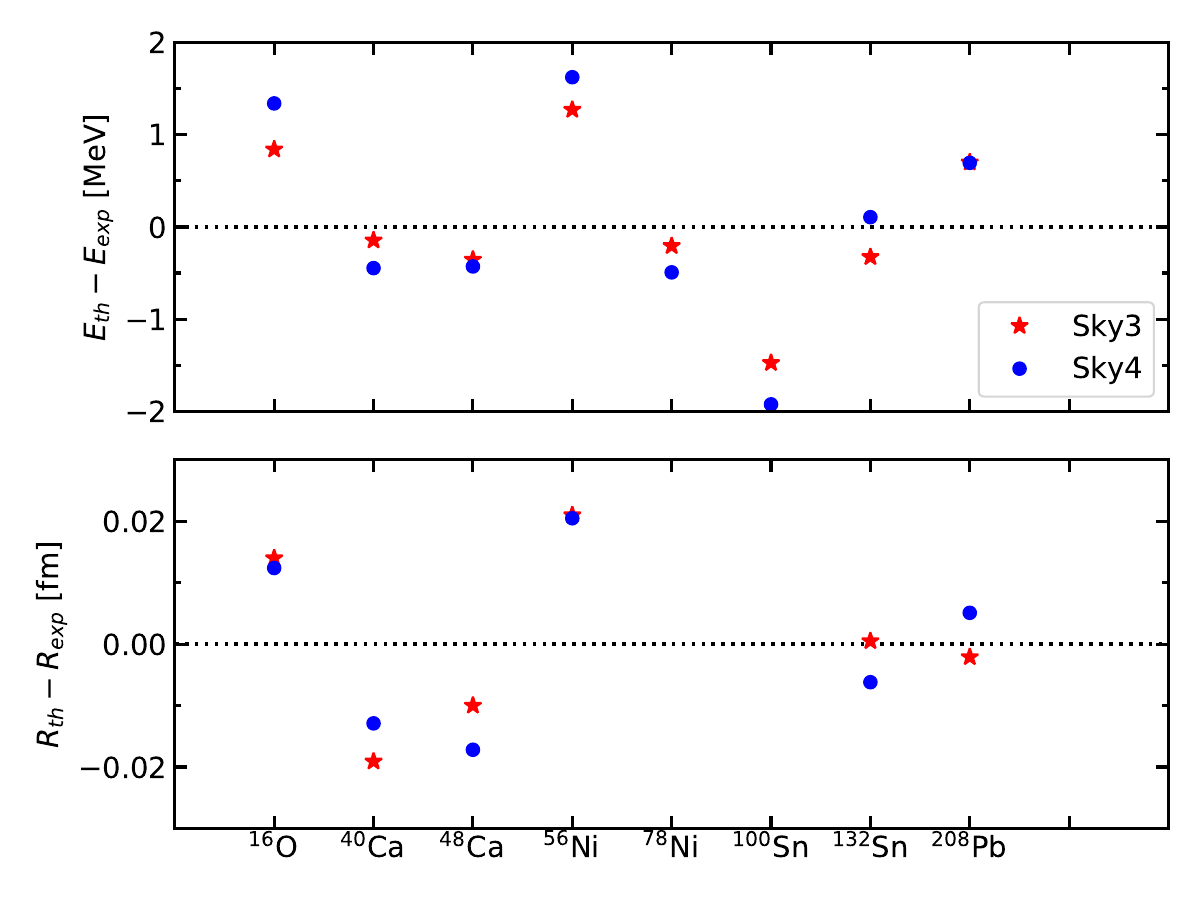} 
\caption{The difference between theoretical calculations and experimental results for doubly closed shell nuclei. Upper panel: binding energies; lower panel: charge radii.}
\label{fig:delta-E-r}
\end{figure}

Now, we present the properties of finite nuclei computed using the two new interactions Sky3 and Sky4. To see how well our new forces can describe doubly closed shell nuclei, we show the difference between theoretical calculations and experimental results in Fig. \ref{fig:delta-E-r}. The results of the binding energies and charge radii are shown in the upper and lower panels of Fig. \ref{fig:delta-E-r}, respectively. The experimental data (cf. Table~\ref{table:binding-energies-and-charge-radii}) is taken from the IAEA data base \cite{IAEA} except for the charge radius of ${}^{56}$Ni which is taken from Ref. \cite{Sommer2022}. As it was already clear from the $\chi^2$ values, the agreement with the binding energies and radii is satisfactory and of a similar quality as in the case of other Skyrme forces such as SLy4. 

However, like previously existing forces, our new forces cannot reproduce simultaneously the experimental results for the neutron-skin thickness of $^{208}$Pb and $^{48}$Ca as measured in the PREX \cite{Adhikari2021} and CREX \cite{Adhikari2022} experiments. In both nuclei, Sky3 and Sky4 give larger neutron skins $R_n-R_p$ than other forces such as SLy4 or BSk19-21. While this improves the agreement with the PREX experiment, it deteriorates the agreement with the CREX experiment: In $^{208}$Pb, we get $R_n-R_p = 0.183$ fm (Sky3) and $0.214$ fm (Sky4), which is still smaller than $0.283 \pm 0.071$ fm from PREX \cite{Adhikari2021} [for comparison, SLy4 and BSk19-21 give only $0.141$ fm (BSk21) - $0.168$ fm (SLy4)]. In $^{48}$Ca, we get $R_n-R_p = 0.175$ fm (Sky3) and $0.185$ fm (Sky4), which is larger than $0.121 \pm 0.026 \pm 0.024$ fm from CREX \cite{Adhikari2022} [for comparison, SLy4 and BSk19-21 give $0.142$ fm (BSk21) -  $0.157$ fm (SLy4)]. Since the extraction of the radii from the experiments is somewhat model dependent, we also compared directly the weak form factors which we computed following \cite{Reinhard2013}, but the conclusion does not change. Even when including the experimental $R_n-R_p$ into $\chi^2$ during the fit, the agreement does not substantially improve.

In summary, as far as we can say from our study that is so far limited to the ground states of doubly magic nuclei, our new forces can be used to describe finite nuclei properties reasonably well. However, we do not know how good the agreement will be for open-shell nuclei and for excited states. This should be investigated in future studies.

\subsection{Description of pure neutron matter and symmetric nuclear matter}
\label{sec:description of PNM and SNM}
We have described the formula of the EoS of pure neutron matter above. The explicit formula for the energy per particle $\frac{E}{A}$ of symmetric nuclear matter can be obtained in the same way and is given in Appendix \ref{app:EoS-uniform-matter}. 
The saturation density $\rho_0$ is defined as the density where $\tfrac{E}{A}(\rho)$ is minimal, i.e.,
\begin{equation}
\frac{\partial}{\partial\rho}\,\frac{E}{A} \bigg|_{\rho_0} = 0.
\end{equation}
Then the incompressibility modulus at zero pressure, i.e., at density $\rho_0$, can be obtained by computing 
\begin{equation}\label{eq:K_inf}
K_{0} =  9 \rho_0^2 \frac{\partial^2}{\partial\rho^2}\, \frac{E}{A}\bigg\vert_{\rho=\rho_0}\,.
\end{equation}
The symmetry energy is defined by
\begin{equation}\label{eq:symmetry energy}
E_{\sym} (\rho) = \frac{1}{2} \frac{\partial^2}{\partial\delta^2}\, \frac{E}{A}\bigg|_{\delta=0}\,,
\end{equation}
where $\delta = (\rho_n-\rho_p)/\rho$.
From $E_{\sym}$ one gets the usual symmetry coefficient (symmetry energy)
\begin{equation}\label{eq:symmetry coefficient}
J_0=E_{\sym}(\rho_0),
\end{equation}
and the density-symmetry coefficient (the slope of the nuclear symmetry energy)
\begin{equation}\label{eq:density-symmetry coefficient}
L_0  = 3 \rho_0 \frac{\partial E_{\sym}}{\partial \rho} \bigg\vert_{\rho=\rho_0}\,.
\end{equation}
The explicit formulas for all these quantities are given in Appendix \ref{app:EoS-uniform-matter}.

\begin{table}
\caption{Parameters of infinite nuclear matter for the new forces. The definitions of the higher-order derivatives of $E/A$ at $\rho_0$ not given here ($K_{\sym}, \dots, Z_{\sym}$) can be found in Ref.~\cite{Margueron2018}.}  
\label{table:parameters of infinite nuclear matter}
\begin{ruledtabular}
\begin{tabular}{lcc}

                        & Sky3      & Sky4     \\
\hline
$\rho_0$ (fm$^{-3}$)    & $0.1545$  & $0.1528$ \\ 
$E_0$ (MeV)             & $-15.92$  & $-15.94$ \\ 
$K_0$ (MeV)             & $252.5$   & $229.4$  \\ 
$J_0$ (MeV)             & $30.3$    & $32.1$   \\ 
$L_0$ (MeV)             & $43.9$    & $44.7$   \\
$K_{\sym}$ (MeV)        & $-104.5$  & $-178.7$ \\
$Q_{\sat}$ (MeV)        & $-288.7$  & $-383.2$ \\
$Q_{\sym}$ (MeV)        & $945.8$   & $635.5$  \\
$Z_{\sat}$ (MeV)        & $1201.9$  & $1665.3$ \\
$Z_{\sym}$ (MeV)        & $-3251.3$ & $-2324.6$\\
$m_{s,0}^{*}/\mnucleon$ & $0.8092$  & $0.8104$ \\ 
$m_{v,0}^{*}/\mnucleon$ & $0.7467$  & $0.7480$ \\ 
\end{tabular}
\end{ruledtabular}
\end{table}
\begin{figure}
\includegraphics[scale=0.44]{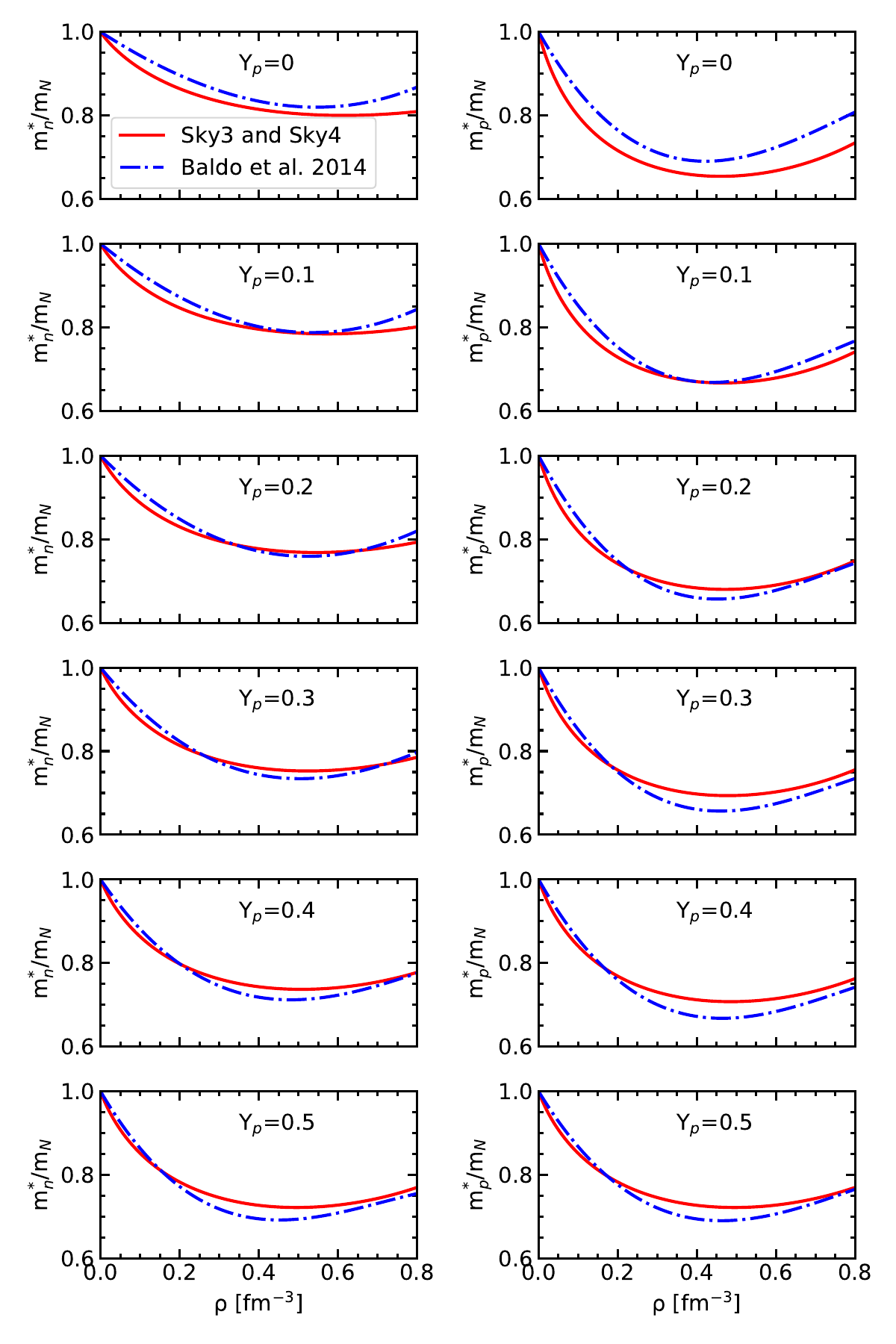} 
\caption{Results of the fits of neutron (left panels) and proton (right panels) effective masses. From top to bottom panels, the proton fractions are $Y_p=0$, 0.1, 0.2, 0.3, 0.4, 0.5.}
\label{fig:effective-masses-PNM-SNM}
\end{figure}
\begin{figure}
\includegraphics[scale=0.42]{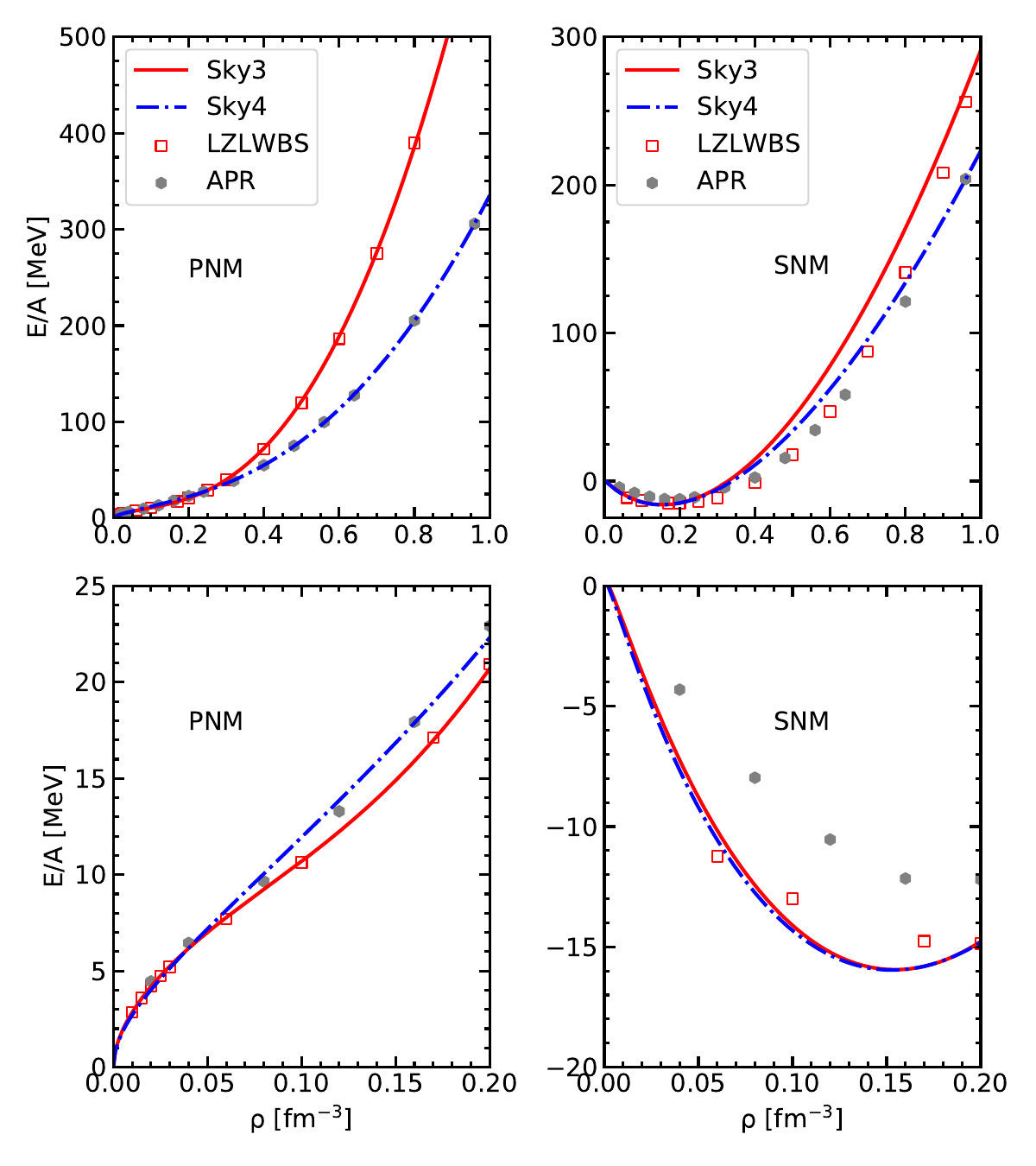} 
\caption{EoS in pure neutron matter and symmetric nuclear matter for the two new interactions Sky3 and Sky4. The upper panels show the evolution of the EoS in a range of densities from 0 to 1 fm$^{-3}$; the lower panels show the low-density EoS (0 - 0.2 fm$^{-3}$). The symbols show the microscopic neutron-matter EoS of \cite{Liu2022} (complemented at $\rho \leq 0.03$ fm$^{-3}$ with results of \cite{Palaniappan2023}) and of \cite{Akmal1998}, used in the determination of our parameters.}
\label{fig:EoS-PNM-SNM}
\end{figure}
We present in Table \ref{table:parameters of infinite nuclear matter} the saturation density $\rho_0$ and some other parameters evaluated at $\rho_0$, such as the energy per nucleon $E_0$, the incompressibility modulus $K_0$, the symmetry coefficient $J_0$, the density-symmetry coefficient $L_0$, the isoscalar effective mass $m_{s,0}^{*}/\mnucleon$, and the isovector effective mass $m_{v,0}^{*}/\mnucleon$. We report also the values of some higher-order derivatives of $\frac{E}{A}$ that are sometimes needed in the meta-modeling framework \cite{Margueron2018}. In Ref. \cite{Dutra2012}, $K_0$ is evaluated in a range of $200-260$ MeV. Fortunately, the values computed using the two new interactions are located in this range. In the recent paper \citep{Lattimer2023}, Lattimer reports that the symmetry energy coefficient and the corresponding slope are $J_0=30.5-33.9$ MeV and $L_0=39.7-66.1$ MeV. Our results have a good agreement with these values, although $J_0$ computed from interaction Sky3 is slightly smaller. According to Li et al. \citep{Li2018}, the nucleon isoscalar and isovector effective masses in nuclear matter at saturation density are $m_{s,0}^{*}/\mnucleon \simeq 0.7-0.9$ and $m_{v,0}^{*}/\mnucleon \simeq 0.6-0.93$, respectively. Our results lie in these ranges.

As described in Sec. \ref{subsec:determination of parameters}, we fit the effective masses computed using the microscopic BHF method to obtain a reasonable behavior of the effective masses for the new Skyrme interactions. Because Sky3 and Sky4 have the same values of $A_1$, $B_1$, $A_2$, and $B_2$, they give the same effective masses. We present the results of the fit in Fig.~\ref{fig:effective-masses-PNM-SNM}. From top to bottom panels, the neutron and proton effective masses for proton fractions $Y_p=0-0.5$ are shown. The red solid curves represent the results computed with our new interactions. Their trends are consistent with the microscopic results from Ref. \cite{Baldo2014} although they do not reproduce them very precisely. In particular, both $m^*_n$ and $m^*_p$ in PNM (two upper panels) are too low, since otherwise we could not have reproduced $\mnucleon^*$ in symmetric matter (bottom panel), as explained below Eq.~\eqref{eq:m_star}. Concerning the form of the density dependence, a better fit could have been obtained with $\beta=\gamma=1$, but this would have resulted in higher powers than $\rho^2$ in $\tfrac{E}{A}(\rho)$, which we wanted to avoid.

In Fig. \ref{fig:EoS-PNM-SNM}, we present the EoS of pure neutron matter and symmetric nuclear matter for the two new interactions Sky3 and Sky4. The evolution of the EoS in the range of densities from 0 to 1 fm$^{-3}$ and a zoom on low densities (0 - 0.2 fm$^{-3}$) are presented in the upper and lower panels, respectively. For comparison, we also show the BHF results in the figure. As we can see in the left panels, the fits according to Eq. \eqref{eq:energy-per-nucleon-PNM} reproduce very well the corresponding BHF EoS of pure neutron matter. From the right panels we can see that, at low densities, the difference between Sky3 and Sky4 in the EoS of SNM is very small, since the saturation properties are determined by the fit of the nuclear binding energies and radii, while the BHF EoS are not fine tuned to correctly reproduce the saturation properties. However, the difference between Sky3 and Sky4 becomes significant with increasing density because they were constrained to approach the respective BHF results at high density.

\begin{figure}
\begin{center}
\includegraphics[scale=0.54]{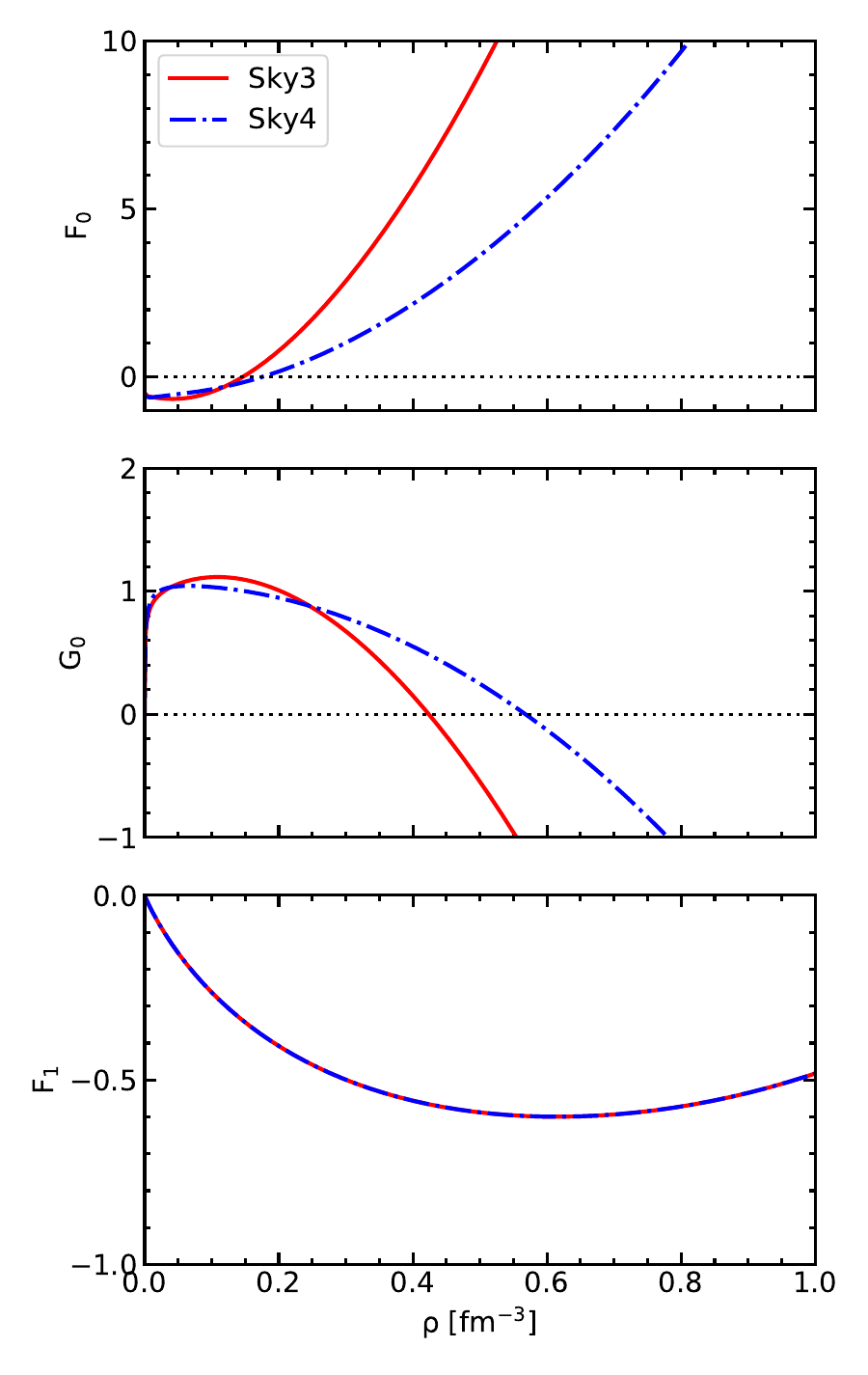} 
\caption{Landau parameters in pure neutron matter for the two new forces Sky3 and Sky4.}
\label{fig:F-G-rho-PNM}
\end{center}
\end{figure}

\begin{figure}
\begin{center}
\includegraphics[scale=0.44]{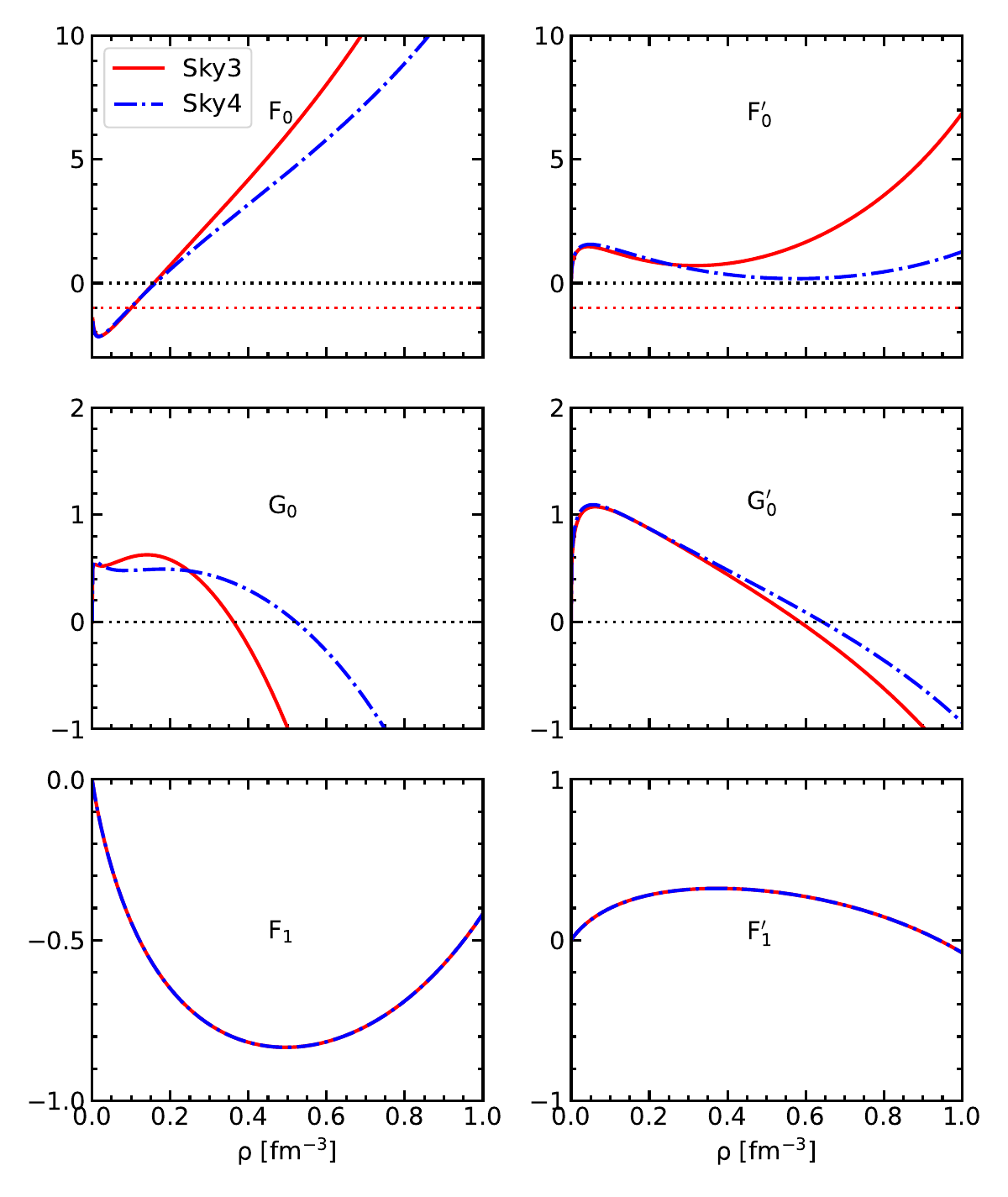} 
\caption{Landau parameters in symmetric nuclear matter for the two new forces Sky3 and Sky4. The region where $F_0$ goes below the red dotted line (i.e., $F_0<-1$) in the upper panel corresponds to the liquid-gas instability.}
\label{fig:F-G-rho-SNM}
\end{center}
\end{figure}
To conclude this subsection, we present the $L=0$ and $L=1$ Landau parameters in pure neutron matter and symmetric nuclear matter for the two new forces Sky3 and Sky4 in Figs. \ref{fig:F-G-rho-PNM} and \ref{fig:F-G-rho-SNM}. Following the usual conventions, $F_L$ denotes the dimensionless Landau parameters relevant for density variations and $G_{L}$ the ones for spin-density variations. In symmetric nuclear matter, $F_L$, $G_L$ are the isoscalar and $F'_L$, $G'_L$ the isovector ones. Notice that $G_1$ and $G_1^{\prime}$ are not included in the figures, because they vanish in our new forces (and also, e.g., in SLy4) due to $\eta_J=0$. The detailed expressions of the Landau parameters can be found in Appendix \ref{app:landau parameters in PNM and SNM}. The Landau parameters play an important role in the response functions determining, e.g., neutrino scattering, but they give also a necessary (but not sufficient, cf. Secs.~\ref{subsec:constraints from finite-size instabilities} and~\ref{subsubsec:spin-0 instabilities}) criterion for the stability of uniform matter. Namely, $F_0$, $F_0^{\prime}$, $G_0$, and $G_0^{\prime}$ should be larger than $-1$; $F_1$, $F_1^{\prime}$, $G_1$, and $G_1^{\prime}$ should be larger than $-3$. In Fig. \ref{fig:F-G-rho-SNM}, we can see that $F_0 <-1$ at subsaturation densities, indicating the well-known liquid-gas instability. $F_1$ in pure neutron matter, $F_0^{\prime}$, $G_0^{\prime}$, $F_1$, and $F_1^{\prime}$ in symmetric nuclear matter do not indicate any instability when the density is below 1 fm$^{-3}$. Since $F_1$ and $F'_1$ are directly related to the effective masses through Galilean invariance \citep{Nozieres}, it is not surprising that Sky3 and Sky4 give identical results for them. However, $G_0$ for the two new interactions is smaller than $-1$ at high densities (above $\sim 3.5 \rho_0$ for Sky3 and $\sim 5\rho_0$ for Sky4) in both pure neutron matter and symmetric nuclear matter, indicating a ferromagnetic instability. We do not think that this is a reason to discard our new interactions. First, as it is already done with the choice $\eta_J=0$, one may think about modifying or adding spin-dependent ($s^2$, $\mathbb{J}^2$, $\vect{s}\cdot\vect{T}$) terms in the functional independently of the spin-independent terms. The spin dependent terms should then be determined by fitting observables that explicitly depend on them, or from microscopic theories, whereas in our present work, they have not been constrained at all. Second, especially in the case of Sky4, the instability appears only at such high densities that one maybe cannot be sure if matter is really unpolarized, and one may also wonder whether one should not replace the purely nucleonic Skyrme theory by something else. 

\subsection{Description of neutron stars}
\label{sec:description of neutron star matter}

\subsubsection{EoS of neutron star matter}\label{subsubsec:EoS}
Neutron stars are composed mainly but not exclusively of neutrons. They also contain protons and electrons for the most simple model of neutron star composition ($npe$ model). Muons are expected to exist at higher densities ($npe \mu$ model). Muons will appear in the internal matter of a neutron star when $\mu_{e} > m_{\mu}$. Generally, the value of the baryon number density is larger than nuclear saturation density in this case \citep{Lattimer1991}. 

The energy density of neutron star matter ($npe \mu$ matter) is a sum of the nucleon and the lepton contributions, and reads as (in $c=1$ units):
\begin{equation}\label{eq:energy-density}
\varepsilon = \varepsilon_{N}(\rho_{n},\rho_{p}) + \rho_b\, m_N + \varepsilon_{e}(\rho_{e}) +\varepsilon_{\mu}(\rho_{\mu}),
\end{equation}
where $\varepsilon_{N}$, $\varepsilon_{e}$, $\varepsilon_{\mu}$ are the energy densities of nucleons, electrons, and muons, respectively; $\rho_{n}$, $\rho_{p}$, $\rho_{e}$, and $\rho_{\mu}$ are number densities of neutrons, protons, electrons, and muons, respectively. The baryon number density is $\rho^{}_b=\rho_{n}+\rho_{p}$.

The required charge neutrality of a neutron star leads to the same number densities of protons and leptons (electrons and muons), i.e., $\rho_p=\rho_e+\rho_{\mu}$. In the case of $npe \mu$ matter, the $\beta$-equilibrium implies that the chemical potentials satisfy
\begin{equation}\label{eq:chemical-potential}
\mu_n=\mu_p+\mu_e, \qquad \mu_\mu=\mu_e,
\end{equation}
where
\begin{equation}\label{eq:chemical-potential1}
\mu^{}_i=\frac{\partial \varepsilon}{\partial \rho^{}_i}, \qquad(i=n,p,e,\mu).
\end{equation}
According to Eq. (\ref{eq:chemical-potential}) combined with the charge neutrality condition, one can determine the fractions $Y_i=\frac{\rho^{}_i}{\rho^{}_b}$. Then a one-parameter EoS of $npe \mu$ matter is given as:
\begin{equation}\label{eq:EOS}
P(\rho^{}_b)=\mu_n \rho^{}_b - \varepsilon (\rho^{}_b).
\end{equation}

\begin{figure}
\begin{center}
\includegraphics[scale=0.53]{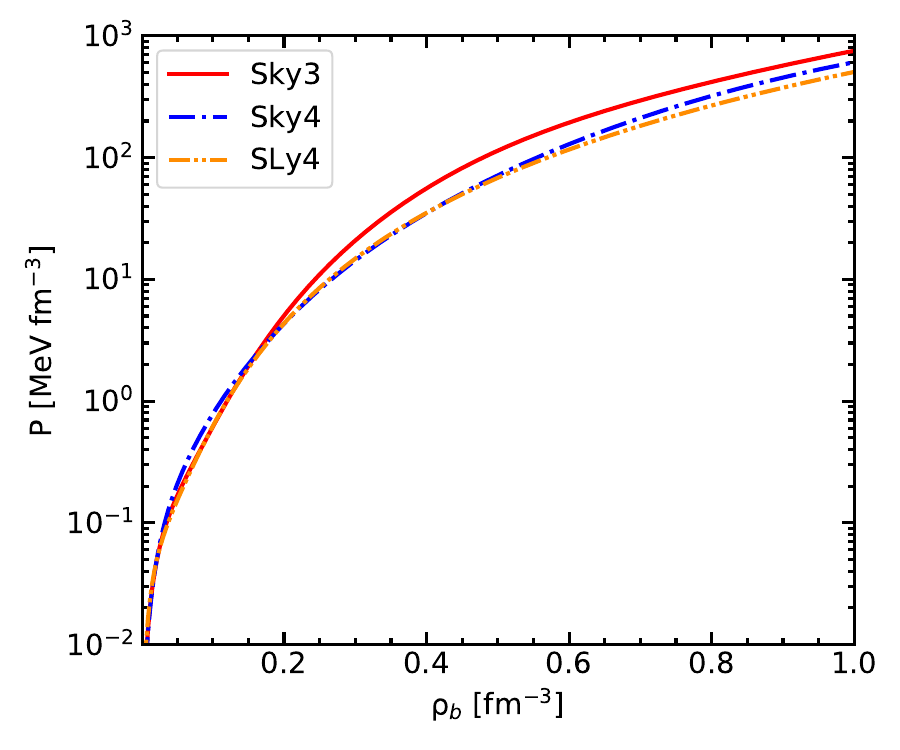}
\caption{EoS of uniform $npe\mu$ matter in $\beta$ equilibrium. Results computed with Sky3, Sky4, and SLy4 are shown.}
\label{fig:EoS-beta-stable}
\end{center}
\end{figure}
\begin{figure}
\begin{center}
\includegraphics[scale=0.53]{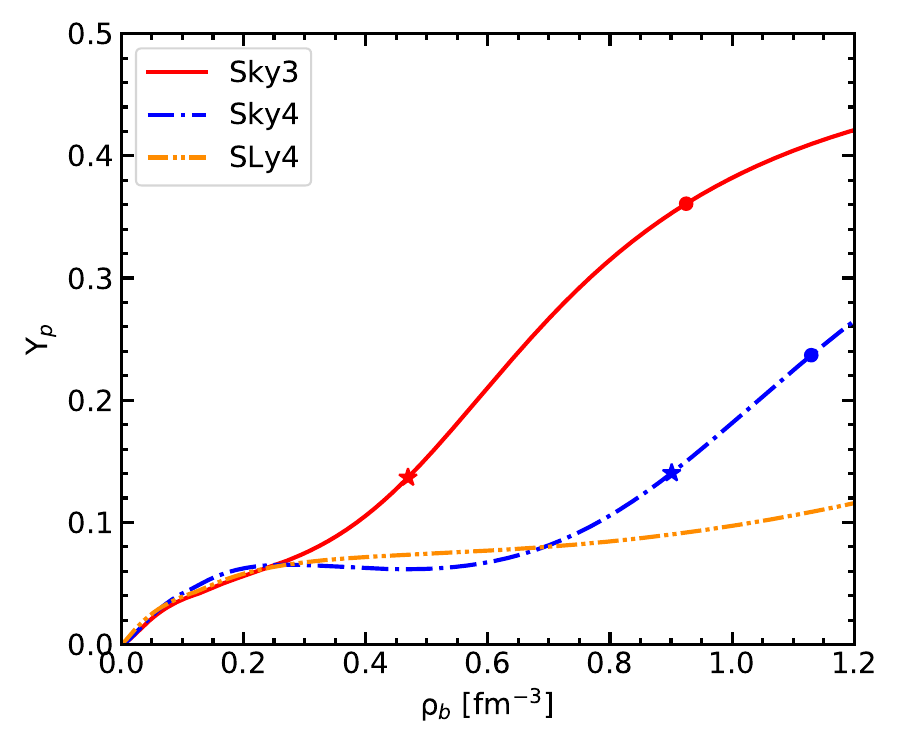} 
\caption{Proton fraction of uniform $npe\mu$ matter in $\beta$ equilibrium as a function of the baryon number density for Sky3 and Sky4. Filled stars correspond to the threshold baryon number density for the direct Urca process, while filled circles denote the maximum central baryon number density of stable neutron stars predicted by the corresponding interaction. The proton fraction in neutron star matter for SLy4 is also shown for comparison.}
\label{fig:Yp}
\end{center}
\end{figure}
In Fig. \ref{fig:EoS-beta-stable}, we show the obtained EoS of neutron star matter computed with the new forces Sky3 and Sky4, and for comparison also the one computed with SLy4. We see that the pressure of Sky4 is slightly higher than the other two below saturation density, while the pressure of Sky3 is higher than the other two above saturation density. 

We present the corresponding proton fractions as functions of baryon number density in Fig. \ref{fig:Yp}. We can see that the proton fractions predicted by the three interactions have only small differences at low densities up to about 0.25 fm$^{-3}$. Above that density, Sky3 can give a much higher proton fraction than the other two interactions. The proton fraction obtained with Sky4 stays close to the SLy4 one until $\rho_b\approx 0.8$ fm$^{-3}$, but then it starts to increase strongly, too, while the one of SLy4 increases only very little. This has important consequences for the possibility of the direct Urca process as discussed below.

So far, we have discussed only uniform matter. However, below some transition density $\rho_t$, the true ground state is inhomogeneous matter, which in a neutron star corresponds to the crust \citep{Chamel2008}. With increasing density, one expects different phases such as a crystal of neutron-rich nuclei in the outer crust and a crystal of neutron rich nuclei, rods, or plates (``pasta phases'') coexisting with a gas of free neutrons in the inner crust, until $\rho_t$ is reached and matter becomes uniform, corresponding to the core. Hence, $\rho_t$ corresponds to the core-crust transition density. We can compute the core-crust transition densities $\rho_t$ using the RPA and Eq.~\eqref{eq:onset-instability}, because it corresponds to the density where the spinodal instability disappears, see Sec.~\ref{subsubsec:spin-0 instabilities}. The results for $\rho_t$ and other relevant neutron-star parameters of this section are listed in Table \ref{table:parameters for neutron star models}.

At densities below $\rho_t$, the crust EoS for our new interactions can be computed, e.g., within the compressible liquid-drop model \cite{Carreau2019} by using the CUTER code \cite{Davis2024}, which needs only the EoS of uniform matter in $\beta$ equilibrium and the nuclear matter parameters of Table~\ref{table:parameters of infinite nuclear matter} as input.The densities where the CUTER code matches the crust EoS with the EoS of uniform matter (0.0859 and 0.0816 fm$^{-3}$ for Sky3 and Sky4, respectively) are in very good agreement with our results for $\rho_t$.
\begin{table}
\centering
\caption{Summary of neutron-star related results obtained with the new interactions Sky3 and Sky4: core-crust transition densities $\rho_t$, pressure at transition densities $P_t$, threshold for the direct Urca cooling $\rho_{\text{Urca}}$, $M_{\text{Urca}}$, and parameters of neutron stars at maximum mass $M_{\text{max}}$ and 1.4$M_{\odot}$: central baryon density $\rho_c$, radius $R$, baryon number $A_b$, binding energy $E_{\text{bind}}$, moment of inertia $I$, and gravitational redshift $z$.}
\label{table:parameters for neutron star models}
\begin{ruledtabular}
\begin{tabular}{lcccc}

                     & Sky3             & Sky4             &Sky3             & Sky4            \\
                     & $M_{\text{max}}$ & $M_{\text{max}}$ & $1.4 M_{\odot}$ & $1.4 M_{\odot}$ \\
\hline
$\rho_t$ (fm$^{-3}$) & $0.0861$         & $0.0825$         & $0.0861$        & $0.0825$        \\
$P_t$ (MeV fm$^{-3}$)& $0.4343$         & $0.5233$         & $0.4343$        & $0.5233$        \\
$\rho_{\text{Urca}}$ (fm$^{-3}$)&$0.4693$& $0.9011$        & $-$             & $-$             \\
$M_{\text{Urca}}$ ($M_{\odot}$)&$1.76$  & $2.10$           & $-$             & $-$             \\
$M$ ($M_{\odot}$)    & $2.35$           & $2.15$           & $1.4$           & $1.4$           \\
$\rho_c$ (fm$^{-3}$) & $0.9248$         & $1.1264$         & $0.3968$        & $0.5248$        \\
$R$ (km)             & $11.32$          & $10.26$          & $12.60$         & $11.75$         \\
$A_b$ ($10^{57}$)    & $3.38$           & $3.10$           & $1.86$          & $1.86$          \\
$E_{\text{bind}}$ ($10^{53}$ erg)&$8.53$& $7.70$           & $2.56$          & $2.67$          \\
$I$ ($10^{45}$ g cm$^2$)& $2.89$        & $2.18$           & $1.60$          & $1.38$          \\
$z$                  & $0.605$          & $0.621$          & $0.221$         & $0.243$         \\
\end{tabular}
\end{ruledtabular}
\end{table}

\subsubsection{Mass-radius relation}\label{subsubsec:mass-radius-relation}
After determining the equation of state of dense $npe\mu$ matter in $\beta$ equilibrium, we can obtain the mass-radius relation of neutron stars by solving the Tolman-Oppenheimer-Volkov (TOV) equation \citep{Tolman1939,Oppenheimer1939,Chabanat1997},  
\begin{align}\label{eq:TOV equation}
 & \frac{dP}{dr} = - \frac{G\includedmass \varepsilon}{r^2} \frac{(1+\frac{P}{\varepsilon})(1+\frac{4\pi r^3 P}{\includedmass})}{1-\frac{2G\includedmass}{r}},  \nonumber \\
 & \includedmass(r) =  \int_{0}^{r} 4\pi r'^{2}\varepsilon(r')dr',
\end{align}
where $\includedmass(r)$ is the mass inside the sphere of radius $r$, with the central boundary conditions: $P(0)=P_c$, $\includedmass(0)=0$, and the outer boundary condition $P(R)=0$. The latter defines the radius $R$ of the neutron star and its total mass $M = \includedmass(R)$. In order to get meaningful radii, it is necessary to take the crust into account \citep{Perot2020}. As mentioned before, we use the CUTER code \citep{Davis2024} to compute crust EoS for our new interactions that match continuously our EoS of the core.

\begin{figure}
\begin{center}
\includegraphics[scale=0.54]{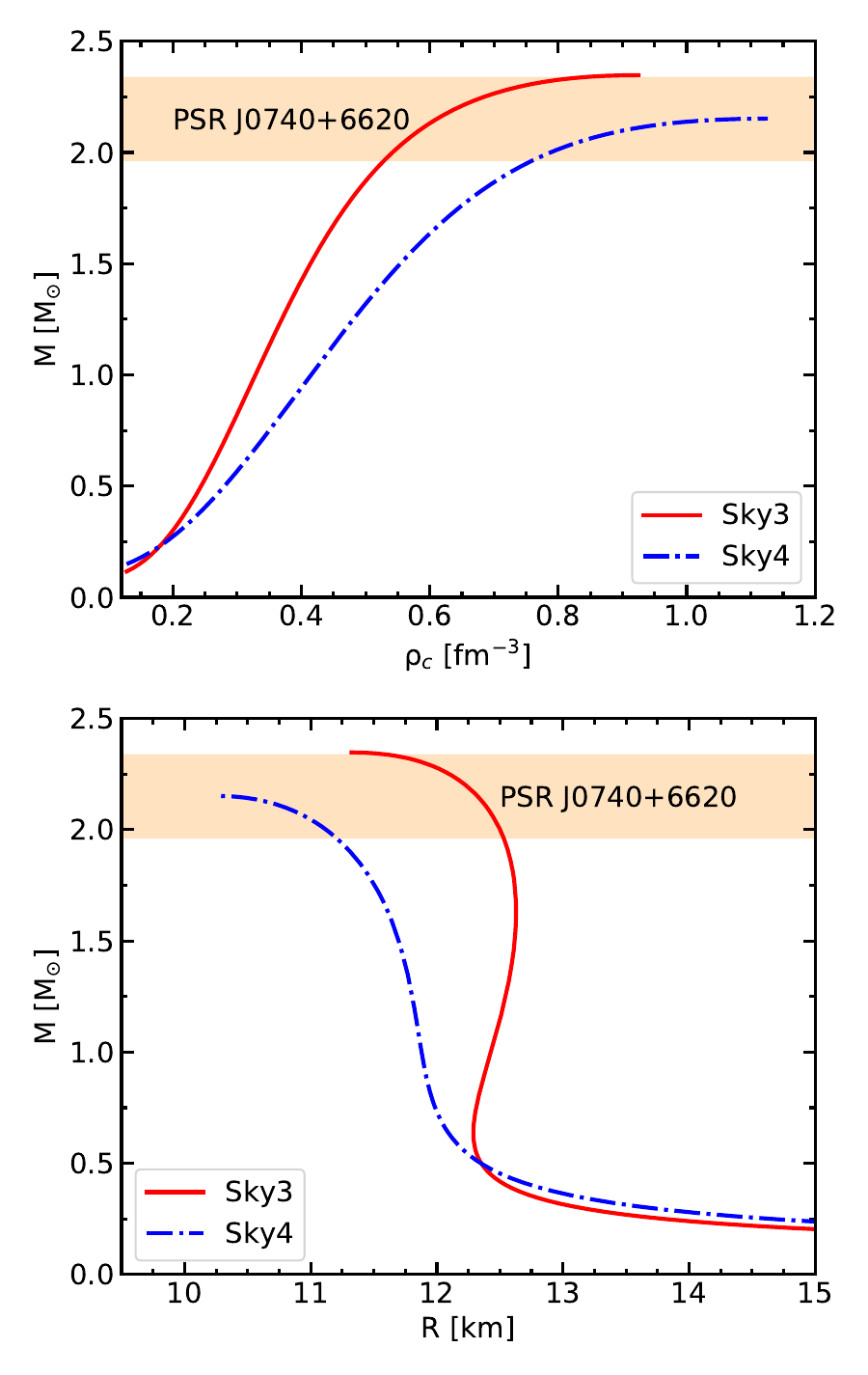} 
\caption{Masses as a function of the radius of the star (lower panel) and masses as a function of the central baryon number density of the star (upper panel) for Sky3 and Sky4.}
\label{fig:M-R-rhoc}
\end{center}
\end{figure}
Neutron star masses can be evaluated using astronomical observational data. Up to now, many neutron stars have been measured. Therefore, measurements that come from observations can be used to make a comparison with the theoretical predictions. The lower and upper panels of Fig. \ref{fig:M-R-rhoc} show the neutron star mass as a function of the radius of the star and as a function of the central baryon number density, respectively. The light orange shaded region corresponds to the measured mass of the millisecond pulsar PSR J0740+6620, i.e., $2.14_{-0.18}^{+0.2}$ $M_\odot$ with a $95.4\%$ credibility interval \citep{Cromartie2020}. We can see that the interactions Sky3 and Sky4 are compatible with this measurement. The estimates of the mass of PSR J0740+6620 were improved in Ref. \cite{Fonseca2021} in 2021. They measured the pulsar mass as $2.08_{-0.07}^{+0.07}$ $M_\odot$ with a $68.3\%$ credibility \citep{Fonseca2021}. Later, Miller et al. \citep{Miller2021} further estimated the radius of PSR J0740+6620 from NICER and XMM-Newton data, and reported that the radius of a 2.08 $M_{\odot}$ neutron star is $12.35_{-0.75}^{+0.75}$ km. To get such a big radius, the EoS must be very stiff. The new interaction Sky3 predicts a radius of about 12.42 km for a 2.08 $M_{\odot}$ neutron star, which is consistent with the observation. 

Other properties of neutron stars, such as binding energy, gravitational redshift, and moment of inertia, can also be obtained as soon as neutron star mass, radius, and baryon number are given. They will be discussed below.

\subsubsection{Other properties of neutron stars}\label{subsubsec:basic properties}
The newly born neutron stars are hot but will rapidly cool down. The dominating cooling mechanism is neutrino emission from the neutron-star interior during about $10^5$ years after its birth \citep{Yakovlev2004}. Especially, the nucleon direct Urca process is the simplest neutrino process and consists, in fact, of the neutron beta decay, electron capture, and nucleon direct Urca process with muons ($npe \mu$ model) \citep{Yakovlev2001}. Fractions of neutrons, protons, electrons, and muons determine whether the direct Urca process can occur. The direct Urca process is allowed when the relative fractions of the components of a neutron star satisfy $Y_n^{1/3} < Y_p^{1/3} + Y_e^{1/3}$ or $Y_n^{1/3} < Y_p^{1/3} + Y_{\mu}^{1/3}$ due to the condition for momentum conservation \citep{Lattimer1991}. The threshold proton fraction for the direct Urca process can be computed using \citep{Alvarez2015}
\begin{equation}\label{eq:direct-Urca}
Y_p^{\text{DU}}= \frac{1}{1+[1+ (\frac{\rho_e}{\rho_e+\rho_{\mu}} )^{1/3}]^3}.
\end{equation}
Filled stars and circles shown in Fig. \ref{fig:Yp} correspond to threshold baryon number density for the direct Urca process and the maximum central baryon number density of stable neutron stars predicted by the corresponding interaction, respectively. In the case of SLy4, the maximum central baryon number density is $\sim 1.21$ fm$^{-3}$ \citep{Fantina2013} and it is impossible to cool down via the direct Urca process for neutron stars described by SLy4 \citep{Chabanat1997}. We can see that the direct Urca process is allowed to occur in neutron star models computed using Sky3 and Sky4. As discussed, e.g., in \cite{Fantina2013}, this feature is desirable, because there are hints that the direct Urca process takes place in massive neutron stars. Combining Fig. \ref{fig:Yp} with the upper panel of Fig. \ref{fig:M-R-rhoc}, we find that our new interactions are consistent with the statement of \cite{Klaehn2006} that the direct Urca process should not take place in stars with masses below $1.5 M_\odot$ (see Table \ref{table:parameters for neutron star models} for actual values).

\begin{figure}
\begin{center}
\includegraphics[scale=0.54]{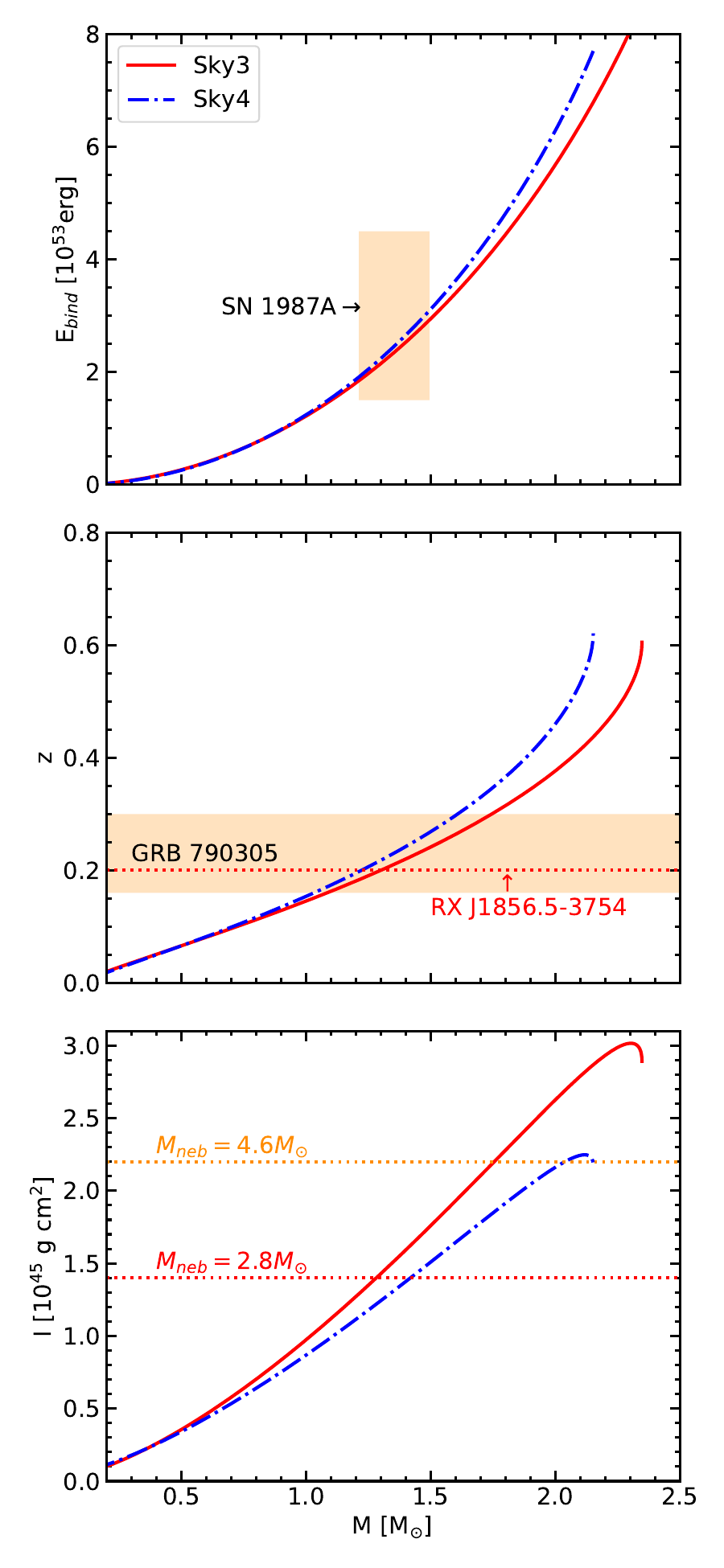} 
\caption{Binding energy (top), gravitational redshift (middle), and moment of inertia (bottom) as functions of neutron star mass for Sky3 and Sky4.}
\label{fig:Ebind-z-I-M}
\end{center}
\end{figure}

During and after a core-collapse supernova explosion, $99\%$ of the binding energy of a neutron star is released in the form of neutrinos \citep{Bersten2017}. In 1987, neutrinos produced by the supernova SN 1987A were detected \citep{Bionta1987,Hirata1987,Alexeyev1988}. This detection allows one to estimate the released energy. The total neutrino energy from SN 1987A is estimated to be $E_{\nu} \simeq (3\pm1.5) \times 10^{53}$ erg, and the mass of the neutron star born in SN 1987A is estimated to be $1.2-1.5$ solar masses \citep{Haensel2007}. To compare, we compute the binding energy in the neutron star models based on interactions Sky3 and Sky4, using the formula \citep{Chabanat1997}
\begin{equation}\label{eq:Ebind}
E_{\text{bind}} = A_bm_0 -M,
\end{equation}
where $m_0$ is the mass of the ${}^{56}$Fe atom divided by 56, $M$ is the mass of the neutron star, and $A_b$ is
the total baryon number of the star, which can be obtained by the integral \citep{Chabanat1997} 
\begin{equation}\label{eq:baryon number}
A_b=\int^{R}_{0} \frac{4\pi r^{2}\rho^{}_{b}(r)dr}{(1-\frac{2G\includedmass(r)}{r})^{1/2}}.
\end{equation}
The results are shown in the top panel of Fig. \ref{fig:Ebind-z-I-M}. The theoretical results are compatible with the estimations from SN 1987A shown as the light orange shaded region.

As mentioned above, many astrophysical phenomena, such as GRBs, GWs, FRBs, and so on, are related to neutron stars \citep{Kumar2015,Abbott2017,Zhang2020}. The central engine of some GRBs is thought to be a millisecond magnetar \citep{Kumar2015}. The detection of GW 170817 shows that binary neutron star mergers can produce gravitational waves \citep{Abbott2017}. Some FRBs are detected as repeaters, and they are confirmed to originate from magnetars \citep{Zhang2020}. In the study of these fields related to neutron stars, the redshift $z$ is an important quantity. Therefore, we compute the gravitational redshift using the relation \citep{Chabanat1997}
\begin{equation}\label{eq:redshift}
z=\left(1- \frac{2GM}{R} \right)^{-1/2}-1,
\end{equation}
and present the results in the central panel of Fig. \ref{fig:Ebind-z-I-M}. The light orange shaded region represents the redshift of GRB 790305, which is from the soft gamma-ray repeater SGR 0526-066 \citep{Higdon1990}. The red dotted line corresponds to the measured gravitational redshift of the isolated neutron star RX J1856.5-3754 \cite{Ho2007}. 
As one can see, these two observations lie in the range of redshifts expected at the surface of a neutron star, and may help to estimate neutron star mass and radius.

The moment of inertia of a neutron star is another important quantity for astrophysical research. For example, the moment of inertia of a pulsar is crucial for the computation of its spindown \citep{ShapiroTeukolsky1983}. It can also be used to study the dominant radiation of the afterglows of magnetar-central-engine GRBs \citep{Zhang2001}. Therefore, we also compute the moment of inertia using the formula \citep{Lattimer2016}
\begin{equation}\label{eq:moment of inertia}
I = \frac{1}{G} \frac{w(R)R^3}{6+2w(R)},
\end{equation}
where $w(R)$ is the solution, at the surface $R$, of the differential equation 
\begin{equation}\label{eq:moment-of-inertia-w}
\frac{dw}{dr}=4\pi G \frac{(\varepsilon + P)(4+w)r}{1-2G\includedmass/r}-\frac{w}{r}(3+w),
\end{equation}
with the boundary condition $w(0)=0$. We solve Eq. (\ref{eq:moment-of-inertia-w}) together with the TOV Eq. (\ref{eq:TOV equation}).
The results are shown in the bottom panel of Fig. \ref{fig:Ebind-z-I-M}.
The red and dark orange dotted lines represent the lower limits for the Crab pulsar moment of inertia for nebula masses $M_{\text{neb}}=2.8 M_{\odot}$ and $M_{\text{neb}}=4.6 M_{\odot}$ (i.e., $I > 1.4 \times 10^{45}$ g cm$^2$ \citep{Fantina2013} and $2.2 \times 10^{45}$ g cm$^2$ \citep{Haensel2007}, respectively). The mass of the Crab pulsar is $M_{\text{Crab}} > 1.2 M_{\odot}$ for $M_{\text{neb}}=2.8 M_{\odot}$ \citep{Bejger2002} and $M_{\text{Crab}} > 1.5 M_{\odot}$ for $M_{\text{neb}}=4.6 M_{\odot}$ \citep{Bejger2003}. As we can see, the corresponding moments of inertia computed with the two new interactions are compatible with the constraint for the nebula mass $M_{\text{neb}}=2.8 M_{\odot}$ and $M_{\text{neb}}=4.6 M_{\odot}$. Neither the prediction of Sky3 nor that of Sky4 could explain a moment of inertia of $3.1 \times 10^{45}$ g cm$^2$ that would be needed if the nebula mass was $M_{\text{neb}}=6.4 M_{\odot}$ \citep{Haensel2007}.
\subsubsection{Instabilities in spin-0 channel for $\beta$-stable neutron star matter}\label{subsubsec:spin-0 instabilities}
\begin{figure}
\begin{center}
\includegraphics[scale=0.54]{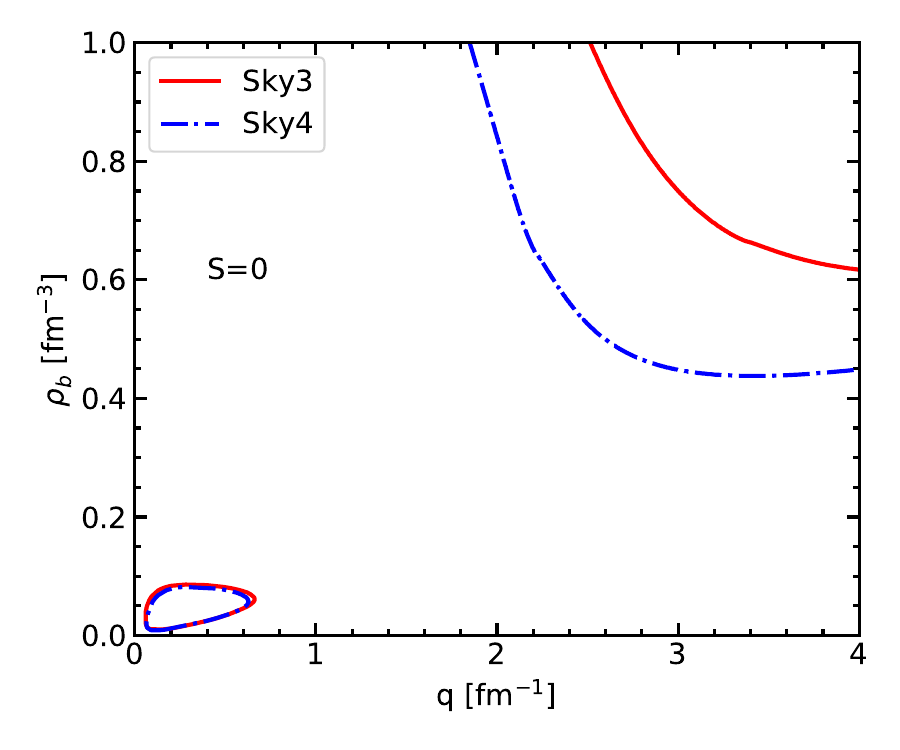} 
\caption{Instabilities in the spin-0 channel for Sky3 and Sky4 for neutron star matter in $\beta$-equilibrium.}
\label{fig:instabilities-s0}
\end{center}
\end{figure}

After determining the proton fraction $Y_p$ through the $\beta$-equilibrium condition at zero temperature, the onset of the instability of neutron star matter can be obtained using Eq.~\eqref{eq:onset-instability}. Figure~\ref{fig:instabilities-s0} shows the critical values of momentum transfer $q$ and baryon density $\rho_b$ for the onset of the instabilities in the spin-0 channel which are computed with the two new Skyrme interactions. We show only the results in the spin-0 channel because our fitting can only provide a good determination of the spin-independent terms.

We see two distinct regions where instabilities appear. The instabilities in the left lower corner of Fig. \ref{fig:instabilities-s0}, i.e., at small $q$ and densities below $\rho_t$, correspond to the spinodal or liquid-gas instability, related to the core-crust transition. To get the correct transition density, it is important to account for the Coulomb interaction and for the electrons. Therefore, in our RPA calculations, we add to the proton-proton particle-hole interaction given in Eq.~(A3) of Ref.~\cite{Duan2023} the screened Coulomb potential \cite{Baym1971b}
\begin{equation}
v^0_{1,pp,\text{Coul}} = \frac{4\pi e^2}{q^2+k_{\text{TF}}^2}
\end{equation}
with
\begin{equation}
k_{\text{TF}}^2 = 4\pi e^2 \frac{d\rho_e}{d\mu_e} = \frac{4\pi}{137}\Big(\frac{3\rho_p}{\pi}\Big)^{2/3}\,.
\end{equation}
Then the values $\rho_t$ given in Table \ref{table:parameters for neutron star models} are obtained as the lowest densities for which the instability is absent for all $q$.

The instabilities in the right upper corner, i.e., at large $q$ and $\rho_b$, are unphysical. However, thanks to the stability constraint in the determination of the $C^{\nabla \rho}_{01}$ and $C^{\nabla \rho}_{11}$ coefficients during the fit, the unphysical instabilities are limited to high densities and momentum transfers. Therefore, they should not pose serious problems for the use of Sky3 and Sky4 in the computation of response functions that are relevant for, e.g., neutrino rates. As explained in Sec.~\ref{subsec:constraints from finite-size instabilities}, the momentum transfer cannot be very large during (proto-)neutron star evolution and supernova explosions. However, in the case of Sky4, there may be a problem at the highest temperatures and densities. Applying the estimates of Sec.~\ref{subsec:constraints from finite-size instabilities} from finite-range instabilites, momentum transfers of $q\approx 2$ fm$^{-1}$, which is where the instabilities set in, can be reached in neutrino scattering at temperatures larger than $40$ MeV for neutrinos that are very far in the tail of the thermal distribution. While this is not the case for Sky3, we were not able to push the instability to higher $q$ in the case of Sky4 without deteriorating the fit of the nuclear radii.

As already mentioned in Sec.~\ref{sec:description of PNM and SNM} in the discussion of the Landau parameter $G_0$, the spin-dependent terms have not been adjusted using an appropriate method. And the fact that $G_0 < -1$ at high densities implies that the $S=1$ instability is even present at $q=0$. The problem of readjusting the spin-dependent terms will be addressed in a separate work.

\subsubsection{Nucleon Fermi velocity and speed of sound}\label{subsubsec:fermiv-soundv}
The Fermi velocity as well as the speed of sound should of course be smaller than the speed of light from the general physical point of view to respect causality. However, since phenomenological Skyrme(-like) energy density functionals are non-relativistic, it is not guaranteed that they fulfil these requirements. Therefore,
we now check the nucleon Fermi velocity and speed of sound described using interactions Sky3 and Sky4.

The fractions of $Y_i$ can be obtained as said in Section \ref{subsubsec:EoS}. Once the neutron and proton fractions are obtained as functions of baryon number density, the corresponding neutron effective mass and Fermi momentum can be given, thus the Fermi velocity $v_F = \hbar k_F/m_N^*$.

Concerning the speed of sound in neutron star matter we follow Ref.~\cite{Goriely2010} and compute
\begin{equation}\label{eq:speed of sound}
\frac{v_s}{c} =  \sqrt{\frac{\partial P}{\partial\varepsilon}\bigg|_{Y_i}} = \sqrt{\frac{1}{\mu_n}\frac{\partial P}{\partial\rho_b}\bigg|_{Y_i}}\,.
\end{equation}
The derivatives have to be taken at frozen composition \cite{Haensel2007,Goriely2010}. Notice that to that end, we cannot use Eq. (\ref{eq:EOS}) which is only valid in $\beta$ equilibrium, but we have to use the more general expression for the pressure
\begin{equation}\label{eq:pressure in ANM}
  P = \mu_n \rho_n+\mu_p \rho_p+\mu_e \rho_e+\mu_\mu \rho_\mu - \varepsilon\,.
\end{equation}
After taking the derivatives, we may use the $\beta$ equilibrium condition which leads to the simplification in the second identity of Eq. (\ref{eq:speed of sound}).
\begin{figure}
\begin{center}
\includegraphics[scale=0.54]{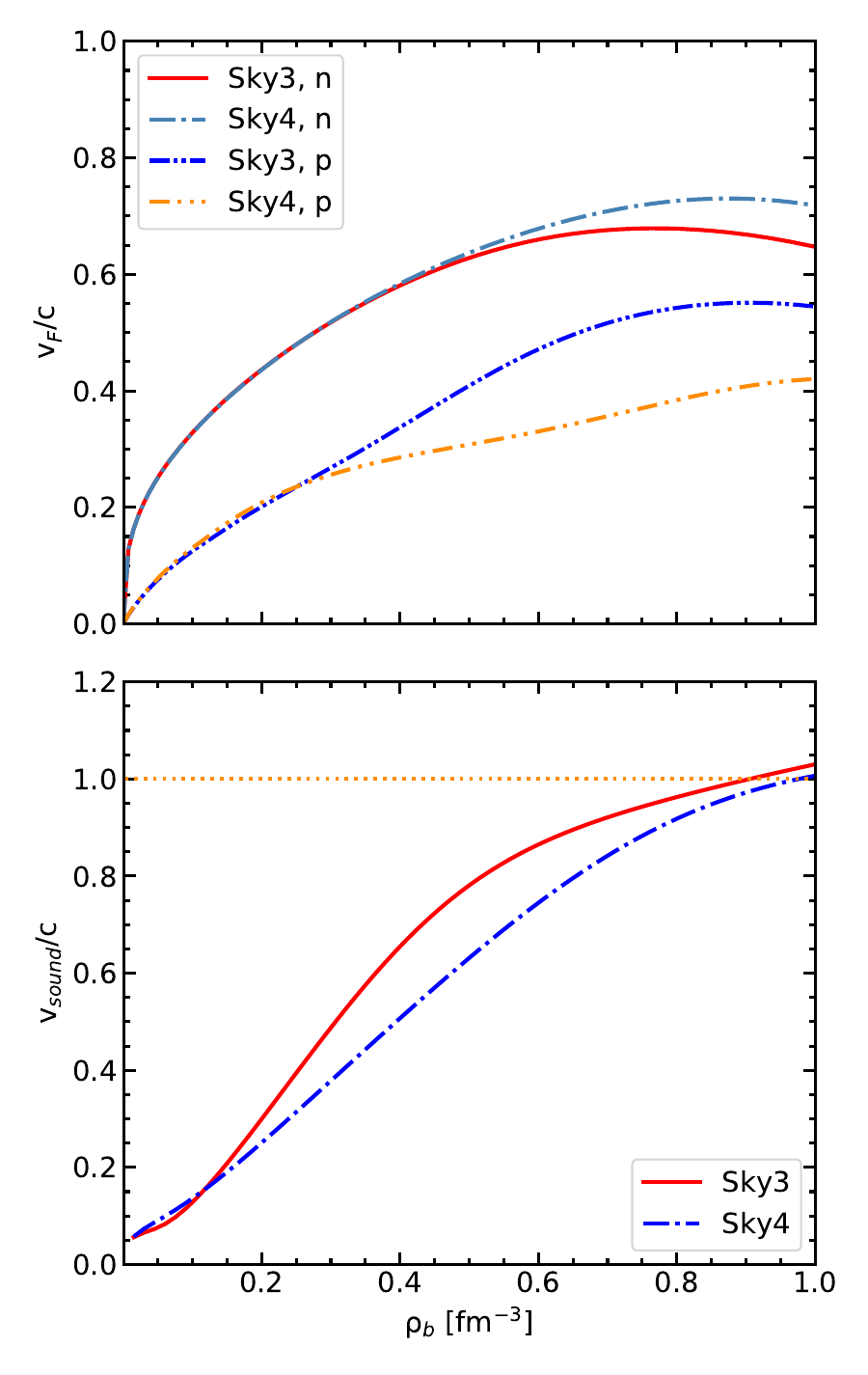} 
\caption{Neutron and proton Fermi velocities (upper panel) and speed of sound (lower panel) as functions of baryon number density for uniform $npe\mu$ matter in $\beta$ equilibrium.}
\label{fig:fermiv-vsound-rho}
\end{center}
\end{figure}

We now present the evolution of the nucleon Fermi velocity and of the speed of sound in Fig.~\ref{fig:fermiv-vsound-rho}. Unlike forces SLy4, BSk19, BSk20, and BSk21, the nucleon Fermi velocity, shown in the upper panel, does not monotonically increase with density for the two new interactions, it increases first, then decreases at higher densities. But it is smaller than the speed of light at all densities, which was one of the motivations for the present work. The reason why Sky3 and Sky4 give different Fermi velocities although they have identical effective masses in PNM and SNM is that the proton fractions in $\beta$-stable matter are not the same.

As we can see in the lower panel of Fig.~\ref{fig:fermiv-vsound-rho}, the speed of sound exceeds the speed of light at densities above $\rho_b \approx $ 0.91 fm$^{-3}$ for Sky3 and $\rho_b \approx$ 0.98 fm$^{-3}$ for Sky4, i.e. already below the maximum central baryon density of a neutron star predicted by Sky3 and Sky4. But this is not a severe problem because the neutron star reaches almost its maximum mass (about 2.35 $M_{\odot}$ for Sky3 and 2.13 $M_{\odot}$ for Sky4) before its central baryon density reaches these critical values (cf. upper panel of Fig.~\ref{fig:M-R-rhoc}).

All in all, the two new Skyrme-like forces seem to be well suited to describe neutron star matter and avoid some of the pathologies of existing interactions in the literature.
\section{Summary and Conclusion} \label{sec:conclusion}
This work aims to get new Skyrme forces that include, in addition to properties of finite nuclei and different microscopic EoS of pure neutron matter, results of microscopic theory for the effective masses \citep{Baldo2014}. This requires to use a functional with many parameters. In practice, we combine the terms that were previously introduced in the BSk \citep{Chamel2009} and KIDS \citep{Gil2019} functionals. We get two sets of parameters in the end: Sky3, which is fitted to the stiffer EoS of neutron matter LZLWBS of \cite{Liu2022}, and Sky4, fitted to the somewhat softer APR EoS \citep{Akmal1998}.

We have shown some properties of finite nuclei computed in Hartree-Fock using the two new forces Sky3 and Sky4, namely binding energies and charge radii of doubly closed shell nuclei that were used in the fit. However, like previously existing forces, our new forces cannot reproduce simultaneously the experimental results for the neutron-skin thickness of $^{208}$Pb and $^{48}$Ca as measured in the PREX \cite{Adhikari2021} and CREX \cite{Adhikari2022} experiments.

Also, we have used the new forces to study properties of nuclear matter, including pure neutron matter, symmetric nuclear matter, and neutron star matter. The effective masses in asymmetric nuclear matter follow roughly those from the microscopic method \citep{Baldo2014}, although the general form of the Skyrme force does not allow one to reproduce the microscopic results very well. This is because the microscopic $1/m^*_n$ and $1/m^*_p$ in asymmetric matter do not split symmetrically around  $1/m^*$ in symmetric matter. To get such a behavior in a Skyrme-like functional, one would have to allow for coupling constants that depend not only on $\rho$ but also on $Y_p$, or for terms of higher order in momentum which could produce such an effect via the different values of $k_{F,n}$ and $k_{F,p}$. In both cases, the number of additional parameters would become very large.

Our new interactions are expected to describe neutron star properties well. The detailed study indicates that the direct Urca process is allowed to occur in neutron star models computed with Sky3 and Sky4. This is one of the advantages compared to previous popular interactions SLy4, BSk20, and BSk21. We also used our new interactions to compute the mass-radius relation, the maximum neutron star mass, neutron-star binding energies, redshifts, and moments of inertia. The results are consistent with available observations. 

Finally, we present the nucleon Fermi velocity and speed of sound in neutron star matter. Our previous paper reported that many Skyrme interactions present the unphysical feature that the neutron Fermi velocity exceeds the speed of light at relatively low densities \citep{Duan2023}. But in Sky3 and Sky4, as a consequence of the behavior of the effective masses, the neutron and proton Fermi velocities stay always below the speed of light at densities that can appear in a neutron star. The speed of sound exceeds the speed of light at densities close to 1 fm$^{-3}$ which, however, do not appear in neutron stars described by Sky3 or Sky4 except very close to the respective maximum masses.

To summarize, we hope that our new Skyrme forces will be useful to compute not only the properties of finite nuclei but also of nuclear matter. Especially, our new forces do not encounter the Fermi velocity problem reported in Ref. \cite{Duan2023}.

However, one should remember that we have not tested our interactions on nuclear excited states or open-shell nuclei. And in the field of neutron stars, we have not yet studied the properties of the inhomogeneous crust. These things remain to be done in the future.

In their present form, our new interactions predict a ferromagnetic instability at densities that may be realized in neutron stars. This is a consequence of the strongly attractive $G_0$ Landau parameter. It is generally assumed that this should not happen. However, if one argues in terms of density-functional theory instead of a density dependent two-body force, the interaction in the spin-1 channel in the functional (responsible, e.g., for the Landau parameters $G_0$, $G'_0$) could be independent from the spin-0 ones \cite{Bender2002}. In some sense, this is already the case if one makes the common choice $\eta_J=0$. Except for the spin-orbit coupling, the spin dependent terms are completely unconstrained by our fit. Their determination should involve experimental data that explicitly depend on them, or microscopic calculations.

Our objective in \cite{Duan2023} was the calculation of response functions of uniform matter in Random-Phase Approximation (RPA) to compute neutrino scattering rates. There we encountered the Fermi velocity problem, which is now solved. However, with the lack of knowledge of the $G_0$ and $G'_0$ Landau parameters or, more generally, of the residual particle-hole interaction in the spin-1 channel, it is still not clear how to make reliable predictions for the neutrino cross sections. Therefore we think that it is urgent to address the problem of the spin-1 particle-hole interaction.

\begin{acknowledgments}
We are grateful to H.-J. Schulze for sending us the data points of the PNM EoS of Ref.~\cite{Liu2022} and to P. Davis and A. Fantina for sending us the results of the CUTER code for the crust. We also thank C. H. Hyun, P. Papakonstantinou, and N. Chamel for discussions about the KIDS and BSk functionals. Mingya Duan is grateful for the support of the China Scholarship Council (CSC No. 202006660002).
\end{acknowledgments}

\appendix


\section{Details of the fits to properties of finite nuclei}
\label{app:details-of-fits}

As explained in Sec.~\ref{sec:fit-finite-nuclei}, the fit to nuclear masses and radii is done in terms of fit variables that are less correlated than the original Skyrme parameters $t_i$ and $y_i$. After setting $\alpha_i = \frac{i}{3}$ and $\beta=\gamma=\frac{1}{3}$ in Eqs.~\eqref{eq:Ccoef}, we
write the $C^{X}_{T}$ coefficients ($X = \rho, \tau, \nabla\rho$, etc.; $T = 0,1$) as
\begin{equation}
  C^{X}_{T}(\rho) = \sum_i C^{X}_{Ti} \rho^{i/3}.
\end{equation}
Then the effective gradient coefficients are defined by
\begin{subequations}
\label{eq:effective gradient C}
\begin{align}
& C^{\nabla \rho}_{0\,\eff} = C^{\nabla\rho}_{00} + C^{\nabla \rho}_{01} \rho^{1/3}_{\eff,0}, \\
& C^{\nabla \rho}_{1\,\eff} = C^{\nabla\rho}_{10} + (C^{\nabla \rho}_{11} - C^{\lap \rho}_{11}) \rho^{1/3}_{\eff,1},
\end{align}   
\end{subequations}
with $\rho_{\eff,0} = 0.074$ fm$^{-3}$ and $\rho_{\eff,1} = 0.069$ fm$^{-3}$ (we found that these values minimize the correlations).
Hence, we can obtain $C^{\nabla \rho}_{T0}$ after determining $C^{\nabla \rho}_{T\,\eff}$ and $C^{\nabla \rho}_{T1}$. The coefficient $C^{\lap \rho}_{11}$ is not an independent parameter since $C^{\lap \rho}_{11}=-C^{\nabla \rho}_{11}-\frac{1}{4} C^\tau_{11}$.

In terms of the nuclear matter properties, namely, $A_1$, $B_1$, $A_2$, $B_2$, $C_1$, $C_2$, $C_3$, $C_4$, $\rho_0$, $E_0$, $K_0$, $\rho_{\high}$ and $E_{\high}$ defined in the main text, we obtain
\begin{subequations}
\label{eq:CofA}
\begin{align}
  C^{\rho}_{03} =&  \frac{3\hbar^2 K_S^2}{10m_N\rho_0 \rho_{\high}^{1/3}}
    -\frac{3}{5} K_S^2 B_1
    -\frac{20 E_0+K_0}{2 D \rho_0^{5/3}}\nonumber\\ &
    +\frac{E_{\high}}{D^3 \rho_{\high}}
    +\frac{5E_0\rho_{\high}^{1/3}}{D^2 \rho_0^{5/3}}
    -\frac{\rho_{\high}^{2/3} E_0}{D^3 \rho_0^{5/3}},\\
  C^{\rho}_{02} =&  -\frac{3\hbar^2 K_S^2}{10m_N \rho_0}
    -\frac{3K_S^2}{5}(A_1+3\rho_0^{1/3} B_1)\nonumber\\ &
    +\frac{12 E_0 + K_0}{2\rho_0^{5/3}}
    -3 \rho_0^{1/3} C^{\rho}_{03},\\
  C^{\rho}_{01} =& \frac{3 \hbar^2 K_S^2}{10m_N \rho_0^{2/3}}
    -\frac{3K_S^2}{5}(2\rho_0^{1/3}A_1+3\rho_0^{2/3} B_1)\nonumber\\ &
    -\frac{3 E_0}{\rho_0^{4/3}}
    -2 \rho_0^{1/3} C^{\rho}_{02} 
    -3 \rho_0^{2/3} C^{\rho}_{03},\\ 
  C^{\rho}_{00} =&  -\frac{\hbar^2 K_S^2}{5m_N \rho_0^{1/3}}
    -\frac{K_S^2}{5}(5\rho_0^{2/3} A_1 + 6\rho_0 B_1)\nonumber\\ &
    -\frac{4}{3} \rho_0^{1/3} C^{\rho}_{01}
    -\frac{5}{3} \rho_0^{2/3} C^{\rho}_{02}
    -2 \rho_0 C^{\rho}_{03},\\
  C^{\rho}_{10} =& C_1-C^{\rho}_{00},\\
  C^{\rho}_{11} =& C_2-C^{\rho}_{01},\\
  C^{\rho}_{12} =& C_3-C^{\rho}_{02}
    +\frac{3K_N^2}{10} (A_2-2A_1),\\
  C^{\rho}_{13} =& C_4-C^{\rho}_{03}
    +\frac{3K_N^2}{10} (B_2-2B_1),\\
  C^{\tau}_{00} =& A_1,\\
  C^{\tau}_{01} =& B_1,\\
  C^{\tau}_{10} =& -\frac{A_2}{2},\\
  C^{\tau}_{11} =& -\frac{B_2}{2},
\end{align}
\end{subequations}
where the following abbreviations were used:
\begin{equation}
  K_S = \Big(\frac{3\pi^2}{2}\Big)^{1/3},\quad 
  K_N = (3\pi^2)^{1/3},\quad
  D = \rho_{\high}^{1/3}-\rho_0^{1/3}.
\end{equation}

Finally, we can solve for the Skyrme parameters:
\begin{subequations}
\label{eq:t0-y5}
\begin{align}
  t_0 &= \tfrac{8}{3} C^{\rho}_{00},\label{tofCfirst}\\
  t_1 &= \tfrac{4}{3} C^{\tau}_{00}+\tfrac{16}{3} C^{\nabla\rho}_{00},\\
  t_2 &= 4 C^{\tau}_{00}-8 C^{\tau}_{10}
    -\tfrac{16}{3}C^{\nabla\rho}_{00}+\tfrac{32}{3} C^{\nabla\rho}_{10},\\
  t_{31} &= 16 C^{\rho}_{01},\\
  t_{32} &= 16 C^{\rho}_{02},\\
  t_{33} &= 16 C^{\rho}_{03},\\
  t_4 &= \tfrac{8}{7} C^{\tau}_{01}+\tfrac{32}{7} C^{\nabla\rho}_{01},\\
  t_5 &= \tfrac{88}{21} C^{\tau}_{01}-\tfrac{16}{3} C^{\tau}_{11}
    +\tfrac{64}{3} C^{\nabla\rho}_{11}-\tfrac{32}{7} C^{\nabla\rho}_{01},\\
  y_0 &= -\tfrac{4}{3} C^{\rho}_{00}-4 C^{\rho}_{10},\\
  y_1 &= -\tfrac{2}{3}C^{\tau}_{00}-2 C^{\tau}_{10} -\tfrac{8}{3} C^{\nabla\rho}_{00}
    -8 C^{\nabla\rho}_{10},\\
  y_2 &= -2 C^{\tau}_{00}+10 C^{\tau}_{10}+\tfrac{8}{3} C^{\nabla\rho}_{00}
    -\tfrac{40}{3} C^{\nabla\rho}_{10},\\    
  y_{31} &= -8 C^{\rho}_{01}-24 C^{\rho}_{11},\\
  y_{32} &= -8 C^{\rho}_{02}-24 C^{\rho}_{12},\\
  y_{33} &= -8 C^{\rho}_{03}-24 C^{\rho}_{13},\\
  y_4 &= -\tfrac{4}{7}C^{\tau}_{01}-4 C^{\tau}_{11}-\tfrac{16}{7}C^{\nabla\rho}_{01}
    -16 C^{\nabla\rho}_{11},\\
  y_5 &= -\tfrac{44}{21} C^{\tau}_{01}+\tfrac{20}{3} C^{\tau}_{11}
    +\tfrac{16}{7} C^{\nabla\rho}_{01}-\tfrac{80}{3} C^{\nabla\rho}_{11}.\label{tofClast}
\end{align}
\end{subequations}

Because they are strongly correlated, error bars of the Skyrme parameters $t_i$ and $y_i$ are not meaningful. But we can quote errors and correlation coefficients for the fit parameters in Tables~\ref{table:parameter-errors}-\ref{table:Sky4-correlations}. Notice that these errors neither include the uncertainties from the BHF calculations in the parameters $A_1$ \dots $C_4$ and $E_{\high}$, nor remaining correlations with the parameters $C^{\nabla\rho}_{T1}$ which are only constrained from the stability requirement.
\begin{table}
    \begin{ruledtabular}
    \begin{tabular}{lcc}
                        & Sky3               & Sky4               \\
    \hline
    $\rho_0$ (fm$^{-3}$)& $0.154\pm 0.003$   & $0.153\pm 0.003$   \\
    $E_0$ (MeV)         & $-15.93\pm 0.05$   & $-15.94\pm 0.05$   \\
    $K_0$ (MeV)         & $252.5\pm 20.2$    & $229.4\pm 20.0$    \\
    $C^{\nabla\rho}_{0\,\eff}$ (MeV fm$^5$) 
                        & $63.1 \pm 2.5$     & $65.4\pm 2.4$      \\
    $C^{\nabla\rho}_{1\,\eff}$ (MeV fm$^5$) 
                        & $26.1 \pm 19.8$    & $-21.1\pm 18.9$    \\
    $W_0$ (MeV fm$^5$)  & $111.5 \pm 5.2$    & $108.2\pm 5.1$
    \end{tabular}
    \end{ruledtabular}
    \caption{Results and parameter uncertainties from the fits of nuclear binding energies and charge radii.}
    \label{table:parameter-errors}
\end{table}
\begin{table}
    \begin{ruledtabular}
    \begin{tabular}{l|cccccc}
    & $\rho_0$ & $E_0$ & $K_0$ & $C^{\nabla\rho}_{0\,\eff}$ & $C^{\nabla\rho}_{1\,\eff}$ & $W_0$\\
    \hline
    $\rho_0$ &$ 1     $&$-0.052 $&$-0.915 $&$ 0.835 $&$-0.297 $&$ 0.023 $\\
    $E_0$    &$-0.052 $&$ 1     $&$-0.257 $&$-0.101 $&$-0.316 $&$ 0.173 $\\
    $K_0$    &$-0.915 $&$-0.257 $&$ 1     $&$-0.844 $&$ 0.312 $&$-0.081 $\\
    $C^{\nabla\rho}_{0\,\eff}$
             &$ 0.835 $&$-0.101 $&$-0.844 $&$ 1     $&$-0.167 $&$ 0.373 $\\
    $C^{\nabla\rho}_{1\,\eff}$ 
             &$-0.297 $&$-0.316 $&$ 0.312 $&$-0.167 $&$ 1     $&$ 0.196 $\\
    $W_0$    &$ 0.023 $&$ 0.173 $&$-0.081 $&$ 0.373 $&$ 0.196 $&$ 1     $
    \end{tabular}
    \end{ruledtabular}
    \caption{Correlation coefficients \cite{James1994} from the fits of nuclear binding energies and charge radii for Sky3.}
    \label{table:Sky3-correlations}
\end{table}
\begin{table}
    \begin{ruledtabular}
    \begin{tabular}{l|cccccc}
    & $\rho_0$ & $E_0$ & $K_0$ & $C^{\nabla\rho}_{0\,\eff}$ & $C^{\nabla\rho}_{1\,\eff}$ & $W_0$\\
    \hline
    $\rho_0$ &$ 1     $&$ 0.012 $&$-0.921 $&$ 0.859 $&$-0.319 $&$-0.020 $\\
    $E_0$    &$ 0.012 $&$ 1     $&$-0.321 $&$-0.039 $&$-0.351 $&$ 0.167 $\\
    $K_0$    &$-0.921 $&$-0.321 $&$ 1     $&$-0.851 $&$ 0.352 $&$-0.030 $\\
    $C^{\nabla\rho}_{0\,\eff}$
             &$ 0.859 $&$-0.039 $&$-0.851 $&$ 1     $&$-0.208 $&$ 0.322 $\\
    $C^{\nabla\rho}_{1\,\eff}$ 
             &$ -0.319$&$-0.351 $&$ 0.352 $&$-0.208 $&$ 1     $&$ 0.208 $\\
    $W_0$    &$ -0.020$&$ 0.167 $&$-0.030 $&$ 0.322 $&$ 0.208 $&$       $1
    \end{tabular}
    \end{ruledtabular}
    \caption{Correlation coefficients \cite{James1994} from the fits of nuclear binding energies and charge radii for Sky4.}
    \label{table:Sky4-correlations}
\end{table}
\section{Details of the Hartree-Fock calculations with the new interactions}
\label{app:hartreefock}
For completeness, let us give the expressions for the effective masses and mean fields in the Hartree-Fock calculations:
\begin{align}\label{eq:nucleon-effective-masses}
\frac{\hbar^2}{2m_{q}^{*}} =\; & \frac{\hbar^2}{2m_N} +\frac{1}{4} \left(t_1+\frac{1}{2} y_1 + t_2 + \frac{1}{2} y_2 \right) \rho \nonumber\\
&+ \frac{1}{4} \left (\frac{1}{2} t_2 + y_2 - \frac{1}{2} t_1 - y_1 \right) \rho_q \notag\\
&+\frac{1}{4} \left [ \left(t_4+\frac12 y_4 \right) \rho^{\beta} + \left(t_5 + \frac{1}{2} y_5\right) \rho^{\gamma} \right ] \rho \nonumber\\
&+ \frac{1}{4} \left [ \left(\frac{1}{2} t_5 + y_5\right) \rho^{\gamma} - \left(\frac{1}{2} t_4 + y_4\right) \rho^{\beta} \right ] \rho_q.
\end{align}
\begin{widetext}
\begin{align}\label{eq:nuclear-central-field-new}
U_q = & \left( t_0 + \frac{1}{2} y_0 \right) \rho - \left( \frac{1}{2} t_0 + y_0 \right) \rho_q \notag \\
& + \frac{1}{4} \left[ \left( t_1 + \frac{1}{2} y_1 \right) \left( \tau - \frac{3}{2} \nabla^2 \rho \right) - \left( \frac{1}{2} t_1 + y_1 \right) \left( \tau_q - \frac{3}{2} \nabla^2 \rho_q \right) \right] \notag \\
& + \frac{1}{4} \left[ \left( t_2 + \frac{1}{2} y_2 \right) \left( \tau + \frac{1}{2} \nabla^2 \rho \right) + \left( \frac{1}{2} t_2 + y_2 \right) \left(\tau_q + \frac{1}{2} \nabla^2 \rho_q \right) \right ] \notag \\
& + \frac{1}{12} \sum_{i=1}^{3} \Bigg[ \left( t_{3i} + \frac{1}{2} y_{3i} \right) (2+\alpha_i) \rho^{\alpha_i+1} - \left( \frac{1}{2} t_{3i} + y_{3i} \right) \Bigg( 2 \rho^{\alpha_i} \rho_q + \alpha_i \rho^{\alpha_i-1} \sum_{q^{\prime}=n,p} \rho_{q^{\prime}}^2 \Bigg) \Bigg] \notag \\
& + \frac{1}{8} \rho^{\beta-1} \left( t_4 + \frac{1}{2} y_4 \right) \left \lbrace 2(1+\beta)\rho \tau -(2 \beta + 3) \left[ \frac{1}{2} \beta (\nabla \rho)^2 + \rho \nabla^2 \rho \right] \right\rbrace \notag \\
& + \frac{1}{8} \rho^{\beta-2} \left( \frac{1}{2} t_4 + y_4 \right) \Bigg\lbrace 3 \beta \rho \nabla \rho \cdot \nabla \rho_q + 3 \rho^2 \nabla^2 \rho_q - 2 \rho^2 \tau_q + \beta (\beta-1) \rho_q (\nabla \rho)^2 \notag \\
& \quad + \beta \rho \rho_q \nabla^2 \rho -\frac{1}{2} \beta \rho \sum_{q^{\prime}=n,p} [ (\nabla \rho_{q^{\prime}})^2 + 4 \rho_{q^{\prime}} \tau_{q^{\prime}} - 2 \rho_{q^{\prime}} \nabla^2 \rho_{q^{\prime}} ] \Bigg\rbrace \notag \\
& + \frac{1}{4}\rho^{\gamma -1} \left( t_5 + \frac{1}{2} y_5 \right) \left[ (1+\gamma) \rho \tau + \frac{1}{4} \gamma (\nabla \rho)^2 + \frac{1}{2} \rho \nabla^2 \rho \right] \notag \\
& + \frac{1}{4} \rho^{\gamma-1} \left ( \frac{1}{2} t_5 + y_5 \right ) \Bigg\lbrace \rho \tau_q + \frac{1}{2} \rho \nabla^2 \rho_q + \gamma \sum_{q^{\prime}=n,p} \left[ \rho_{q^{\prime}} \tau_{q^{\prime}} - \frac{1}{4} (\nabla \rho_{q^{\prime}})^2 \right ] + \frac{1}{2} \gamma \nabla \rho \cdot \nabla \rho_q \Bigg\rbrace \notag \\
& +\frac{\eta_J}{16}\Bigg[- (y_4 \beta \rho^{\beta -1} + y_5 \gamma \rho^{\gamma -1}) J^2 + (t_4 \beta \rho^{\beta -1} - t_5 \gamma \rho^{\gamma -1}) \sum_{q^{\prime}=n,p} J_{q^{\prime}}^2\Bigg]\notag \\
& - \frac{1}{2} W_0 (\nabla \cdot \vect{J} + \nabla \cdot \vect{J}_{q}) + \delta_{q,p} V^{\text{Coul}}.
\end{align}
\end{widetext}

\begin{align}\label{eq:spin-orbit-field-new}
\vect{W}_q =\;& \frac{1}{2} W_0 \nabla (\rho + \rho_q)
  - \frac{\eta_J}{8} (y_1+y_2+y_4 \rho^{\beta} + y_5 \rho^{\gamma}) \vect{J} \nonumber\\
  &+ \frac{\eta_J}{8} (t_1-t_2+t_4 \rho^{\beta} - t_5 \rho^{\gamma}) \vect{J}_q.
\end{align}

Although they are not needed for our interactions that use $\eta_J=0$ (\texttt{iftm=0} in the SKHAFO code \cite{Reinhard1991}), we have also written the contributions of the $\mathbb{J}^2$ terms in the functional (proportional to $\eta_J$) to the central and spin-orbit fields, because they are needed for some other parametrizations that use $\eta_J=1$ (e.g., SLy5, BSk functionals up to BSk18, KIDS).

The expression for the rearrangement energy used in the Hartree-Fock code to compute the total energy (cf. Eqs. (2.13) and (2.15) in \cite{Reinhard1991}), has to be modified as follows to account for the additional terms:
\begin{widetext}
\begin{align}\label{eq:rearrangement energy}
E_{\text{rearr}} =   -  \int_{0}^{\infty} d^3 r & \Bigg\{ \sum_{i=1}^{3} \frac{\alpha_i}{24}  \rho^{\alpha_i} \left[ \left( t_{3i}+\frac{1}{2} y_{3i} \right) \rho^2 - \left( \frac{1}{2} t_{3i} +y_{3i} \right) \sum_{q=n,p}\rho_q^2 \right] \notag \\
& + \frac{1}{16}  \left( t_4+ \frac{1}{2} y_4 \right) \beta \rho^{\beta} \left[ 2 \rho \tau +\left( \beta + \frac{3}{2} \right) (\nabla \rho)^2 \right] \notag \\
& + \frac{1}{16}  \left( \frac{1}{2} t_4 + y_4 \right) \rho^{\beta-1} \sum_{q=n,p}\left( -\frac{1}{2} \beta \rho \left[ 3(\nabla \rho_q)^2 + 4 \rho_q \tau_q \right] -\beta^2 \rho_q (\nabla \rho)\cdot (\nabla \rho_q) \right) \notag \\
& + \frac{1}{8}  \left( t_5+\frac12 y_5 \right) \gamma \rho^{\gamma} \left[ \rho \tau - \frac{1}{4} (\nabla \rho)^2 \right] + \frac{1}{8}  \left( \frac{1}{2} t_5 + y_5 \right) \gamma \rho^{\gamma} \sum_{q=n,p}\left( \rho_q \tau_q -\frac{1}{4} (\nabla \rho_q)^2 \right) \notag \\
& + \frac{\eta_J}{32} \left[ - (y_4 \beta \rho^{\beta} + y_5 \gamma \rho^{\gamma}) J^2 + (t_4 \beta \rho^{\beta} - t_5 \gamma \rho^{\gamma}) \sum_{q=n,p} J_q^2 \right]  \Bigg \} + E_{\text{Coul,rearr}}.
\end{align}
\end{widetext}
The center-of-mass correction is expressed as (\texttt{ifzpe=0} in the code)
\begin{equation}\label{eq:center-of-mass}
\frac{1}{2\mnucleon} \rightarrow \frac{1}{2\mnucleon} \left(1- \frac{1}{A} \right)
\end{equation}
($A$ is the nucleon number).
The proton density is used in the Coulomb potential (\texttt{ifrhoc=0}) and the Coulomb exchange is taken into account (\texttt{ifex=1}).

\section{EoS of uniform matter and related quantities}
\label{app:EoS-uniform-matter}
Using in $\mathcal{H}$ the expression $\tau_q=\frac{3}{5} (3\pi^2)^{2/3} \rho_q^{5/3}$ and discarding all terms containing $\nabla\rho$ or $\vect{J}$, as well as the Coulomb energy, one obtains the energy density of uniform matter. Expressing $\rho_n$ and $\rho_p$ in terms of $\rho = \rho_n+\rho_p$ and $Y_p = \rho_p/\rho$, the corresponding energy per particle can be written as
\begin{align}\label{eq:energy-density-infinite-ANM}
\frac{E}{A} = & \frac{3\hbar^2}{10\mnucleon} \Big(\frac{3\pi^2}{2} \Big)^{\!2/3} \rho^{2/3} F_{5/3}(Y_p) \notag \\
& + \frac{1}{8} \rho [ 2(y_0+2t_0) - (2y_0 +t_0) F_2(Y_p) ] \notag \\
& + \frac{1}{48} \sum_{i=1}^{3} \rho^{\alpha_i +1} [ 2(y_{3i} +2t_{3i}) - (2y_{3i} +t_{3i})F_2(Y_p) ] \notag \\
& + \frac{3}{40} \Big( \frac{3\pi^2}{2} \Big)^{\!2/3} \rho^{5/3} ( y_1 +2t_1 +y_2+2t_2 )F_{5/3}(Y_p) \notag \\
& +\frac{3}{80} \Big( \frac{3\pi^2}{2} \Big)^{\!2/3} \rho^{5/3} (2y_2 +t_2-2y_1-t_1)F_{8/3}(Y_p) \notag \\
& + \frac{3}{40} \Big( \frac{3\pi^2}{2} \Big)^{\!2/3} \rho^{\beta+5/3} (y_4+2t_4) F_{5/3}(Y_p) \notag \\
& - \frac{3}{80} \Big( \frac{3\pi^2}{2} \Big)^{\!2/3} \rho^{\beta+5/3} (2y_4+t_4) F_{8/3}(Y_p) \notag \\
& + \frac{3}{40} \Big( \frac{3\pi^2}{2} \Big)^{\!2/3} \rho^{\gamma+5/3} (y_5+2t_5) F_{5/3}(Y_p) \notag \\
& + \frac{3}{80} \Big( \frac{3\pi^2}{2} \Big)^{\!2/3} \rho^{\gamma+5/3} (2y_5+t_5) F_{8/3}(Y_p),
\end{align}
where $F_m(Y_p) = 2^{m-1} [Y_p^m + (1-Y_p)^m]$. The special cases of pure neutron matter and symmetric nuclear matter are readily obtained from this general equation by setting
\begin{equation}
    F_m(Y_p) = 
    \begin{cases} 
      2^{m-1}& \mbox{for PNM $(Y_p=0)$,}\\ 1& \mbox{for SNM $(Y_p=\tfrac{1}{2})$.} 
    \end{cases}
\end{equation}

Thus, the energy per nucleon for symmetric nuclear matter is
\begin{align}\label{eq:energy per nucleon}
\frac{E}{A} (\rho) = & \frac{3\hbar^2}{10\mnucleon} \Big( \frac{3\pi^2}{2} \Big)^{\!2/3} \rho^{2/3} + \frac{3}{8} t_0 \rho \notag \\
& +\frac{3}{80} (3t_1+5t_2+4y_2)\Big( \frac{3\pi^2}{2} \Big)^{\!2/3} \rho^{5/3} \notag \\
& + \frac{1}{16} \sum_{i=1}^{3} t_{3i} \rho^{\alpha_i+1} +\frac{9}{80} t_4 \Big( \frac{3\pi^2}{2} \Big)^{\!2/3} \rho^{\beta + 5/3} \notag \\
& + \frac{3}{80} (5t_5+4y_5) \Big( \frac{3\pi^2}{2} \Big)^{\!2/3} \rho^{\gamma + 5/3}.
\end{align}

With this formula and the definition Eq.~\eqref{eq:K_inf}, it is straight-forward to compute the incompressibility modulus
\begin{align}\label{eq:K_inf-explicit}
K_{0} & = -\frac{3\hbar^2}{5\mnucleon} \Big( \frac{3\pi^2}{2} \Big)^{\!2/3} \rho_0^{2/3} \notag \\
& \quad + \frac{3}{8} (3 t_1 + 5t_2+4y_2) \Big( \frac{3\pi^2}{2} \Big)^{\!2/3} \rho_0^{5/3} \notag \\
& \quad + \frac{9}{16} \sum_{i=1}^{3} t_{3i} (\alpha_i+1) \alpha_i \rho_0^{\alpha_i+1} \notag \\
& \quad + \frac{81}{80} t_4 \Big( \frac{3\pi^2}{2} \Big)^{\!2/3} \left (\beta+\tfrac{5}{3} \right ) \left (\beta +\tfrac{2}{3} \right ) \rho_0^{\beta + 5/3} \notag\\
& \quad + \frac{27}{80}(5 t_5+4y_5) \Big( \frac{3\pi^2}{2} \Big)^{\!2/3} \left ( \gamma + \tfrac{5}{3} \right ) \left( \gamma+ \tfrac{2}{3} \right ) \rho_0^{\gamma + 5/3}.
\end{align}

To compute the symmetry energy Eq. (\ref{eq:symmetry energy}), notice that $\delta = 1-2Y_p$. Setting $Y_p=\frac{1}{2}$ after taking the second derivative one gets
\begin{align}\label{eq:symmetry energy-explicit}
E_{\sym} (\rho) = & \frac{\hbar^2}{6\mnucleon} \Big( \frac{3\pi^2}{2} \Big)^{\!2/3} \rho^{2/3} - \frac{1}{8} (2y_0+t_0) \rho \notag \\
& +\frac{1}{24} (-3y_1 + 4t_2+5y_2) \Big( \frac{3\pi^2}{2} \Big)^{\!2/3} \rho^{5/3} \notag \\
& -\frac{1}{48} \sum_{i=1}^{3} (t_{3i}+2y_{3i}) \rho^{\alpha_i +1} \notag\\
& -\frac{1}{8} y_4 \Big( \frac{3\pi^2}{2} \Big)^{\!2/3} \rho^{\beta+5/3} \notag \\
& + \frac{1}{24}(4t_5+5y_5) \Big( \frac{3\pi^2}{2} \Big)^{\!2/3} \rho^{\gamma + 5/3}.
\end{align}

Finally, the explicit formula for $L_0$ defined in Eq. (\ref{eq:density-symmetry coefficient}) reads:
\begin{align}\label{eq:density-symmetry coefficient-explicit}
L_0 & = \frac{\hbar^2}{3\mnucleon} \Big( \frac{3\pi^2}{2} \Big)^{\!2/3} \rho_0^{2/3} - \frac{3}{8} (2y_0+t_0) \rho_0 \notag \\
& \quad + \frac{5}{24} (-3y_1 + 4t_2+ 5y_2) \Big( \frac{3\pi^2}{2} \Big)^{\!2/3} \rho_0^{5/3} \notag \\
& \quad -\frac{1}{16} \sum_{i=1}^{3} (t_{3i}+2y_{3i}) (\alpha_i+1) \rho_0^{\alpha_i+1} \notag \\
& \quad - \frac{3}{8} y_4 \Big( \frac{3\pi^2}{2} \Big)^{\!2/3} \left(\beta + \tfrac{5}{3}\right) \rho_0^{\beta +5/3} \notag \\
& \quad + \frac{1}{8} (4t_5+5y_5) \Big( \frac{3\pi^2}{2} \Big)^{\!2/3} \left(\gamma+\tfrac{5}{3}\right) \rho_0^{\gamma+ 5/3}.
\end{align}

\section{Landau parameters in pure neutron matter and symmetric nuclear matter}
\label{app:landau parameters in PNM and SNM}

The Landau parameters for BSk interactions were given in the appendix of Ref. \cite{Chamel2009}. It is straight-forward to generalize the $C^\rho$ and $C^s$ coefficients to the case of our new Skyrme forces. For completeness, we give here the expressions:

The dimensionless Landau parameters in pure neutron matter for our new Skyrme forces are
\begin{subequations}
\label{eq:landau-parameters-PNM}
\begin{align}
 F_0^{\text{PNM}} = & N_0 \lbrace 2(C_0^{\rho} + C_1^{\rho}) +2k_{F}^2(C_0^{\tau} +C_1^{\tau}) \notag \\
&  + 4 \rho[(C_0^{\rho})^{\prime} + (C_1^{\rho})^{\prime} ] + \rho^2[(C_0^{\rho})^{\prime \prime} + (C_1^{\rho})^{\prime \prime}] \notag \\
&  + \rho \tau [(C_0^{\tau})^{\prime \prime} + (C_1^{\tau})^{\prime \prime}] \notag \\
&  + 2[(C_0^{\tau})^{\prime} + (C_1^{\tau})^{\prime}](\tau+\rho k_{F}^2) \rbrace,\\
 F_1^{\text{PNM}}= & -2N_0(C_0^{\tau} +C_1^{\tau}) k_{F}^2,\\
 G_0^{\text{PNM}}= &2N_0[C_0^s+C_1^s+k_{F}^2(C_0^{sT} +C_1^{sT})], \\
 G_1^{\text{PNM}}= &-2N_0k_{F}^2(C_0^{sT}+C_1^{sT}),
\end{align}
\end{subequations}
while the Landau parameters in symmetric nuclear matter for our new Skyrme forces are
\begin{subequations}
\label{eq:landau-parameters-SNM}
\begin{align}
 F_0= & 2N_0[2(C_0^{\rho} +C_0^{\tau} k_F^2) + 4\rho (C_0^{\rho})^{\prime} + \rho^2 (C_0^{\rho})^{\prime \prime} \notag \\
&  + \rho \tau (C_0^{\tau})^{\prime \prime} + 2(C_0^{\tau})^{\prime}(\tau+\rho k_F^2)], \\
 F_0^{\prime}= & 4N_0(C_1^{\rho} + C_1^{\tau} k_F^2), \\
 F_1=& -4N_0C_0^{\tau} k_F^2, \\
 F_1^{\prime}=& -4N_0C_1^{\tau} k_F^2,\\
 G_0=& 4N_0(C_0^s+C_0^{sT}k_F^2), \\
 G_0^{\prime}=& 4N_0(C_1^s+C_1^{sT}k_F^2),\\
 G_1=& -4N_0C_0^{sT}k_F^2,\\
 G_1^{\prime}=& -4N_0 C_1^{sT}k_F^2,
\end{align}
\end{subequations}
where $N_0=\frac{m^{*}k_F}{\pi^2\hbar^2}$, $\tau=\frac{3}{5} \rho k_F^2$, and $k_F=(3 \pi^2 \rho)^{1/3}$ for pure neutron matter and $k_F=(\frac{3\pi^2 \rho}{2})^{1/3}$ for symmetric nuclear matter.

The $C^X_i$ coefficients entering the above equations are given in Eqs.~\eqref{eq:Ccoef} and
\begin{align}\label{eq:Cs0}
C_0^s=& -\frac{1}{8} (t_0-2y_0) - \sum_{i=1}^3 \frac{1}{48} (t_{3i} - 2y_{3i})\rho^{\alpha_i}, \\
C_1^s=& -\frac{1}{8} t_0-\sum_{i=1}^3 \frac{1}{48} t_{3i} \rho^{\alpha_i},\label{eq:Cs1}
\end{align}
Additionally, $C_i^{\prime}=\dfrac{dC_i}{d\rho}$ and $C_i^{\prime \prime}=\dfrac{d^2C_i}{d\rho^2}$.

\bibliography{references}

\begin{thebibliography}{78}
\expandafter\ifx\csname natexlab\endcsname\relax\def\natexlab#1{#1}\fi
\expandafter\ifx\csname bibnamefont\endcsname\relax
  \def\bibnamefont#1{#1}\fi
\expandafter\ifx\csname bibfnamefont\endcsname\relax
  \def\bibfnamefont#1{#1}\fi
\expandafter\ifx\csname citenamefont\endcsname\relax
  \def\citenamefont#1{#1}\fi
\expandafter\ifx\csname url\endcsname\relax
  \def\url#1{\texttt{#1}}\fi
\expandafter\ifx\csname urlprefix\endcsname\relax\def\urlprefix{URL }\fi
\providecommand{\bibinfo}[2]{#2}
\providecommand{\eprint}[2][]{\url{#2}}

\bibitem[{\citenamefont{{Cerda-Duran} and
  {Elias-Rosa}}(2018)}]{Cerda-DuranElias-Rosa2018}
\bibinfo{author}{\bibfnamefont{P.}~\bibnamefont{{Cerda-Duran}}}
  \bibnamefont{and}
  \bibinfo{author}{\bibfnamefont{N.}~\bibnamefont{{Elias-Rosa}}}, in
  \emph{\bibinfo{booktitle}{Astrophysics and Space Science Library}}, edited by
  \bibinfo{editor}{\bibfnamefont{L.}~\bibnamefont{{Rezzolla}}},
  \bibinfo{editor}{\bibfnamefont{P.}~\bibnamefont{{Pizzochero}}},
  \bibinfo{editor}{\bibfnamefont{D.~I.} \bibnamefont{{Jones}}},
  \bibinfo{editor}{\bibfnamefont{N.}~\bibnamefont{{Rea}}}, \bibnamefont{and}
  \bibinfo{editor}{\bibfnamefont{I.}~\bibnamefont{{Vida{\~n}a}}}
  (\bibinfo{year}{2018}), vol. \bibinfo{volume}{457} of
  \emph{\bibinfo{series}{Astrophysics and Space Science Library}},
  p.~\bibinfo{pages}{1}.

\bibitem[{\citenamefont{{Kumar} and {Zhang}}(2015)}]{Kumar2015}
\bibinfo{author}{\bibfnamefont{P.}~\bibnamefont{{Kumar}}} \bibnamefont{and}
  \bibinfo{author}{\bibfnamefont{B.}~\bibnamefont{{Zhang}}},
  \bibinfo{journal}{Phys. Rep.} \textbf{\bibinfo{volume}{561}},
  \bibinfo{pages}{1} (\bibinfo{year}{2015}).

\bibitem[{\citenamefont{{Abbott} et~al.}(2017)\citenamefont{{Abbott}, {Abbott},
  {Abbott}, {Acernese}, {Ackley}, {Adams}, {Adams}, {Addesso}, {Adhikari},
  {Adya} et~al.}}]{Abbott2017}
\bibinfo{author}{\bibfnamefont{B.~P.} \bibnamefont{{Abbott}}},
  \bibinfo{author}{\bibfnamefont{R.}~\bibnamefont{{Abbott}}},
  \bibinfo{author}{\bibfnamefont{T.~D.} \bibnamefont{{Abbott}}},
  \bibinfo{author}{\bibfnamefont{F.}~\bibnamefont{{Acernese}}},
  \bibinfo{author}{\bibfnamefont{K.}~\bibnamefont{{Ackley}}},
  \bibinfo{author}{\bibfnamefont{C.}~\bibnamefont{{Adams}}},
  \bibinfo{author}{\bibfnamefont{T.}~\bibnamefont{{Adams}}},
  \bibinfo{author}{\bibfnamefont{P.}~\bibnamefont{{Addesso}}},
  \bibinfo{author}{\bibfnamefont{R.~X.} \bibnamefont{{Adhikari}}},
  \bibinfo{author}{\bibfnamefont{V.~B.} \bibnamefont{{Adya}}},
  \bibnamefont{et~al.}, \bibinfo{journal}{Phys. Rev. Lett.}
  \textbf{\bibinfo{volume}{119}}, \bibinfo{eid}{161101} (\bibinfo{year}{2017}).

\bibitem[{\citenamefont{{Zhang}}(2020)}]{Zhang2020}
\bibinfo{author}{\bibfnamefont{B.}~\bibnamefont{{Zhang}}},
  \bibinfo{journal}{Nature} \textbf{\bibinfo{volume}{587}}, \bibinfo{pages}{45}
  (\bibinfo{year}{2020}).

\bibitem[{\citenamefont{{Shklovsky}}(1967)}]{Shklovsky1967}
\bibinfo{author}{\bibfnamefont{I.~S.} \bibnamefont{{Shklovsky}}},
  \bibinfo{journal}{Astrophys. J. Lett.} \textbf{\bibinfo{volume}{148}},
  \bibinfo{pages}{L1} (\bibinfo{year}{1967}).

\bibitem[{\citenamefont{{Yakovlev} and {Pethick}}(2004)}]{Yakovlev2004}
\bibinfo{author}{\bibfnamefont{D.~G.} \bibnamefont{{Yakovlev}}}
  \bibnamefont{and} \bibinfo{author}{\bibfnamefont{C.~J.}
  \bibnamefont{{Pethick}}}, \bibinfo{journal}{Ann. Rev. Astron. Astrophys.}
  \textbf{\bibinfo{volume}{42}}, \bibinfo{pages}{169} (\bibinfo{year}{2004}).

\bibitem[{\citenamefont{{Vautherin} and {Brink}}(1972)}]{Vautherin1972}
\bibinfo{author}{\bibfnamefont{D.}~\bibnamefont{{Vautherin}}} \bibnamefont{and}
  \bibinfo{author}{\bibfnamefont{D.~M.} \bibnamefont{{Brink}}},
  \bibinfo{journal}{Phys. Rev. C} \textbf{\bibinfo{volume}{5}},
  \bibinfo{pages}{626} (\bibinfo{year}{1972}).

\bibitem[{\citenamefont{{Beiner} et~al.}(1975)\citenamefont{{Beiner},
  {Flocard}, {Van Giai}, and {Quentin}}}]{Beiner1975}
\bibinfo{author}{\bibfnamefont{M.}~\bibnamefont{{Beiner}}},
  \bibinfo{author}{\bibfnamefont{H.}~\bibnamefont{{Flocard}}},
  \bibinfo{author}{\bibfnamefont{N.}~\bibnamefont{{Van Giai}}},
  \bibnamefont{and}
  \bibinfo{author}{\bibfnamefont{P.}~\bibnamefont{{Quentin}}},
  \bibinfo{journal}{Nucl. Phys. A} \textbf{\bibinfo{volume}{238}},
  \bibinfo{pages}{29} (\bibinfo{year}{1975}).

\bibitem[{\citenamefont{{K{\"o}hler}}(1976)}]{Kohler1976}
\bibinfo{author}{\bibfnamefont{H.~S.} \bibnamefont{{K{\"o}hler}}},
  \bibinfo{journal}{Nucl. Phys. A} \textbf{\bibinfo{volume}{258}},
  \bibinfo{pages}{301} (\bibinfo{year}{1976}).

\bibitem[{\citenamefont{{Chabanat} et~al.}(1997)\citenamefont{{Chabanat},
  {Bonche}, {Haensel}, {Meyer}, and {Schaeffer}}}]{Chabanat1997}
\bibinfo{author}{\bibfnamefont{E.}~\bibnamefont{{Chabanat}}},
  \bibinfo{author}{\bibfnamefont{P.}~\bibnamefont{{Bonche}}},
  \bibinfo{author}{\bibfnamefont{P.}~\bibnamefont{{Haensel}}},
  \bibinfo{author}{\bibfnamefont{J.}~\bibnamefont{{Meyer}}}, \bibnamefont{and}
  \bibinfo{author}{\bibfnamefont{R.}~\bibnamefont{{Schaeffer}}},
  \bibinfo{journal}{Nucl. Phys. A} \textbf{\bibinfo{volume}{627}},
  \bibinfo{pages}{710} (\bibinfo{year}{1997}).

\bibitem[{\citenamefont{{Chabanat} et~al.}(1998)\citenamefont{{Chabanat},
  {Bonche}, {Haensel}, {Meyer}, and {Schaeffer}}}]{Chabanat1998}
\bibinfo{author}{\bibfnamefont{E.}~\bibnamefont{{Chabanat}}},
  \bibinfo{author}{\bibfnamefont{P.}~\bibnamefont{{Bonche}}},
  \bibinfo{author}{\bibfnamefont{P.}~\bibnamefont{{Haensel}}},
  \bibinfo{author}{\bibfnamefont{J.}~\bibnamefont{{Meyer}}}, \bibnamefont{and}
  \bibinfo{author}{\bibfnamefont{R.}~\bibnamefont{{Schaeffer}}},
  \bibinfo{journal}{Nucl. Phys. A} \textbf{\bibinfo{volume}{635}},
  \bibinfo{pages}{231} (\bibinfo{year}{1998}).

\bibitem[{\citenamefont{Goriely et~al.}(2010)\citenamefont{Goriely, Chamel, and
  Pearson}}]{Goriely2010}
\bibinfo{author}{\bibfnamefont{S.}~\bibnamefont{Goriely}},
  \bibinfo{author}{\bibfnamefont{N.}~\bibnamefont{Chamel}}, \bibnamefont{and}
  \bibinfo{author}{\bibfnamefont{J.~M.} \bibnamefont{Pearson}},
  \bibinfo{journal}{Phys. Rev. C} \textbf{\bibinfo{volume}{82}},
  \bibinfo{pages}{035804} (\bibinfo{year}{2010}).

\bibitem[{\citenamefont{{Goriely} et~al.}(2013)\citenamefont{{Goriely},
  {Chamel}, and {Pearson}}}]{Goriely2013}
\bibinfo{author}{\bibfnamefont{S.}~\bibnamefont{{Goriely}}},
  \bibinfo{author}{\bibfnamefont{N.}~\bibnamefont{{Chamel}}}, \bibnamefont{and}
  \bibinfo{author}{\bibfnamefont{J.~M.} \bibnamefont{{Pearson}}},
  \bibinfo{journal}{\prc} \textbf{\bibinfo{volume}{88}}, \bibinfo{eid}{024308}
  (\bibinfo{year}{2013}).

\bibitem[{\citenamefont{{Goriely} et~al.}(2016)\citenamefont{{Goriely},
  {Chamel}, and {Pearson}}}]{Goriely2016}
\bibinfo{author}{\bibfnamefont{S.}~\bibnamefont{{Goriely}}},
  \bibinfo{author}{\bibfnamefont{N.}~\bibnamefont{{Chamel}}}, \bibnamefont{and}
  \bibinfo{author}{\bibfnamefont{J.~M.} \bibnamefont{{Pearson}}},
  \bibinfo{journal}{\prc} \textbf{\bibinfo{volume}{93}}, \bibinfo{eid}{034337}
  (\bibinfo{year}{2016}).

\bibitem[{\citenamefont{{Dutra} et~al.}(2012)\citenamefont{{Dutra},
  {Louren{\c{c}}o}, {S{\'a} Martins}, {Delfino}, {Stone}, and
  {Stevenson}}}]{Dutra2012}
\bibinfo{author}{\bibfnamefont{M.}~\bibnamefont{{Dutra}}},
  \bibinfo{author}{\bibfnamefont{O.}~\bibnamefont{{Louren{\c{c}}o}}},
  \bibinfo{author}{\bibfnamefont{J.~S.} \bibnamefont{{S{\'a} Martins}}},
  \bibinfo{author}{\bibfnamefont{A.}~\bibnamefont{{Delfino}}},
  \bibinfo{author}{\bibfnamefont{J.~R.} \bibnamefont{{Stone}}},
  \bibnamefont{and} \bibinfo{author}{\bibfnamefont{P.~D.}
  \bibnamefont{{Stevenson}}}, \bibinfo{journal}{Phys. Rev. C}
  \textbf{\bibinfo{volume}{85}}, \bibinfo{eid}{035201} (\bibinfo{year}{2012}).

\bibitem[{\citenamefont{{Baldo} et~al.}(2014)\citenamefont{{Baldo}, {Burgio},
  {Schulze}, and {Taranto}}}]{Baldo2014}
\bibinfo{author}{\bibfnamefont{M.}~\bibnamefont{{Baldo}}},
  \bibinfo{author}{\bibfnamefont{G.~F.} \bibnamefont{{Burgio}}},
  \bibinfo{author}{\bibfnamefont{H.~J.} \bibnamefont{{Schulze}}},
  \bibnamefont{and}
  \bibinfo{author}{\bibfnamefont{G.}~\bibnamefont{{Taranto}}},
  \bibinfo{journal}{Phys. Rev. C} \textbf{\bibinfo{volume}{89}},
  \bibinfo{eid}{048801} (\bibinfo{year}{2014}).

\bibitem[{\citenamefont{van Dalen et~al.}(2005)\citenamefont{van Dalen, Fuchs,
  and Faessler}}]{vanDalen2005}
\bibinfo{author}{\bibfnamefont{E.~N.~E.} \bibnamefont{van Dalen}},
  \bibinfo{author}{\bibfnamefont{C.}~\bibnamefont{Fuchs}}, \bibnamefont{and}
  \bibinfo{author}{\bibfnamefont{A.}~\bibnamefont{Faessler}},
  \bibinfo{journal}{Phys. Rev. Lett.} \textbf{\bibinfo{volume}{95}},
  \bibinfo{pages}{022302} (\bibinfo{year}{2005}).

\bibitem[{\citenamefont{Wang et~al.}(2023)\citenamefont{Wang, Tong, Zhao, Wang,
  Ring, and Meng}}]{Wang2023}
\bibinfo{author}{\bibfnamefont{S.}~\bibnamefont{Wang}},
  \bibinfo{author}{\bibfnamefont{H.}~\bibnamefont{Tong}},
  \bibinfo{author}{\bibfnamefont{Q.}~\bibnamefont{Zhao}},
  \bibinfo{author}{\bibfnamefont{C.}~\bibnamefont{Wang}},
  \bibinfo{author}{\bibfnamefont{P.}~\bibnamefont{Ring}}, \bibnamefont{and}
  \bibinfo{author}{\bibfnamefont{J.}~\bibnamefont{Meng}},
  \bibinfo{journal}{Phys. Rev. C} \textbf{\bibinfo{volume}{108}},
  \bibinfo{pages}{L031303} (\bibinfo{year}{2023}).

\bibitem[{\citenamefont{{Brueckner}}(1955)}]{Brueckner1955}
\bibinfo{author}{\bibfnamefont{K.~A.} \bibnamefont{{Brueckner}}},
  \bibinfo{journal}{Physical Review} \textbf{\bibinfo{volume}{97}},
  \bibinfo{pages}{1353} (\bibinfo{year}{1955}).

\bibitem[{\citenamefont{Jaminon and Mahaux}(1989)}]{Jaminon1989}
\bibinfo{author}{\bibfnamefont{M.}~\bibnamefont{Jaminon}} \bibnamefont{and}
  \bibinfo{author}{\bibfnamefont{C.}~\bibnamefont{Mahaux}},
  \bibinfo{journal}{Phys. Rev. C} \textbf{\bibinfo{volume}{40}},
  \bibinfo{pages}{354} (\bibinfo{year}{1989}).

\bibitem[{\citenamefont{{Duan} and {Urban}}(2023)}]{Duan2023}
\bibinfo{author}{\bibfnamefont{M.}~\bibnamefont{{Duan}}} \bibnamefont{and}
  \bibinfo{author}{\bibfnamefont{M.}~\bibnamefont{{Urban}}},
  \bibinfo{journal}{Phys. Rev. C} \textbf{\bibinfo{volume}{108}},
  \bibinfo{eid}{025813} (\bibinfo{year}{2023}).

\bibitem[{\citenamefont{{Hutauruk} et~al.}(2022)\citenamefont{{Hutauruk},
  {Gil}, {Nam}, and {Hyun}}}]{Hutauruk2022}
\bibinfo{author}{\bibfnamefont{P.~T.~P.} \bibnamefont{{Hutauruk}}},
  \bibinfo{author}{\bibfnamefont{H.}~\bibnamefont{{Gil}}},
  \bibinfo{author}{\bibfnamefont{S.-I.} \bibnamefont{{Nam}}}, \bibnamefont{and}
  \bibinfo{author}{\bibfnamefont{C.~H.} \bibnamefont{{Hyun}}},
  \bibinfo{journal}{Phys. Rev. C} \textbf{\bibinfo{volume}{106}},
  \bibinfo{eid}{035802} (\bibinfo{year}{2022}).

\bibitem[{\citenamefont{Baldo et~al.}(2017)\citenamefont{Baldo, Robledo,
  Schuck, and Vi\~nas}}]{Baldo2017}
\bibinfo{author}{\bibfnamefont{M.}~\bibnamefont{Baldo}},
  \bibinfo{author}{\bibfnamefont{L.~M.} \bibnamefont{Robledo}},
  \bibinfo{author}{\bibfnamefont{P.}~\bibnamefont{Schuck}}, \bibnamefont{and}
  \bibinfo{author}{\bibfnamefont{X.}~\bibnamefont{Vi\~nas}},
  \bibinfo{journal}{Phys. Rev. C} \textbf{\bibinfo{volume}{95}},
  \bibinfo{pages}{014318} (\bibinfo{year}{2017}).

\bibitem[{\citenamefont{{Akmal} et~al.}(1998)\citenamefont{{Akmal},
  {Pandharipande}, and {Ravenhall}}}]{Akmal1998}
\bibinfo{author}{\bibfnamefont{A.}~\bibnamefont{{Akmal}}},
  \bibinfo{author}{\bibfnamefont{V.~R.} \bibnamefont{{Pandharipande}}},
  \bibnamefont{and} \bibinfo{author}{\bibfnamefont{D.~G.}
  \bibnamefont{{Ravenhall}}}, \bibinfo{journal}{Phys. Rev. C}
  \textbf{\bibinfo{volume}{58}}, \bibinfo{pages}{1804} (\bibinfo{year}{1998}).

\bibitem[{\citenamefont{{Liu} et~al.}(2022)\citenamefont{{Liu}, {Zhang}, {Li},
  {Wei}, {Burgio}, and {Schulze}}}]{Liu2022}
\bibinfo{author}{\bibfnamefont{H.-M.} \bibnamefont{{Liu}}},
  \bibinfo{author}{\bibfnamefont{J.}~\bibnamefont{{Zhang}}},
  \bibinfo{author}{\bibfnamefont{Z.-H.} \bibnamefont{{Li}}},
  \bibinfo{author}{\bibfnamefont{J.-B.} \bibnamefont{{Wei}}},
  \bibinfo{author}{\bibfnamefont{G.~F.} \bibnamefont{{Burgio}}},
  \bibnamefont{and} \bibinfo{author}{\bibfnamefont{H.-J.}
  \bibnamefont{{Schulze}}}, \bibinfo{journal}{Phys. Rev. C}
  \textbf{\bibinfo{volume}{106}}, \bibinfo{eid}{025801} (\bibinfo{year}{2022}).

\bibitem[{\citenamefont{{Palaniappan} et~al.}(2023)\citenamefont{{Palaniappan},
  {Ramanan}, and {Urban}}}]{Palaniappan2023}
\bibinfo{author}{\bibfnamefont{V.}~\bibnamefont{{Palaniappan}}},
  \bibinfo{author}{\bibfnamefont{S.}~\bibnamefont{{Ramanan}}},
  \bibnamefont{and} \bibinfo{author}{\bibfnamefont{M.}~\bibnamefont{{Urban}}},
  \bibinfo{journal}{Phys. Rev. C} \textbf{\bibinfo{volume}{107}},
  \bibinfo{eid}{025804} (\bibinfo{year}{2023}).

\bibitem[{\citenamefont{{Skyrme}}(1956)}]{Skyrme1956}
\bibinfo{author}{\bibfnamefont{T.~H.~R.} \bibnamefont{{Skyrme}}},
  \bibinfo{journal}{Philosophical Magazine} \textbf{\bibinfo{volume}{1}},
  \bibinfo{pages}{1043} (\bibinfo{year}{1956}).

\bibitem[{\citenamefont{{Skyrme}}(1958)}]{Skyrme1958}
\bibinfo{author}{\bibfnamefont{T.~H.~R.} \bibnamefont{{Skyrme}}},
  \bibinfo{journal}{Nuclear Physics} \textbf{\bibinfo{volume}{9}},
  \bibinfo{pages}{615} (\bibinfo{year}{1958}).

\bibitem[{\citenamefont{{Giannoni} and {Quentin}}(1980)}]{Giannoni1980}
\bibinfo{author}{\bibfnamefont{M.~J.} \bibnamefont{{Giannoni}}}
  \bibnamefont{and}
  \bibinfo{author}{\bibfnamefont{P.}~\bibnamefont{{Quentin}}},
  \bibinfo{journal}{Phys. Rev. C} \textbf{\bibinfo{volume}{21}},
  \bibinfo{pages}{2076} (\bibinfo{year}{1980}).

\bibitem[{\citenamefont{{Farine} et~al.}(1997)\citenamefont{{Farine},
  {Pearson}, and {Tondeur}}}]{Farine1997}
\bibinfo{author}{\bibfnamefont{M.}~\bibnamefont{{Farine}}},
  \bibinfo{author}{\bibfnamefont{J.~M.} \bibnamefont{{Pearson}}},
  \bibnamefont{and}
  \bibinfo{author}{\bibfnamefont{F.}~\bibnamefont{{Tondeur}}},
  \bibinfo{journal}{Nucl. Phys. A} \textbf{\bibinfo{volume}{615}},
  \bibinfo{pages}{135} (\bibinfo{year}{1997}).

\bibitem[{\citenamefont{{Cochet} et~al.}(2004)\citenamefont{{Cochet},
  {Bennaceur}, {Bonche}, {Duguet}, and {Meyer}}}]{Cochet2004}
\bibinfo{author}{\bibfnamefont{B.}~\bibnamefont{{Cochet}}},
  \bibinfo{author}{\bibfnamefont{K.}~\bibnamefont{{Bennaceur}}},
  \bibinfo{author}{\bibfnamefont{P.}~\bibnamefont{{Bonche}}},
  \bibinfo{author}{\bibfnamefont{T.}~\bibnamefont{{Duguet}}}, \bibnamefont{and}
  \bibinfo{author}{\bibfnamefont{J.}~\bibnamefont{{Meyer}}},
  \bibinfo{journal}{Nucl. Phys. A} \textbf{\bibinfo{volume}{731}},
  \bibinfo{pages}{34} (\bibinfo{year}{2004}).

\bibitem[{\citenamefont{{Chamel} et~al.}(2009)\citenamefont{{Chamel},
  {Goriely}, and {Pearson}}}]{Chamel2009}
\bibinfo{author}{\bibfnamefont{N.}~\bibnamefont{{Chamel}}},
  \bibinfo{author}{\bibfnamefont{S.}~\bibnamefont{{Goriely}}},
  \bibnamefont{and} \bibinfo{author}{\bibfnamefont{J.~M.}
  \bibnamefont{{Pearson}}}, \bibinfo{journal}{Phys. Rev. C}
  \textbf{\bibinfo{volume}{80}}, \bibinfo{eid}{065804} (\bibinfo{year}{2009}).

\bibitem[{\citenamefont{{Gil} et~al.}(2019)\citenamefont{{Gil},
  {Papakonstantinou}, {Hyun}, and {Oh}}}]{Gil2019}
\bibinfo{author}{\bibfnamefont{H.}~\bibnamefont{{Gil}}},
  \bibinfo{author}{\bibfnamefont{P.}~\bibnamefont{{Papakonstantinou}}},
  \bibinfo{author}{\bibfnamefont{C.~H.} \bibnamefont{{Hyun}}},
  \bibnamefont{and} \bibinfo{author}{\bibfnamefont{Y.}~\bibnamefont{{Oh}}},
  \bibinfo{journal}{Phys. Rev. C} \textbf{\bibinfo{volume}{99}},
  \bibinfo{eid}{064319} (\bibinfo{year}{2019}).

\bibitem[{\citenamefont{Bender et~al.}(2003)\citenamefont{Bender, Heenen, and
  Reinhard}}]{Bender2003}
\bibinfo{author}{\bibfnamefont{M.}~\bibnamefont{Bender}},
  \bibinfo{author}{\bibfnamefont{P.-H.} \bibnamefont{Heenen}},
  \bibnamefont{and} \bibinfo{author}{\bibfnamefont{P.-G.}
  \bibnamefont{Reinhard}}, \bibinfo{journal}{Rev. Mod. Phys.}
  \textbf{\bibinfo{volume}{75}}, \bibinfo{pages}{121} (\bibinfo{year}{2003}).

\bibitem[{\citenamefont{Zhang and Chen}(2016)}]{Zhang2016}
\bibinfo{author}{\bibfnamefont{Z.}~\bibnamefont{Zhang}} \bibnamefont{and}
  \bibinfo{author}{\bibfnamefont{L.-W.} \bibnamefont{Chen}},
  \bibinfo{journal}{Phys. Rev. C} \textbf{\bibinfo{volume}{94}},
  \bibinfo{pages}{064326} (\bibinfo{year}{2016}).

\bibitem[{\citenamefont{Lesinski et~al.}(2007)\citenamefont{Lesinski, Bender,
  Bennaceur, Duguet, and Meyer}}]{Lesinski2007}
\bibinfo{author}{\bibfnamefont{T.}~\bibnamefont{Lesinski}},
  \bibinfo{author}{\bibfnamefont{M.}~\bibnamefont{Bender}},
  \bibinfo{author}{\bibfnamefont{K.}~\bibnamefont{Bennaceur}},
  \bibinfo{author}{\bibfnamefont{T.}~\bibnamefont{Duguet}}, \bibnamefont{and}
  \bibinfo{author}{\bibfnamefont{J.}~\bibnamefont{Meyer}},
  \bibinfo{journal}{Phys. Rev. C} \textbf{\bibinfo{volume}{76}},
  \bibinfo{pages}{014312} (\bibinfo{year}{2007}).

\bibitem[{\citenamefont{{Wang} et~al.}(2024)\citenamefont{{Wang}, {Wang}, {Ye},
  and {Chen}}}]{Wang2024}
\bibinfo{author}{\bibfnamefont{S.-P.} \bibnamefont{{Wang}}},
  \bibinfo{author}{\bibfnamefont{R.}~\bibnamefont{{Wang}}},
  \bibinfo{author}{\bibfnamefont{J.-T.} \bibnamefont{{Ye}}}, \bibnamefont{and}
  \bibinfo{author}{\bibfnamefont{L.-W.} \bibnamefont{{Chen}}},
  \bibinfo{journal}{Phys. Rev. C} \textbf{\bibinfo{volume}{109}},
  \bibinfo{eid}{054623} (\bibinfo{year}{2024}).

\bibitem[{\citenamefont{Reinhard}(1991)}]{Reinhard1991}
\bibinfo{author}{\bibfnamefont{P.-G.} \bibnamefont{Reinhard}}, in
  \emph{\bibinfo{booktitle}{Computational Nuclear Physics 1: Nuclear
  Structure}}, edited by
  \bibinfo{editor}{\bibfnamefont{K.}~\bibnamefont{Langanke}},
  \bibinfo{editor}{\bibfnamefont{J.~A.} \bibnamefont{Maruhn}},
  \bibnamefont{and} \bibinfo{editor}{\bibfnamefont{S.~E.} \bibnamefont{Koonin}}
  (\bibinfo{publisher}{Springer}, \bibinfo{address}{Berlin},
  \bibinfo{year}{1991}), pp. \bibinfo{pages}{28--59}.

\bibitem[{IAE()}]{IAEA}
\emph{\bibinfo{title}{{IAEA Live Chart of Nuclides}}},
  \bibinfo{howpublished}{\url{https://www-nds.iaea.org}},
  \bibinfo{note}{updated Nov. 2023}.

\bibitem[{\citenamefont{{Sommer} et~al.}(2022)\citenamefont{{Sommer},
  {K{\"o}nig}, {Rossi}, {Everett}, {Garand}, {de Groote}, {Holt}, {Imgram},
  {Incorvati}, {Kalman} et~al.}}]{Sommer2022}
\bibinfo{author}{\bibfnamefont{F.}~\bibnamefont{{Sommer}}},
  \bibinfo{author}{\bibfnamefont{K.}~\bibnamefont{{K{\"o}nig}}},
  \bibinfo{author}{\bibfnamefont{D.~M.} \bibnamefont{{Rossi}}},
  \bibinfo{author}{\bibfnamefont{N.}~\bibnamefont{{Everett}}},
  \bibinfo{author}{\bibfnamefont{D.}~\bibnamefont{{Garand}}},
  \bibinfo{author}{\bibfnamefont{R.~P.} \bibnamefont{{de Groote}}},
  \bibinfo{author}{\bibfnamefont{J.~D.} \bibnamefont{{Holt}}},
  \bibinfo{author}{\bibfnamefont{P.}~\bibnamefont{{Imgram}}},
  \bibinfo{author}{\bibfnamefont{A.}~\bibnamefont{{Incorvati}}},
  \bibinfo{author}{\bibfnamefont{C.}~\bibnamefont{{Kalman}}},
  \bibnamefont{et~al.}, \bibinfo{journal}{Phys. Rev. Lett.}
  \textbf{\bibinfo{volume}{129}}, \bibinfo{eid}{132501} (\bibinfo{year}{2022}).

\bibitem[{\citenamefont{Hellemans et~al.}(2013)\citenamefont{Hellemans,
  Pastore, Duguet, Bennaceur, Davesne, Meyer, Bender, and
  Heenen}}]{Hellemans2013}
\bibinfo{author}{\bibfnamefont{V.}~\bibnamefont{Hellemans}},
  \bibinfo{author}{\bibfnamefont{A.}~\bibnamefont{Pastore}},
  \bibinfo{author}{\bibfnamefont{T.}~\bibnamefont{Duguet}},
  \bibinfo{author}{\bibfnamefont{K.}~\bibnamefont{Bennaceur}},
  \bibinfo{author}{\bibfnamefont{D.}~\bibnamefont{Davesne}},
  \bibinfo{author}{\bibfnamefont{J.}~\bibnamefont{Meyer}},
  \bibinfo{author}{\bibfnamefont{M.}~\bibnamefont{Bender}}, \bibnamefont{and}
  \bibinfo{author}{\bibfnamefont{P.-H.} \bibnamefont{Heenen}},
  \bibinfo{journal}{Phys. Rev. C} \textbf{\bibinfo{volume}{88}},
  \bibinfo{pages}{064323} (\bibinfo{year}{2013}).

\bibitem[{\citenamefont{{Pastore} et~al.}(2014)\citenamefont{{Pastore},
  {Martini}, {Davesne}, {Navarro}, {Goriely}, and {Chamel}}}]{Pastore2014}
\bibinfo{author}{\bibfnamefont{A.}~\bibnamefont{{Pastore}}},
  \bibinfo{author}{\bibfnamefont{M.}~\bibnamefont{{Martini}}},
  \bibinfo{author}{\bibfnamefont{D.}~\bibnamefont{{Davesne}}},
  \bibinfo{author}{\bibfnamefont{J.}~\bibnamefont{{Navarro}}},
  \bibinfo{author}{\bibfnamefont{S.}~\bibnamefont{{Goriely}}},
  \bibnamefont{and} \bibinfo{author}{\bibfnamefont{N.}~\bibnamefont{{Chamel}}},
  \bibinfo{journal}{\prc} \textbf{\bibinfo{volume}{90}}, \bibinfo{eid}{025804}
  (\bibinfo{year}{2014}).

\bibitem[{\citenamefont{Adhikari et~al.}(2021)\citenamefont{Adhikari,
  Albataineh, Androic, Aniol, Armstrong, Averett, Ayerbe~Gayoso, Barcus,
  Bellini, Beminiwattha et~al.}}]{Adhikari2021}
\bibinfo{author}{\bibfnamefont{D.}~\bibnamefont{Adhikari}},
  \bibinfo{author}{\bibfnamefont{H.}~\bibnamefont{Albataineh}},
  \bibinfo{author}{\bibfnamefont{D.}~\bibnamefont{Androic}},
  \bibinfo{author}{\bibfnamefont{K.}~\bibnamefont{Aniol}},
  \bibinfo{author}{\bibfnamefont{D.~S.} \bibnamefont{Armstrong}},
  \bibinfo{author}{\bibfnamefont{T.}~\bibnamefont{Averett}},
  \bibinfo{author}{\bibfnamefont{C.}~\bibnamefont{Ayerbe~Gayoso}},
  \bibinfo{author}{\bibfnamefont{S.}~\bibnamefont{Barcus}},
  \bibinfo{author}{\bibfnamefont{V.}~\bibnamefont{Bellini}},
  \bibinfo{author}{\bibfnamefont{R.~S.} \bibnamefont{Beminiwattha}},
  \bibnamefont{et~al.} (\bibinfo{collaboration}{PREX Collaboration}),
  \bibinfo{journal}{Phys. Rev. Lett.} \textbf{\bibinfo{volume}{126}},
  \bibinfo{pages}{172502} (\bibinfo{year}{2021}).

\bibitem[{\citenamefont{Adhikari et~al.}(2022)\citenamefont{Adhikari,
  Albataineh, Androic, Aniol, Armstrong, Averett, Ayerbe~Gayoso, Barcus,
  Bellini, Beminiwattha et~al.}}]{Adhikari2022}
\bibinfo{author}{\bibfnamefont{D.}~\bibnamefont{Adhikari}},
  \bibinfo{author}{\bibfnamefont{H.}~\bibnamefont{Albataineh}},
  \bibinfo{author}{\bibfnamefont{D.}~\bibnamefont{Androic}},
  \bibinfo{author}{\bibfnamefont{K.~A.} \bibnamefont{Aniol}},
  \bibinfo{author}{\bibfnamefont{D.~S.} \bibnamefont{Armstrong}},
  \bibinfo{author}{\bibfnamefont{T.}~\bibnamefont{Averett}},
  \bibinfo{author}{\bibfnamefont{C.}~\bibnamefont{Ayerbe~Gayoso}},
  \bibinfo{author}{\bibfnamefont{S.~K.} \bibnamefont{Barcus}},
  \bibinfo{author}{\bibfnamefont{V.}~\bibnamefont{Bellini}},
  \bibinfo{author}{\bibfnamefont{R.~S.} \bibnamefont{Beminiwattha}},
  \bibnamefont{et~al.} (\bibinfo{collaboration}{CREX Collaboration}),
  \bibinfo{journal}{Phys. Rev. Lett.} \textbf{\bibinfo{volume}{129}},
  \bibinfo{pages}{042501} (\bibinfo{year}{2022}).

\bibitem[{\citenamefont{Reinhard et~al.}(2013)\citenamefont{Reinhard,
  Piekarewicz, Nazarewicz, Agrawal, Paar, and Roca-Maza}}]{Reinhard2013}
\bibinfo{author}{\bibfnamefont{P.-G.} \bibnamefont{Reinhard}},
  \bibinfo{author}{\bibfnamefont{J.}~\bibnamefont{Piekarewicz}},
  \bibinfo{author}{\bibfnamefont{W.}~\bibnamefont{Nazarewicz}},
  \bibinfo{author}{\bibfnamefont{B.~K.} \bibnamefont{Agrawal}},
  \bibinfo{author}{\bibfnamefont{N.}~\bibnamefont{Paar}}, \bibnamefont{and}
  \bibinfo{author}{\bibfnamefont{X.}~\bibnamefont{Roca-Maza}},
  \bibinfo{journal}{Phys. Rev. C} \textbf{\bibinfo{volume}{88}},
  \bibinfo{pages}{034325} (\bibinfo{year}{2013}).

\bibitem[{\citenamefont{Margueron et~al.}(2018)\citenamefont{Margueron,
  Hoffmann~Casali, and Gulminelli}}]{Margueron2018}
\bibinfo{author}{\bibfnamefont{J.}~\bibnamefont{Margueron}},
  \bibinfo{author}{\bibfnamefont{R.}~\bibnamefont{Hoffmann~Casali}},
  \bibnamefont{and}
  \bibinfo{author}{\bibfnamefont{F.}~\bibnamefont{Gulminelli}},
  \bibinfo{journal}{Phys. Rev. C} \textbf{\bibinfo{volume}{97}},
  \bibinfo{pages}{025805} (\bibinfo{year}{2018}).

\bibitem[{\citenamefont{{Lattimer}}(2023)}]{Lattimer2023}
\bibinfo{author}{\bibfnamefont{J.~M.} \bibnamefont{{Lattimer}}}, in
  \emph{\bibinfo{booktitle}{Journal of Physics Conference Series}}
  (\bibinfo{year}{2023}), vol. \bibinfo{volume}{2536} of
  \emph{\bibinfo{series}{Journal of Physics Conference Series}}, p.
  \bibinfo{pages}{012009}.

\bibitem[{\citenamefont{{Li} et~al.}(2018)\citenamefont{{Li}, {Cai}, {Chen},
  and {Xu}}}]{Li2018}
\bibinfo{author}{\bibfnamefont{B.-A.} \bibnamefont{{Li}}},
  \bibinfo{author}{\bibfnamefont{B.-J.} \bibnamefont{{Cai}}},
  \bibinfo{author}{\bibfnamefont{L.-W.} \bibnamefont{{Chen}}},
  \bibnamefont{and} \bibinfo{author}{\bibfnamefont{J.}~\bibnamefont{{Xu}}},
  \bibinfo{journal}{Progress in Particle and Nuclear Physics}
  \textbf{\bibinfo{volume}{99}}, \bibinfo{pages}{29} (\bibinfo{year}{2018}).

\bibitem[{\citenamefont{Nozi{\`e}res}(1964)}]{Nozieres}
\bibinfo{author}{\bibfnamefont{P.}~\bibnamefont{Nozi{\`e}res}},
  \emph{\bibinfo{title}{{Theory of interaction Fermi systems}}}
  (\bibinfo{publisher}{Benjamin}, \bibinfo{address}{New York},
  \bibinfo{year}{1964}).

\bibitem[{\citenamefont{{Lattimer} et~al.}(1991)\citenamefont{{Lattimer},
  {Pethick}, {Prakash}, and {Haensel}}}]{Lattimer1991}
\bibinfo{author}{\bibfnamefont{J.~M.} \bibnamefont{{Lattimer}}},
  \bibinfo{author}{\bibfnamefont{C.~J.} \bibnamefont{{Pethick}}},
  \bibinfo{author}{\bibfnamefont{M.}~\bibnamefont{{Prakash}}},
  \bibnamefont{and}
  \bibinfo{author}{\bibfnamefont{P.}~\bibnamefont{{Haensel}}},
  \bibinfo{journal}{Phys. Rev. Lett.} \textbf{\bibinfo{volume}{66}},
  \bibinfo{pages}{2701} (\bibinfo{year}{1991}).

\bibitem[{\citenamefont{Chamel and Haensel}(2008)}]{Chamel2008}
\bibinfo{author}{\bibfnamefont{N.}~\bibnamefont{Chamel}} \bibnamefont{and}
  \bibinfo{author}{\bibfnamefont{P.}~\bibnamefont{Haensel}},
  \bibinfo{journal}{Living Rev. Rel.} \textbf{\bibinfo{volume}{11}}
  (\bibinfo{year}{2008}).

\bibitem[{\citenamefont{Carreau et~al.}(2019)\citenamefont{Carreau, Gulminelli,
  and Margueron}}]{Carreau2019}
\bibinfo{author}{\bibfnamefont{T.}~\bibnamefont{Carreau}},
  \bibinfo{author}{\bibfnamefont{F.}~\bibnamefont{Gulminelli}},
  \bibnamefont{and}
  \bibinfo{author}{\bibfnamefont{J.}~\bibnamefont{Margueron}},
  \bibinfo{journal}{Eur. Phys. J. A} \textbf{\bibinfo{volume}{55}},
  \bibinfo{pages}{188} (\bibinfo{year}{2019}).

\bibitem[{\citenamefont{Davis et~al.}(2024)\citenamefont{Davis, Dinh~Thi,
  Fantina, Gulminelli, Oertel, and Suleiman}}]{Davis2024}
\bibinfo{author}{\bibfnamefont{P.~J.} \bibnamefont{Davis}},
  \bibinfo{author}{\bibfnamefont{H.}~\bibnamefont{Dinh~Thi}},
  \bibinfo{author}{\bibfnamefont{A.~F.} \bibnamefont{Fantina}},
  \bibinfo{author}{\bibfnamefont{F.}~\bibnamefont{Gulminelli}},
  \bibinfo{author}{\bibfnamefont{M.}~\bibnamefont{Oertel}}, \bibnamefont{and}
  \bibinfo{author}{\bibfnamefont{L.}~\bibnamefont{Suleiman}},
  \bibinfo{journal}{Astron. Astrophys.} \textbf{\bibinfo{volume}{687}},
  \bibinfo{pages}{A44} (\bibinfo{year}{2024}).

\bibitem[{\citenamefont{{Tolman}}(1939)}]{Tolman1939}
\bibinfo{author}{\bibfnamefont{R.~C.} \bibnamefont{{Tolman}}},
  \bibinfo{journal}{Physical Review} \textbf{\bibinfo{volume}{55}},
  \bibinfo{pages}{364} (\bibinfo{year}{1939}).

\bibitem[{\citenamefont{{Oppenheimer} and {Volkoff}}(1939)}]{Oppenheimer1939}
\bibinfo{author}{\bibfnamefont{J.~R.} \bibnamefont{{Oppenheimer}}}
  \bibnamefont{and} \bibinfo{author}{\bibfnamefont{G.~M.}
  \bibnamefont{{Volkoff}}}, \bibinfo{journal}{Physical Review}
  \textbf{\bibinfo{volume}{55}}, \bibinfo{pages}{374} (\bibinfo{year}{1939}).

\bibitem[{\citenamefont{Perot et~al.}(2020)\citenamefont{Perot, Chamel, and
  Sourie}}]{Perot2020}
\bibinfo{author}{\bibfnamefont{L.}~\bibnamefont{Perot}},
  \bibinfo{author}{\bibfnamefont{N.}~\bibnamefont{Chamel}}, \bibnamefont{and}
  \bibinfo{author}{\bibfnamefont{A.}~\bibnamefont{Sourie}},
  \bibinfo{journal}{Phys. Rev. C} \textbf{\bibinfo{volume}{101}},
  \bibinfo{pages}{015806} (\bibinfo{year}{2020}).

\bibitem[{\citenamefont{{Cromartie} et~al.}(2020)\citenamefont{{Cromartie},
  {Fonseca}, {Ransom}, {Demorest}, {Arzoumanian}, {Blumer}, {Brook}, {DeCesar},
  {Dolch}, {Ellis} et~al.}}]{Cromartie2020}
\bibinfo{author}{\bibfnamefont{H.~T.} \bibnamefont{{Cromartie}}},
  \bibinfo{author}{\bibfnamefont{E.}~\bibnamefont{{Fonseca}}},
  \bibinfo{author}{\bibfnamefont{S.~M.} \bibnamefont{{Ransom}}},
  \bibinfo{author}{\bibfnamefont{P.~B.} \bibnamefont{{Demorest}}},
  \bibinfo{author}{\bibfnamefont{Z.}~\bibnamefont{{Arzoumanian}}},
  \bibinfo{author}{\bibfnamefont{H.}~\bibnamefont{{Blumer}}},
  \bibinfo{author}{\bibfnamefont{P.~R.} \bibnamefont{{Brook}}},
  \bibinfo{author}{\bibfnamefont{M.~E.} \bibnamefont{{DeCesar}}},
  \bibinfo{author}{\bibfnamefont{T.}~\bibnamefont{{Dolch}}},
  \bibinfo{author}{\bibfnamefont{J.~A.} \bibnamefont{{Ellis}}},
  \bibnamefont{et~al.}, \bibinfo{journal}{Nature Astronomy}
  \textbf{\bibinfo{volume}{4}}, \bibinfo{pages}{72} (\bibinfo{year}{2020}).

\bibitem[{\citenamefont{{Fonseca} et~al.}(2021)\citenamefont{{Fonseca},
  {Cromartie}, {Pennucci}, {Ray}, {Kirichenko}, {Ransom}, {Demorest}, {Stairs},
  {Arzoumanian}, {Guillemot} et~al.}}]{Fonseca2021}
\bibinfo{author}{\bibfnamefont{E.}~\bibnamefont{{Fonseca}}},
  \bibinfo{author}{\bibfnamefont{H.~T.} \bibnamefont{{Cromartie}}},
  \bibinfo{author}{\bibfnamefont{T.~T.} \bibnamefont{{Pennucci}}},
  \bibinfo{author}{\bibfnamefont{P.~S.} \bibnamefont{{Ray}}},
  \bibinfo{author}{\bibfnamefont{A.~Y.} \bibnamefont{{Kirichenko}}},
  \bibinfo{author}{\bibfnamefont{S.~M.} \bibnamefont{{Ransom}}},
  \bibinfo{author}{\bibfnamefont{P.~B.} \bibnamefont{{Demorest}}},
  \bibinfo{author}{\bibfnamefont{I.~H.} \bibnamefont{{Stairs}}},
  \bibinfo{author}{\bibfnamefont{Z.}~\bibnamefont{{Arzoumanian}}},
  \bibinfo{author}{\bibfnamefont{L.}~\bibnamefont{{Guillemot}}},
  \bibnamefont{et~al.}, \bibinfo{journal}{Astrophys. J. Lett}
  \textbf{\bibinfo{volume}{915}}, \bibinfo{eid}{L12} (\bibinfo{year}{2021}).

\bibitem[{\citenamefont{{Miller} et~al.}(2021)\citenamefont{{Miller}, {Lamb},
  {Dittmann}, {Bogdanov}, {Arzoumanian}, {Gendreau}, {Guillot}, {Ho},
  {Lattimer}, {Loewenstein} et~al.}}]{Miller2021}
\bibinfo{author}{\bibfnamefont{M.~C.} \bibnamefont{{Miller}}},
  \bibinfo{author}{\bibfnamefont{F.~K.} \bibnamefont{{Lamb}}},
  \bibinfo{author}{\bibfnamefont{A.~J.} \bibnamefont{{Dittmann}}},
  \bibinfo{author}{\bibfnamefont{S.}~\bibnamefont{{Bogdanov}}},
  \bibinfo{author}{\bibfnamefont{Z.}~\bibnamefont{{Arzoumanian}}},
  \bibinfo{author}{\bibfnamefont{K.~C.} \bibnamefont{{Gendreau}}},
  \bibinfo{author}{\bibfnamefont{S.}~\bibnamefont{{Guillot}}},
  \bibinfo{author}{\bibfnamefont{W.~C.~G.} \bibnamefont{{Ho}}},
  \bibinfo{author}{\bibfnamefont{J.~M.} \bibnamefont{{Lattimer}}},
  \bibinfo{author}{\bibfnamefont{M.}~\bibnamefont{{Loewenstein}}},
  \bibnamefont{et~al.}, \bibinfo{journal}{Astrophys. J. Lett}
  \textbf{\bibinfo{volume}{918}}, \bibinfo{eid}{L28} (\bibinfo{year}{2021}).

\bibitem[{\citenamefont{{Yakovlev} et~al.}(2001)\citenamefont{{Yakovlev},
  {Kaminker}, {Gnedin}, and {Haensel}}}]{Yakovlev2001}
\bibinfo{author}{\bibfnamefont{D.~G.} \bibnamefont{{Yakovlev}}},
  \bibinfo{author}{\bibfnamefont{A.~D.} \bibnamefont{{Kaminker}}},
  \bibinfo{author}{\bibfnamefont{O.~Y.} \bibnamefont{{Gnedin}}},
  \bibnamefont{and}
  \bibinfo{author}{\bibfnamefont{P.}~\bibnamefont{{Haensel}}},
  \bibinfo{journal}{Phys. Rep.} \textbf{\bibinfo{volume}{354}},
  \bibinfo{pages}{1} (\bibinfo{year}{2001}).

\bibitem[{\citenamefont{{Alvarez Castillo} and {Blaschke}}(2015)}]{Alvarez2015}
\bibinfo{author}{\bibfnamefont{D.~E.} \bibnamefont{{Alvarez Castillo}}}
  \bibnamefont{and}
  \bibinfo{author}{\bibfnamefont{D.}~\bibnamefont{{Blaschke}}}, in
  \emph{\bibinfo{booktitle}{Proceedings of The Modern Physics of Compact Stars
  2015 (MPCS2015). 30 September 2015 - 3 October 2015. Yerevan}}
  (\bibinfo{year}{2015}), p.~\bibinfo{pages}{26}.

\bibitem[{\citenamefont{{Fantina} et~al.}(2013)\citenamefont{{Fantina},
  {Chamel}, {Pearson}, and {Goriely}}}]{Fantina2013}
\bibinfo{author}{\bibfnamefont{A.~F.} \bibnamefont{{Fantina}}},
  \bibinfo{author}{\bibfnamefont{N.}~\bibnamefont{{Chamel}}},
  \bibinfo{author}{\bibfnamefont{J.~M.} \bibnamefont{{Pearson}}},
  \bibnamefont{and}
  \bibinfo{author}{\bibfnamefont{S.}~\bibnamefont{{Goriely}}},
  \bibinfo{journal}{Astron. Astrophys.} \textbf{\bibinfo{volume}{559}},
  \bibinfo{eid}{A128} (\bibinfo{year}{2013}).

\bibitem[{\citenamefont{Kl{\"a}hn et~al.}(2006)\citenamefont{Kl{\"a}hn,
  Blaschke, Typel, van Dalen, Faessler, Fuchs, Gaitanos, Grigorian, Ho,
  Kolomeitsev et~al.}}]{Klaehn2006}
\bibinfo{author}{\bibfnamefont{T.}~\bibnamefont{Kl{\"a}hn}},
  \bibinfo{author}{\bibfnamefont{D.}~\bibnamefont{Blaschke}},
  \bibinfo{author}{\bibfnamefont{S.}~\bibnamefont{Typel}},
  \bibinfo{author}{\bibfnamefont{E.~N.~E.} \bibnamefont{van Dalen}},
  \bibinfo{author}{\bibfnamefont{A.}~\bibnamefont{Faessler}},
  \bibinfo{author}{\bibfnamefont{C.}~\bibnamefont{Fuchs}},
  \bibinfo{author}{\bibfnamefont{T.}~\bibnamefont{Gaitanos}},
  \bibinfo{author}{\bibfnamefont{H.}~\bibnamefont{Grigorian}},
  \bibinfo{author}{\bibfnamefont{A.}~\bibnamefont{Ho}},
  \bibinfo{author}{\bibfnamefont{E.~E.} \bibnamefont{Kolomeitsev}},
  \bibnamefont{et~al.}, \bibinfo{journal}{Phys. Rev. C}
  \textbf{\bibinfo{volume}{74}}, \bibinfo{pages}{035802}
  (\bibinfo{year}{2006}).

\bibitem[{\citenamefont{{Bersten} and {Mazzali}}(2017)}]{Bersten2017}
\bibinfo{author}{\bibfnamefont{M.~C.} \bibnamefont{{Bersten}}}
  \bibnamefont{and} \bibinfo{author}{\bibfnamefont{P.~A.}
  \bibnamefont{{Mazzali}}}, in \emph{\bibinfo{booktitle}{Handbook of
  Supernovae}}, edited by \bibinfo{editor}{\bibfnamefont{A.~W.}
  \bibnamefont{{Alsabti}}} \bibnamefont{and}
  \bibinfo{editor}{\bibfnamefont{P.}~\bibnamefont{{Murdin}}}
  (\bibinfo{publisher}{Springer}, \bibinfo{year}{2017}), p.
  \bibinfo{pages}{723}.

\bibitem[{\citenamefont{{Bionta} et~al.}(1987)\citenamefont{{Bionta},
  {Blewitt}, {Bratton}, {Casper}, {Ciocio}, {Claus}, {Cortez}, {Crouch}, {Dye},
  {Errede} et~al.}}]{Bionta1987}
\bibinfo{author}{\bibfnamefont{R.~M.} \bibnamefont{{Bionta}}},
  \bibinfo{author}{\bibfnamefont{G.}~\bibnamefont{{Blewitt}}},
  \bibinfo{author}{\bibfnamefont{C.~B.} \bibnamefont{{Bratton}}},
  \bibinfo{author}{\bibfnamefont{D.}~\bibnamefont{{Casper}}},
  \bibinfo{author}{\bibfnamefont{A.}~\bibnamefont{{Ciocio}}},
  \bibinfo{author}{\bibfnamefont{R.}~\bibnamefont{{Claus}}},
  \bibinfo{author}{\bibfnamefont{B.}~\bibnamefont{{Cortez}}},
  \bibinfo{author}{\bibfnamefont{M.}~\bibnamefont{{Crouch}}},
  \bibinfo{author}{\bibfnamefont{S.~T.} \bibnamefont{{Dye}}},
  \bibinfo{author}{\bibfnamefont{S.}~\bibnamefont{{Errede}}},
  \bibnamefont{et~al.}, \bibinfo{journal}{Phys. Rev. Lett.}
  \textbf{\bibinfo{volume}{58}}, \bibinfo{pages}{1494} (\bibinfo{year}{1987}).

\bibitem[{\citenamefont{{Hirata} et~al.}(1987)\citenamefont{{Hirata}, {Kajita},
  {Koshiba}, {Nakahata}, {Oyama}, {Sato}, {Suzuki}, {Takita}, {Totsuka},
  {Kifune} et~al.}}]{Hirata1987}
\bibinfo{author}{\bibfnamefont{K.}~\bibnamefont{{Hirata}}},
  \bibinfo{author}{\bibfnamefont{T.}~\bibnamefont{{Kajita}}},
  \bibinfo{author}{\bibfnamefont{M.}~\bibnamefont{{Koshiba}}},
  \bibinfo{author}{\bibfnamefont{M.}~\bibnamefont{{Nakahata}}},
  \bibinfo{author}{\bibfnamefont{Y.}~\bibnamefont{{Oyama}}},
  \bibinfo{author}{\bibfnamefont{N.}~\bibnamefont{{Sato}}},
  \bibinfo{author}{\bibfnamefont{A.}~\bibnamefont{{Suzuki}}},
  \bibinfo{author}{\bibfnamefont{M.}~\bibnamefont{{Takita}}},
  \bibinfo{author}{\bibfnamefont{Y.}~\bibnamefont{{Totsuka}}},
  \bibinfo{author}{\bibfnamefont{T.}~\bibnamefont{{Kifune}}},
  \bibnamefont{et~al.}, \bibinfo{journal}{Phys. Rev. Lett.}
  \textbf{\bibinfo{volume}{58}}, \bibinfo{pages}{1490} (\bibinfo{year}{1987}).

\bibitem[{\citenamefont{{Alexeyev} et~al.}(1988)\citenamefont{{Alexeyev},
  {Alexeyeva}, {Krivosheina}, and {Volchenko}}}]{Alexeyev1988}
\bibinfo{author}{\bibfnamefont{E.~N.} \bibnamefont{{Alexeyev}}},
  \bibinfo{author}{\bibfnamefont{L.~N.} \bibnamefont{{Alexeyeva}}},
  \bibinfo{author}{\bibfnamefont{I.~V.} \bibnamefont{{Krivosheina}}},
  \bibnamefont{and} \bibinfo{author}{\bibfnamefont{V.~I.}
  \bibnamefont{{Volchenko}}}, \bibinfo{journal}{Phys. Lett. B}
  \textbf{\bibinfo{volume}{205}}, \bibinfo{pages}{209} (\bibinfo{year}{1988}).

\bibitem[{\citenamefont{{Haensel} et~al.}(2007)\citenamefont{{Haensel},
  {Potekhin}, and {Yakovlev}}}]{Haensel2007}
\bibinfo{author}{\bibfnamefont{P.}~\bibnamefont{{Haensel}}},
  \bibinfo{author}{\bibfnamefont{A.~Y.} \bibnamefont{{Potekhin}}},
  \bibnamefont{and} \bibinfo{author}{\bibfnamefont{D.~G.}
  \bibnamefont{{Yakovlev}}}, \emph{\bibinfo{title}{{Neutron Stars 1 : Equation
  of State and Structure}}}, vol. \bibinfo{volume}{326} of
  \emph{\bibinfo{series}{{Astrophysics and Space Science Library}}}
  (\bibinfo{publisher}{Springer}, \bibinfo{year}{2007}).

\bibitem[{\citenamefont{{Higdon} and {Lingenfelter}}(1990)}]{Higdon1990}
\bibinfo{author}{\bibfnamefont{J.~C.} \bibnamefont{{Higdon}}} \bibnamefont{and}
  \bibinfo{author}{\bibfnamefont{R.~E.} \bibnamefont{{Lingenfelter}}},
  \bibinfo{journal}{Ann. Rev. Astron. Astrophys.}
  \textbf{\bibinfo{volume}{28}}, \bibinfo{pages}{401} (\bibinfo{year}{1990}).

\bibitem[{\citenamefont{Ho}(2007)}]{Ho2007}
\bibinfo{author}{\bibfnamefont{W.~C.~G.} \bibnamefont{Ho}},
  \bibinfo{journal}{Monthly Notices of the Royal Astronomical Society}
  \textbf{\bibinfo{volume}{380}}, \bibinfo{pages}{71} (\bibinfo{year}{2007}).

\bibitem[{\citenamefont{{Shapiro} and
  {Teukolsky}}(1983)}]{ShapiroTeukolsky1983}
\bibinfo{author}{\bibfnamefont{S.~L.} \bibnamefont{{Shapiro}}}
  \bibnamefont{and} \bibinfo{author}{\bibfnamefont{S.~A.}
  \bibnamefont{{Teukolsky}}}, \emph{\bibinfo{title}{{Black Holes, White Dwarfs,
  and Neutron Stars. The Physics of Compact Objects}}}
  (\bibinfo{publisher}{Wiley}, \bibinfo{year}{1983}).

\bibitem[{\citenamefont{{Zhang} and {M{\'e}sz{\'a}ros}}(2001)}]{Zhang2001}
\bibinfo{author}{\bibfnamefont{B.}~\bibnamefont{{Zhang}}} \bibnamefont{and}
  \bibinfo{author}{\bibfnamefont{P.}~\bibnamefont{{M{\'e}sz{\'a}ros}}},
  \bibinfo{journal}{Astrophys. J. Lett} \textbf{\bibinfo{volume}{552}},
  \bibinfo{pages}{L35} (\bibinfo{year}{2001}).

\bibitem[{\citenamefont{{Lattimer} and {Prakash}}(2016)}]{Lattimer2016}
\bibinfo{author}{\bibfnamefont{J.~M.} \bibnamefont{{Lattimer}}}
  \bibnamefont{and}
  \bibinfo{author}{\bibfnamefont{M.}~\bibnamefont{{Prakash}}},
  \bibinfo{journal}{Phys. Rep.} \textbf{\bibinfo{volume}{621}},
  \bibinfo{pages}{127} (\bibinfo{year}{2016}).

\bibitem[{\citenamefont{{Bejger} and {Haensel}}(2002)}]{Bejger2002}
\bibinfo{author}{\bibfnamefont{M.}~\bibnamefont{{Bejger}}} \bibnamefont{and}
  \bibinfo{author}{\bibfnamefont{P.}~\bibnamefont{{Haensel}}},
  \bibinfo{journal}{Astron. Astrophys.} \textbf{\bibinfo{volume}{396}},
  \bibinfo{pages}{917} (\bibinfo{year}{2002}).

\bibitem[{\citenamefont{{Bejger} and {Haensel}}(2003)}]{Bejger2003}
\bibinfo{author}{\bibfnamefont{M.}~\bibnamefont{{Bejger}}} \bibnamefont{and}
  \bibinfo{author}{\bibfnamefont{P.}~\bibnamefont{{Haensel}}},
  \bibinfo{journal}{Astron. Astrophys.} \textbf{\bibinfo{volume}{405}},
  \bibinfo{pages}{747} (\bibinfo{year}{2003}).

\bibitem[{\citenamefont{{Baym} et~al.}(1971)\citenamefont{{Baym}, {Bethe}, and
  {Pethick}}}]{Baym1971b}
\bibinfo{author}{\bibfnamefont{G.}~\bibnamefont{{Baym}}},
  \bibinfo{author}{\bibfnamefont{H.~A.} \bibnamefont{{Bethe}}},
  \bibnamefont{and} \bibinfo{author}{\bibfnamefont{C.~J.}
  \bibnamefont{{Pethick}}}, \bibinfo{journal}{Nucl. Phys. A}
  \textbf{\bibinfo{volume}{175}}, \bibinfo{pages}{225} (\bibinfo{year}{1971}).

\bibitem[{\citenamefont{{Bender} et~al.}(2002)\citenamefont{{Bender},
  {Dobaczewski}, {Engel}, and {Nazarewicz}}}]{Bender2002}
\bibinfo{author}{\bibfnamefont{M.}~\bibnamefont{{Bender}}},
  \bibinfo{author}{\bibfnamefont{J.}~\bibnamefont{{Dobaczewski}}},
  \bibinfo{author}{\bibfnamefont{J.}~\bibnamefont{{Engel}}}, \bibnamefont{and}
  \bibinfo{author}{\bibfnamefont{W.}~\bibnamefont{{Nazarewicz}}},
  \bibinfo{journal}{\prc} \textbf{\bibinfo{volume}{65}}, \bibinfo{eid}{054322}
  (\bibinfo{year}{2002}).

\bibitem[{\citenamefont{James}(1994)}]{James1994}
\bibinfo{author}{\bibfnamefont{F.}~\bibnamefont{James}},
  \emph{\bibinfo{title}{MINUIT - Function Minimization and Error Analysis}},
  \bibinfo{organization}{CERN}, \bibinfo{address}{Geneva, Switzerland}
  (\bibinfo{year}{1994}).

\end{thebibliography}

\end{document}